%% file: paper.tex
\RequirePackage{lineno}
\documentclass[twocolumn,letterpaper,showpacs,aps,floatfix,prd]{revtex4}

\usepackage{graphicx}
\usepackage{dcolumn}
\usepackage{amsmath}
\usepackage{epsfig}

\input babarsym.tex

\newcommand{\BABARPubYear}    {10}
\newcommand{\BABARPubNumber}  {021}

\newcommand{\SLACPubNumber} {14329 }

\def\figurebox#1#2#3{%
    \def\arg{#3}%
    \ifx\arg\empty
    {\hfill\vbox{\hsize#2\hrule\hbox to #2{\vrule\hfill\vbox to #1{\hsize#2\vfill}\vrule}\hrule}\hfill}%
    \else
    {\hfill\epsfbox{#3}\hfill}%
    \fi}

\newcommand{\kst}{K^*(892)^0}
\newcommand{\akst}{\overline{K}^*(892)^0}
\newcommand{\kstp}{K^*(1410)^0}
\newcommand{\akstp}{\overline{K}^*(1410)^0}

\newcommand{\akstd}{\overline{K}^*_2(1430)^0}
\newcommand{\ksts}{K^*_0(1430)^0}

\newcommand{\ba}{\begin{array}}
\newcommand{\ea}{\end{array}}
\newcommand{\bc}{\begin{center}}
\newcommand{\ec}{\end{center}}
\newcommand{\beq}{\begin{eqnarray}}
\newcommand{\eeq}{\end{eqnarray}}
\newcommand{\bes}{\begin{eqnarray*}}
\newcommand{\ees}{\end{eqnarray*}}
\newcommand{\Zz}{\ifmmode {\rm Z} \else ${\rm Z } $ \fi}
\newcommand{\xxbar}{\ifmmode {\rm x\bar{x}} \else ${\rm x\bar{x}} $ \fi}
\newcommand{\rphi}{\ifmmode {\rm R\phi} \else ${\rm R\phi} $ \fi}

\long\def\inst#1{\par\nobreak\kern 4pt\nobreak
  {\it #1}\par\vskip 10pt plus 3pt minus 3pt}

\begin{document}
%\linenumbers
%\preprint{\hbox to\hsize{{\babar\ }Analysis Document \#2239, Version 14\hfil}}

\preprint{\babar-PUB-\BABARPubYear/\BABARPubNumber} 
\preprint{SLAC-PUB-\SLACPubNumber} 

\begin{flushleft}
%{\babar\ }Analysis Document \#2239, Version 14\\
\babar-PUB-\BABARPubYear/\BABARPubNumber\\
SLAC-PUB-\SLACPubNumber\\
%hep-ex/\LANLNumber\\[10mm]
\end{flushleft}

\title{

{\large \bf
Analysis of the {\boldmath{$\Dp \rightarrow \Km \pip \ep \nue$}} decay channel}
}

\input authors_jul2010.tex

\date{\today} 

\begin{abstract}
Using $347.5 \invfb$ of data recorded by the \babar\ detector at the PEP-II 
electron-positron collider, $244\times10^3$ signal events for the
$\Dp  \rightarrow \ \Km  \pip  \ep  \nue $ decay channel are analyzed.
This decay mode is dominated by the $\akst$ contribution. 
We determine the $\akst$ parameters:
$m_{\kst}=(895.4\pm0.2\pm0.2)$ $\mevcc$, $\Gamma^0_{\kst}=(46.5\pm0.3\pm0.2)$ $\mevcc$ and the Blatt-Weisskopf parameter
$r_{BW}=2.1\pm0.5\pm0.5$ $(\gevc)^{-1}$ where the first uncertainty comes from
statistics and the second from systematic uncertainties.
We also measure the parameters defining the corresponding hadronic form factors at $q^{2}=0$ ($r_{V}= \frac{V(0)}{A_{1}(0)}=1.463\pm0.017\pm0.031$, 
$r_{2}=\frac{A_{2}(0)}{A_{1}(0)}=0.801\pm0.020\pm0.020$) and the value of the axial-vector pole mass parameterizing 
the $q^2$ variation of $A_{1}$ and $A_{2}$:
$m_{A}=(2.63 \pm0.10\pm0.13)$ $\gevcc$. 
The $S$-wave fraction
is equal to $(5.79\pm0.16\pm0.15)\%$. Other signal components correspond to fractions below $1\%$.
Using the $\Dp  \rightarrow \Km  \pip \pip $ channel as a normalization, 
we measure the $\Dp $ semileptonic branching fraction:
$\BR(\Dp  \rightarrow \ \Km  \pip  \ep  \nue )$ = $(4.00 \pm0.03 \pm0.04 \pm0.09)\times10^{-2}$ where the third uncertainty comes from
external inputs. 
We then obtain the value of the hadronic form factor $A_1$ at $q^{2}=0$: $A_{1}(0)=0.6200\pm0.0056 \pm0.0065 \pm0.0071$.
Fixing the $P$-wave parameters we measure the phase of the $S$-wave 
for several values of the $K\pi$ mass.
These results confirm those obtained with $K\pi$ production at small 
momentum transfer in fixed target experiments.

\end{abstract}

% PACS, the Physics and Astronomy Classification Scheme.

\maketitle

%Text goes here

\section{Introduction}
\label{sec:Introduction}

Detailed study of the $\Dp  \rightarrow \ \Km  \pip  \ep  \nue $
decay channel is of interest for three main reasons:
\begin{itemize}
\item it allows measurements of the different $K\pi$ resonant and 
non-resonant amplitudes that contribute to this decay.
In this respect, we have measured the $S$-wave contribution and 
searched for radially excited $P$-wave 
and $D$-wave components. Accurate measurements of the various contributions 
can serve as useful guidelines to $B$-meson semileptonic decays where there 
are still missing exclusive final states with mass higher than the $D^*$ mass.
\item High statistics in this decay allows accurate measurements of 
the properties of the $\akst$ meson, the main contribution to the decay. Both
resonance parameters and hadronic transition form factors can be precisely
measured. The latter can be compared with hadronic model
expectations and Lattice QCD computations.
\item Variation of the $K\pi$ $S$-wave phase versus the $K\pi$ mass can be 
determined, and compared with other experimental determinations.
\end{itemize}

Meson-meson interactions are basic processes in QCD that deserve
accurate measurements. Unfortunately, meson targets do not exist
in nature and studies of these interactions usually require extrapolations
to the physical region. 

In the $K\pi$ system, $S$-wave interactions proceeding through 
isospin equal to $1/2$ states
are of particular interest
because, contrary to exotic $I=3/2$ final states, they depend on 
the presence of scalar resonances.
Studies of the candidate scalar meson $\kappa~\equiv~K^*_0(800)$ 
 can thus benefit from more accurate measurements
of the $I=1/2$ $S$-wave phase below  $m_{K\pi}=1~\gevcc$ \cite{ref:seb0}. 
The phase variation 
of this amplitude with the $K\pi$ mass also enters
in integrals which allow the determination of the strange quark mass in the QCD sum rule
approach \cite{ref:jamin1,ref:colangelo1}.

Information on the $K\pi$ $S$-wave phase in the isospin states $I=1/2$
and $I=3/2$ originates from various experimental situations, such as
kaon scattering, $D \rightarrow K\pi\pi$ Dalitz plot analyses,
and semileptonic decays of charm mesons and $\tau$ leptons. 
In kaon scattering fixed target experiments \cite{ref:easta1,ref:lass1},
measurements from LASS (Large Aperture Solenoid Spectrometer)
\cite{ref:lass1} start at $m_{K\pi}=0.825~\gevcc$, 
a value which is $0.192~\gevcc$
above threshold. Results from Ref. \cite{ref:easta1} start
at $0.7~\gevcc$ but are less accurate.
More recently, several high statistics  3-body Dalitz plot analyses 
of charm meson hadronic decays have become available
\cite{ref:e791kpipi,ref:focuskpipi,ref:focuskpipi2,ref:kpipi_cleoc}.
They provide values starting at threshold and can complement 
results from $K$ scattering, but
in the overlap region, they obtain somewhat different results. 
It is tempting to attribute these differences to the presence of 
an additional hadron
in the final state. 
The first indication in this
direction was obtained from the measurement of 
the phase difference between $S$- and $P$-waves versus $m_{K\pi}$
in $\Bzb \rightarrow \jpsi \Km \pip $ \cite{ref:babarpsikpi} which agrees with
LASS results apart from a relative sign between the two amplitudes.
In this channel, the $\jpsi$ meson in the final state is not expected
to interact with the $K\pi$ system.

In $\tau$ decays into $K \pi \nut$ there is no additional hadron in the  
final state and only the $I=1/2$ amplitude contributes.
A study of the different partial waves requires separation  
of the $\tau$ polarization components using, for instance, 
information from the decay of the
other $\tau$ lepton. No result is available yet on the phase of the 
$K\pi$ $S$-wave \cite{ref:taubelle} from these analyses. 
In $\Dp  \rightarrow \Km \pip  \ep  \nue $ there is also
no additional hadron in the final state. All needed 
information to separate the different
hadronic angular momentum components can be obtained 
through correlations between the leptonic 
and hadronic systems. This requires 
measurement of the complete dependence of the differential decay rate
on the five-dimensional phase space.
Because of limited statistics previous experiments 
\cite{ref:e687,ref:focus1,ref:cleoc1} 
have measured an $S$-wave component
but were unable to study its properties as a function of the $K\pi$ mass.
We present the first semileptonic charm decay analysis which measures
the phase of the $I=1/2$ $K \pi$ $S$-wave
as a function of $m_{K\pi}$
from threshold up to 1.5 $\gevcc$.

\begin{table}[h]
\begin{center}
  \caption {{ Possible resonances contributing to Cabibbo-favored $\Dp $ semileptonic decays \cite{ref:pdg10}.}}
  \label{tab:kpistates}
  \begin{tabular}{c c c c c }
    \hline
\hline\noalign{\vskip1pt}
resonance & $J^P$ & $\BR(X \to K\pi)$ & mass & width\\
 X  & & $\%$ & $\mevcc$ & $\mevcc$\\
\hline
$K^*_0(800)$ (?) & $0^+$ & $100(?)$ &$672\pm 40$ & $550\pm34$ \\
$K^*(892)$ & $1^-$ & $100$ &$895.94\pm 0.22$ & $48.7\pm0.8$ \\
$K_1(1270)$ & $1^+$ & $0$ &$1272\pm 7$ & $90\pm 20$ \\
$K_1(1400)$ & $1^+$ & $0$ &$1403\pm 7$ & $174\pm 13$ \\
$K^*(1410)$ & $1^-$ & $6.6 \pm 1.3$ &$1414\pm 15$ & $ 232\pm 21$ \\
$K^*_0(1430)$ & $0^+$ & $93 \pm 10$ & $1425\pm 50$ & $270 \pm 80$ \\
$K^*_2(1430)$ & $2^+$ & $49.9 \pm 1.2$ &$1432.4\pm 1.3$ & $109\pm 5$ \\
$K^*(1680)$ & $1^-$ & $38.7 \pm 2.5$ &$1717\pm 27$ & $322\pm110$ \\
\hline
\hline
  \end{tabular}
\end{center}
\end{table}

Table \ref{tab:kpistates} lists strange particle resonances that can appear in
Cabibbo-favored $\Dp $ semileptonic decays.
$J^P=1^+$ states 
do not decay into $K\pi$ and cannot be observed 
in the present analysis.
The $K^*(1410)$ is a $1^-$ radial excitation and has a small branching fraction into
$K\pi$. The  $K^*(1680)$ has a mass close to the kinematic limit and its production is disfavored
by the available phase space. Above the $K^*(892)$ one is thus left with possible contributions from 
the $K^*_0(1430)$, $K^*(1410)$ and $K^*_2(1430)$ which decay into 
$K\pi$ through $S$-, $P$- and $D$-waves, respectively.
At low $K\pi$ mass values one also expects an $S$-wave contribution which can be
resonant ($\kappa$) or not. A question mark is placed 
after the $\kappa~\equiv~K^*_0(800)$ as this state is not well established.

This paper is organized in the following way.
In Section \ref{sec:kpielast} general aspects of the $K\pi$ system in the
elastic regime, which are relevant to present measurements, are explained.
In particular the Watson theorem, 
which allows the relating of the values
of the hadronic phase measured in various processes, 
is introduced. In
Section \ref{sec:previous}, previous measurements of the $S$-wave
$K\pi$ system are explained and compared. The differential
decay distribution used to analyze the data is detailed in 
Section~\ref{sec:decayslbr}. In Section \ref{sec:detector} a short
description of the detector components which are important in this measurement
is given. The selection of signal events, the background rejection, the
tuning of the simulation and the fitting procedure are then considered 
in Section \ref{sec:analysis}. Results of a fit which includes the
$S$-wave and $\akst$ signal components are given in Section \ref{sec:LASSAMP_sec}.
Since the fit model with only $S$- and $P$-wave components does not
seem to be adequate at large $K\pi$ mass, fit results for
signal models which comprise $S+\akst+\akstp$ and $S+\akst+\akstp+D$
components are given in Section \ref{sec:SPPrime_1}. In the same section,
fixing the parameters of the $\akst$ component,
measurements of the phase difference between $S$ and $P$ waves are obtained, 
for several values of the $K\pi$ mass. In Section \ref{sec:decayrate},
measurements of the studied
semileptonic decay channel branching fraction, relative
to the $\Dp  \rightarrow \Km  \pip  \pip $ channel, and of its 
different components are obtained. This allows one to extract
an absolute normalization for the hadronic form factors.
Finally in Section \ref{sec:summary} results obtained in this analysis
are summarized.

\section{The $K\pi$ system in the elastic regime region}
\label{sec:kpielast}

The $K\pi$ scattering amplitude ($T_{K\pi}$) has two isospin components denoted $T^{1/2}$ 
and 
$T^{3/2}$. Depending on the channel studied, measurements are sensitive to 
different linear combinations of these components. In
 $\Dp  \rightarrow \Km \pip  \ep  \nue $, $\tau^- \rightarrow \KS \pim  \nut$
and $\Bzb \rightarrow \jpsi \Km \pip $ decays, only 
the $I=1/2$ component contributes. The 
$I=3/2$ component was  measured in 
$\Kp  \proton \rightarrow \Kp  \pip  \neutron$
reactions \cite{ref:easta1} whereas 
$\Km  \proton \rightarrow \Km  \pip  \neutron$ depends on
the two isospin amplitudes:
$T_{\Km \pip }=\frac{1}{3} (2T^{1/2}+T^{3/2})$. In Dalitz plot analyses 
of 3-body charm meson decays, the relative
importance of the two components has to be determined from data.

A given $K\pi$ scattering isospin amplitude can be expanded into partial waves:
\begin{equation} 
T^I(s,t,u) = 16\pi \sum_{\ell=0}^{\infty}(2\ell+1)P_{\ell}(\cos{\theta})t^I_{\ell}(s)
\end{equation}
where the normalization is such that the differential $K\pi$ scattering cross-section is equal to:
\begin{equation}
\frac{{\rm d}\sigma^I}{{\rm d}\Omega} = \frac{4}{s} \frac{\left | T^I(s,t,u)\right |^2}{(16 \pi)^2},
\end{equation}
\noindent where $s,~t$ and $u$ are the Mandelstam variables, $\theta$ is the scattering angle and $P_{\ell}(\cos{\theta})$ is the Legendre polynomial of order $\ell$. 

Close to threshold, the amplitudes $t^I_{\ell}(s)$ can be expressed as Taylor series:
\begin{eqnarray}
{\rm Re}\, t^I_{\ell}(s) = \frac{1}{2} \sqrt{s} \left ( p^*\right )^{2\ell}\left (a^I_{\ell} + b^I_{\ell}\left ( p^*\right )^{2} +{\mathcal O} \left ( p^*\right )^{4} \right ),\label{eq:defasbs}
\end{eqnarray}
where $a^I_{\ell}$  and $b^I_{\ell}$ are, respectively, the scattering length 
and the effective range
parameters, $p^*$ is the $K$ or $\pi$ momentum in the $K\pi$
center-of-mass (CM). This expansion is valid close to threshold
for $p^*<m_{\pi}$.
Values of $a^I_{\ell}$  and $b^I_{\ell}$ are
obtained from Chiral Perturbation Theory
\cite{ref:meiss1,ref:bijn1}. In Table \ref{tab:predchi} these
predictions are compared with a determination \cite{ref:seb1} of these
quantities obtained from an analysis of experimental data on
$K\pi$ scattering and $\pi\pi \rightarrow K \overline{K}$. 
Constraints from analyticity and unitarity of the amplitude are used
to obtain its behavior close to threshold. The similarity
between predicted and fitted values of $a^{1/2}_0$ and $b^{1/2}_0$
is  a non-trivial test of Chiral Perturbation Theory \cite{ref:bijn1}.

\begin{table}[h]
\begin{center}
 \caption {{ Predicted values for scattering length and effective range 
parameters.}
  \label{tab:predchi}}
  \begin{tabular}{c c c }
    \hline
\hline\noalign{\vskip1pt}
Parameter & \cite{ref:bijn1}&  \cite{ref:seb1}\\
\noalign{\vskip1pt}
\hline\noalign{\vskip1pt}
$a^{1/2}_0$ $(\gev^{-1})$& $1.52$ & $1.60 \pm 0.16$ \\
$b^{1/2}_0$ $(\gev^{-3})$& $47.0$ & $31.2\pm1.5$ \\
\noalign{\vskip1pt}
\hline\noalign{\vskip1pt}
$a^{1/2}_1$ $(\gev^{-3})$& $5.59$ & $7.0\pm 0.4$\\
\noalign{\vskip1pt}
\hline
\hline
  \end{tabular}
\end{center}
\end{table}

The complex amplitude $t^I_{\ell}(s)$ can be also expressed in terms
of its magnitude and phase. If 
the process remains elastic, this gives:
\begin{equation} 
t^I_{\ell}(s) = \frac{\sqrt{s}}{2 p^*} \frac{1}{2i}
\left ( e^{2i\delta^I_{\ell}(s)} -1\right )
=\frac{\sqrt{s}}{2 p^*} \sin{\delta^I_{\ell}(s)}e^{i\delta^I_{\ell}(s)}.
\label{eq:amplielast}
\end{equation}
Using the expansion given in Eq.~(\ref{eq:defasbs}), close to the threshold
the phase $\delta^I_{\ell}(s)$ is expected to satisfy
the following expression:

\begin{equation}
\delta^I_{\ell}(s)= (p^*)^{2l+1} \left ( \alpha + \beta \,(p^*)^2 \right ).
\label{eq:phasechiral}
\end{equation}
Using Eq.~(\ref{eq:defasbs}), (\ref{eq:amplielast}) and (\ref{eq:phasechiral})
one can relate $\alpha$ and  $\beta$ to $a^I_{\ell}$ and $b^I_{\ell}$:
\begin{equation}
\alpha = a^I_{\ell}\,{\rm and}\,\beta=b^I_{\ell}+\frac{2}{3}(a^I_{\ell})^3\delta_{l0}.
\label{eq:alphatoa}
\end{equation}
\noindent
In Eq. \ref{eq:alphatoa}, the symbol $\delta_{l0}$ is the Kronecker $\delta$
function: $\delta_{00}=1$, $\delta_{l0}=0$ for $l \neq 0$.

The Watson theorem \cite{ref:watson} implies that, in this elastic regime,
phases measured in $K\pi$ elastic scattering and in a decay channel
in which the $K\pi$ system has no strong interaction with other hadrons 
are equal modulo $\pi$ radians \cite{ref:leyaou} for the same values of
isospin and angular momentum. In this analysis, this
ambiguity is solved by determining the sign
of the $S$-wave amplitude from data. This theorem does not provide any constraint
on the corresponding amplitude moduli. In particular, it is 
not legitimate (though nonetheless frequently done) to assume 
that the $S$-wave amplitude in 
a decay is proportional to the elastic amplitude $t^I_{\ell}(s)$.
 The $K\pi$ scattering $S$-wave, $I=1/2$, remains elastic up to the $K\eta$
threshold, but since the coupling to this channel is weak \cite{ref:keta}, 
it is considered
in practice to be elastic up to the $K\eta^{\prime}$
threshold. 

Even if the $K\pi$ system is studied without any accompanying hadron,
the $S$- or $P$-waves amplitudes cannot be measured in an absolute way. 
Phase measurements
are obtained through interference between different waves. As a result,
values quoted by an experiment for the phase of the $S$-wave depend on the
parameters used to determine the $P$-wave.
For the $P$-wave, the validity domain of the Watson theorem is 
a-priori more restricted
because the coupling to $K\eta$ is no longer suppressed. 
However the  $p^{*3}$ dependence of the decay width implies that this
contribution is an order of magnitude smaller than $K\pi$ for
$m_{K\pi}<1.2~\gevcc$.
 
For pseudoscalar-meson elastic scattering at threshold all phases 
are expected to be equal to zero (see Eq. (\ref{eq:phasechiral})).
This is another
important difference as compared with Dalitz plot analyses where
arbitrary phases exist between the different contributing
waves due to interaction with the spectator hadron.
It is thus important to verify if apart from a global constant 
$S$-wave phases
measured versus $m_{K\pi}$, in 3-body $D\rightarrow K\pi\pi$ Dalitz plot 
analyses,  
depend on the presence of the
third hadron.
Comparison between present measurements and those obtained in three-body
Dalitz plot analyses are given in Section \ref{sec:MIAMP_SPP}.

\section{Previous measurements}
\label{sec:previous}
In the following sections, we describe previous measurements of the phase 
and magnitude of the
$K\pi$ $S$-wave amplitude obtained in $K^{\pm}\proton$ scattering at small 
transfer, in $\tau$ semileptonic decays, $D$ meson three-body decays,
and in charm semileptonic decays.

\subsection{$K\pi$ production at small momentum transfer}
\begin{figure}[!htb]
  \begin{center}
\includegraphics[width=8cm]{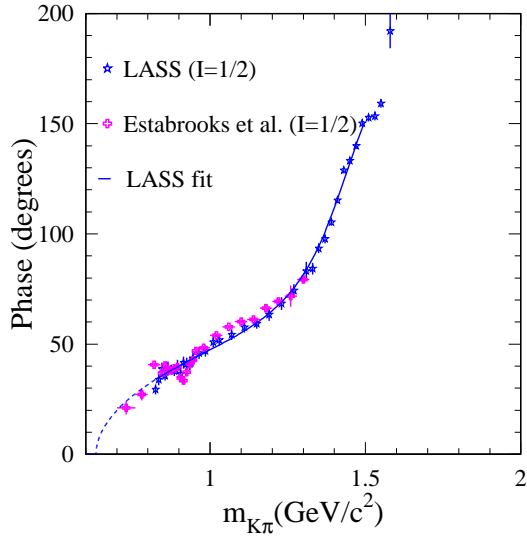}
  \end{center}
  \caption[]{{(color online) Comparison between the $I=1/2$ $S$-wave phase measured in 
$K\pi$ production at small transfer for several values of
the $K\pi$ mass. Results from Ref. \cite{ref:easta1} are limited
to $m_{K\pi}<1.3~\gevcc$ to remain in the elastic regime
where there is a single solution for the amplitude. 
The curve corresponds to the
fit given in the second column of Table \ref{tab:lassfit}.
}
   \label{fig:lasseasta}}
\end{figure}

A $K\pi$ partial wave analysis of high statistics data
for the reactions $K^{\pm}\proton \rightarrow K^{\pm}\pip \neutron$
and $K^{\pm} \proton \rightarrow K^{\pm}\pim \Delta^{++}$ at 13 $\gev$,
on events selected at small momentum transfer \cite{ref:easta1}, 
provided information on $K\pi$ scattering for 
$m_{K\pi}$ in the range $[0.7,~1.9]~\gevcc$.
The $I=3/2$ $K\pi$ scattering was studied directly from the 
analyses of  $\Kp \proton \rightarrow \Kp \pip \neutron$  and 
$\Km \proton \rightarrow \Km  \pim \Delta^{++}$ reactions.
The phase of the elastic amplitude $(\delta_S^{3/2})$
was measured and was used to extract the phase of the $I=1/2$
amplitude from measurements of $\Km \pip $ scattering.
Values obtained for $\delta_S^{1/2}$ are displayed in 
Fig.~\ref{fig:lasseasta} for 
$m_{K\pi}<1.3 ~\gevcc$, a mass range in which the interaction
is expected to remain elastic. Above $1.46~\gevcc$ there were
several solutions for the amplitude.

A few years later, the LASS  experiment analyzed
data from 11 $\gevc$ kaon scattering  on hydrogen:
$\Km  \proton \rightarrow \Km  \pip  \neutron$ \cite{ref:lass1}. 
They performed a partial wave analysis of 
$1.5\times10^5$  events which satisfied cuts 
to ensure $K\pi$ production dominated by pion exchange 
and no excitation of the target into baryon resonances.

The $K\pi$, $I=1/2$, $S$-wave was parameterized as the sum of a background 
term $(BG)$ and the 
$K^*_0(1430)$, 
which were combined such that the resulting amplitude satisfied unitarity:
\begin{eqnarray}
A_S^{1/2} &=&\sin{\delta_{BG}^{1/2}}\, e^{i\delta_{BG}^{1/2}}\\
&+&e^{2i\delta_{BG}^{1/2}}\sin{\delta_{K^*_{0}(1430)}}\, e^{i\delta_{K^*_{0}(1430)}}\nonumber \\
&=&\sin{\left ( \delta_{BG}^{1/2}+\delta_{K^*_{0}(1430)}\right )}e^{i \left ( \delta_{BG}^{1/2}+\delta_{K^*_{0}(1430)}\right )}\nonumber,
\label{eq:aslass}
\end{eqnarray}
where $\delta_{BG}^{1/2}$
and $\delta_{K^*_{0}(1430)}$ depended on the $K\pi$ mass.

The mass dependence of $\delta_{BG}^{1/2}$ was described by means of an effective range parameterization:
\begin{equation}
\cot{(\delta_{BG}^{1/2})}=\frac{1}{a_{S,BG}^{1/2}~p^*}+\frac{b_{S,BG}^{1/2}~p^*}{2},
\label{eq:phasebg}
\end{equation}
where $a_{S,BG}^{1/2}$ is the scattering length and $b_{S,BG}^{1/2}$ is the effective range. 
Note that these two parameters are different from $a_{\ell}^I$ and $\b_{\ell}^I$
introduced in Eq.~(\ref{eq:defasbs}) as the latter referred to the total amplitude
and also because Eq.~(\ref{eq:phasebg}) corresponds to an 
expansion near threshold which differs from Eq.~(\ref{eq:phasechiral}).
The mass dependence of $\delta_{K^*_{0}(1430)}$ was obtained assuming that
the $K^*_{0}(1430)$ decay amplitude obeys a Breit-Wigner distribution:

\begin{equation}
\cot{(\delta_{K^*_{0}(1430)} )}=\frac{m_{K^*_{0}(1430)}^2-m_{K\pi}^2}{m_{K^*_{0}(1430)}  \Gamma_{K^*_{0}(1430)}(m_{K\pi})},
\label{eq:phasek1430}
\end{equation}
where $m_{K^*_{0}(1430)}$ is the pole mass of the resonance and 
$\Gamma_{K^*_{0}(1430)}(m_{K\pi})$ its mass-dependent total width.

The total $I=1/2$ $S$-wave phase was then:
\begin{equation}
\delta_{LASS}^{1/2}=\delta_{BG}^{1/2}+\delta_{K^*_{0}(1430)}
\label{eq:phaselass}
\end{equation}

The LASS measurements were based on fits to moments of angular distributions which depended on the interference 
between $S$-, $P$-, $D$-...waves. 
To obtain the $I=1/2$ $\Km \pip $ $S$-wave amplitude, 
the measured $I=3/2$ component \cite{ref:easta1} was subtracted from the 
LASS measurement of $T_{\Km \pip }$ and the resulting values
were fitted using Eq.~(\ref{eq:phaselass}).
The corresponding results \cite{ref:Dunwoodie}
are given in Table \ref{tab:lassfit} and displayed 
in Fig.~\ref{fig:lasseasta}.

\begin{table}[h]
\begin{center}
  \caption {{ Fit results to LASS data \cite{ref:Dunwoodie}
for two mass intervals.}
  \label{tab:lassfit}}
  \begin{tabular}{c c c }
    \hline
\hline\noalign{\vskip1pt}
Parameter & $m_{K\pi}\in [0.825,~1.52]$ & $m_{K\pi}\in [0.825,~1.60]$\\
           & $ \gevcc$  & $ \gevcc$ \\
\noalign{\vskip1pt}
\hline\noalign{\vskip1pt}
$m_{K^*_0(1430)}~(\mevcc)$ & $1435\pm5$ & $1415\pm3$ \\
$\Gamma_{K^*_0(1430)}~(\mevcc)$ & $279\pm6$ & $300\pm6$ \\
$a_{S,BG}^{1/2}~(\gev^{-1})$ & $1.95\pm0.09$ & $2.07\pm0.10$ \\
$b_{S,BG}^{1/2}~(\gev^{-1})$ & $1.76\pm0.36$ & $3.32\pm0.34$ \\
\noalign{\vskip1pt}
\hline
\hline
  \end{tabular}
\end{center}
\end{table}

\subsection{$\taum \rightarrow K\pi \nut$ decays}

The \babar\ and Belle collaborations \cite{ref:taubabar, ref:taubelle} 
measured the $\KS\pi$ 
mass distribution
in $\taum \rightarrow \KS\pim  \nut$.
Results from Belle were analyzed in Ref. \cite{ref:taupich1} 
using, in addition to the $K^*(892)$:
\begin{itemize}
\item a contribution from the $K^*(1410)$ to the vector form factor;

\item a scalar contribution, with a mass dependence compatible with LASS
measurements but whose branching fraction was not provided.
\end{itemize}
Another interpretation of these data was given in Ref.~\cite{ref:mouss2}.
Using the value of the rate determined from Belle data,
for the $K^*(1410)$, its relative contribution to the 
$\Dp  \rightarrow \Km  \pip  \ep  \nue $ channel was evaluated 
to be of the order of $0.5\%$.

\subsection{Hadronic $D$ meson decays}
$K\pi$ interactions were studied in several Dalitz plot analyses of
three-body $D$ decays
and we consider only 
$\Dp  \rightarrow \Km  \pip  \pip $  as measured by the E791 \cite{ref:e791kpipi}, FOCUS \cite{ref:focuskpipi,ref:focuskpipi2}, 
and CLEO-c \cite{ref:kpipi_cleoc} collaborations.
This final state is known to have a large $S$-wave component because 
there is no resonant contribution to the 
$\pip \pip $ system. 
In practice each collaboration has developed various approaches and results
are difficult to compare.

\begin{figure*}[!htb]
  \begin{center}
    \mbox{\epsfig{file=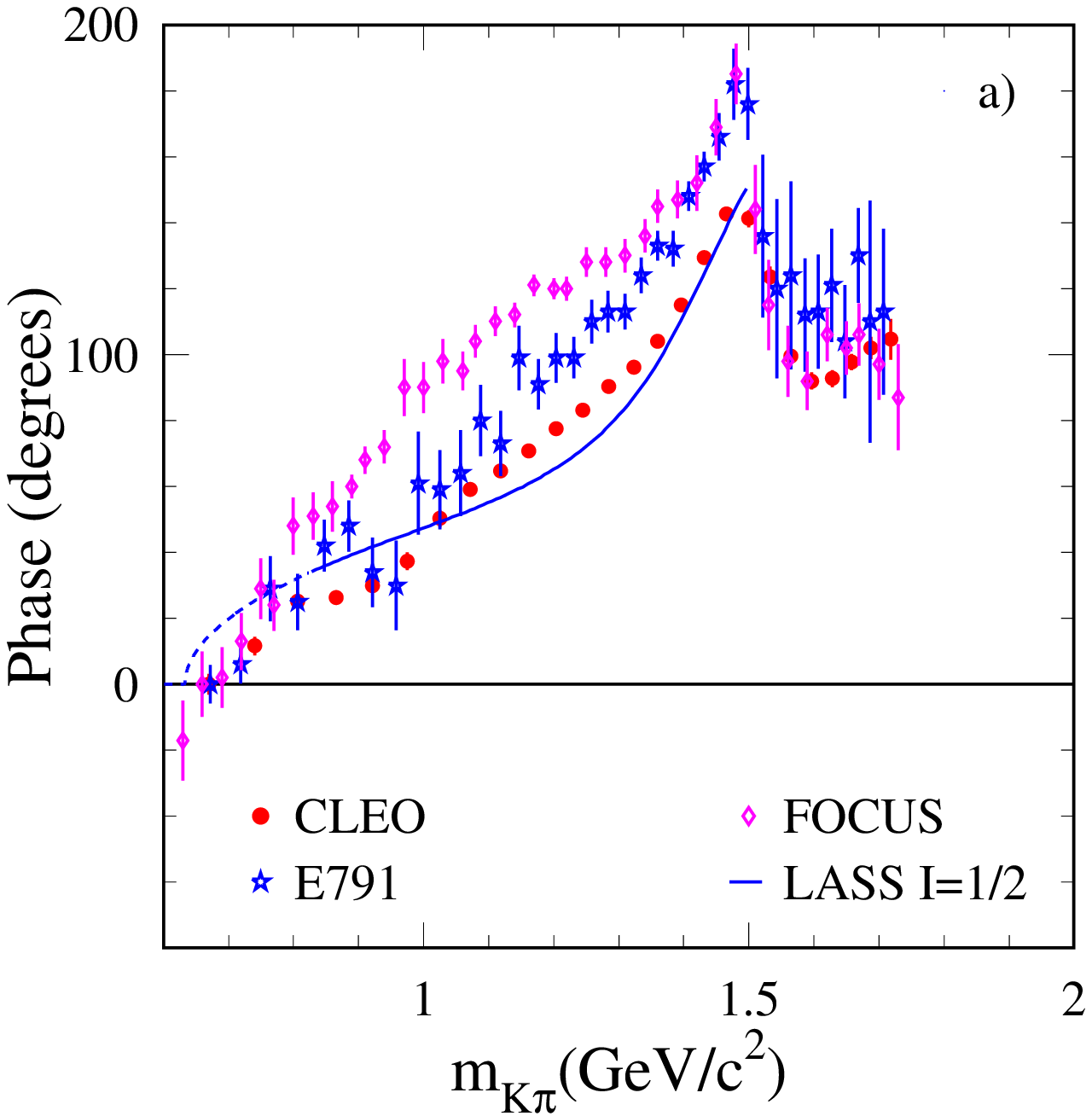,width=.5\textwidth}
\epsfig{file=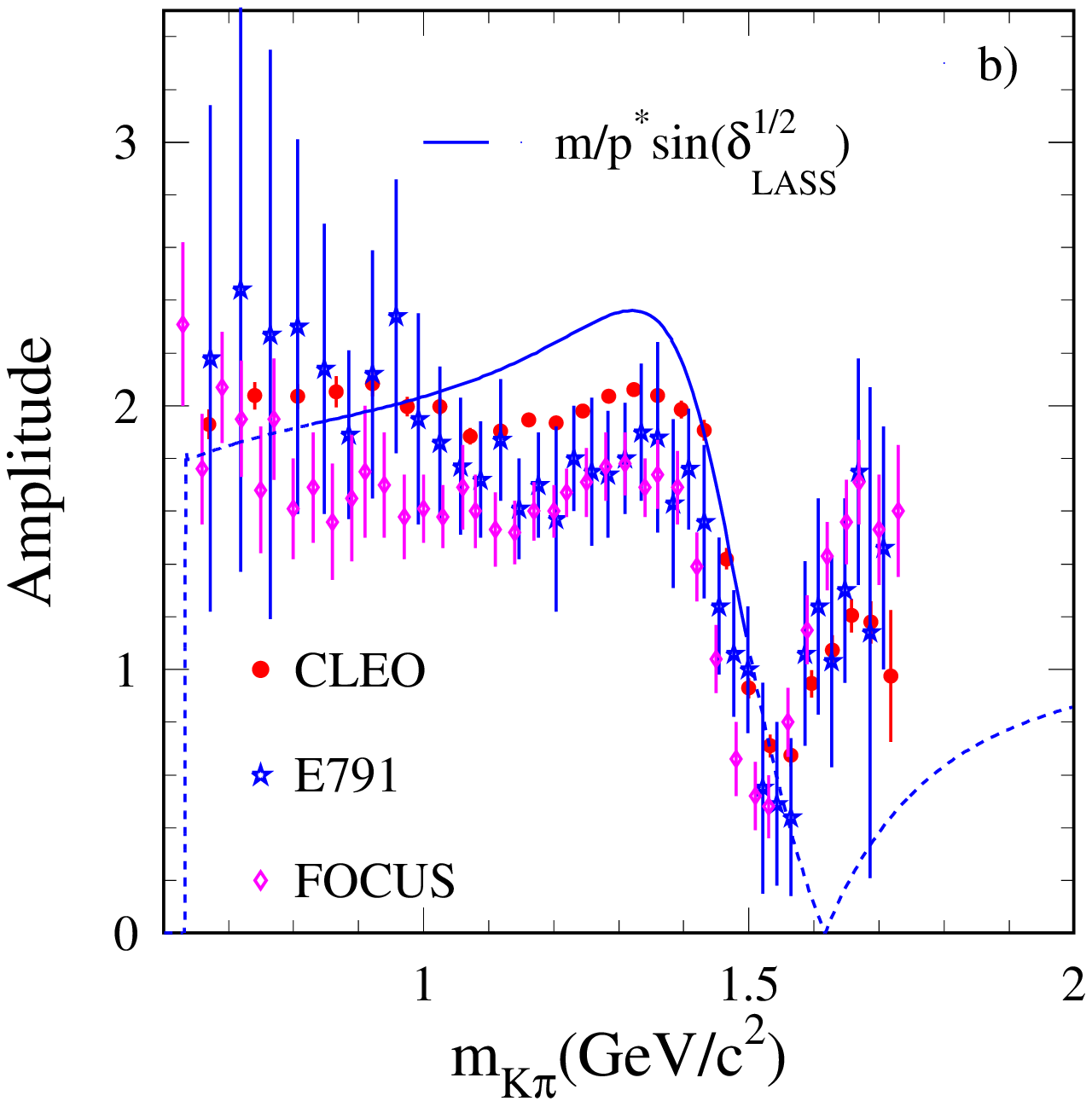,width=.5\textwidth}
}
  \end{center}
  \caption[]{{(color online) a) Comparison between the $S$-wave phase measured in
various experiments analyzing the $\Dp  \rightarrow \Km  \pip  \pip $ 
channel (E791 \cite{ref:e791kpipi}, 
FOCUS \cite{ref:focuskpipi,ref:focuskpipi2} 
and CLEO \cite{ref:kpipi_cleoc}) and a fit to LASS data (continuous line). The dashed line 
corresponds to the extrapolation of the fitted curve. Phase measurements
from $\Dp $ decays are shifted to be equal to zero at 
$m_{K\pi}=0.67~\gevcc$. b) The $S$-wave amplitude magnitude measured in
various experiments
is compared with the elastic expression. 
Normalization is arbitrary between the various distributions. 
}
   \label{fig:e791}}
\end{figure*}

The $S$-wave phase measured by these collaborations is
compared in Fig.~\ref{fig:e791}-a with the phase of
the ($I=1/2$) amplitude determined from LASS data. Measurements
from $\Dp $ decays are shifted so that the phase is equal
to zero for $m_{K\pi}=0.67~\gevcc$.
The magnitude of the amplitude obtained 
in Dalitz plot analyses is compared 
in Fig.~\ref{fig:e791}-b 
with the ``naive'' estimate 
given in Eq.~(\ref{eq:amplielast}), which is derived
from the elastic ($I=1/2$) amplitude fitted to LASS data.

By comparing results obtained by the three experiments analyzing
$\Dp \rightarrow \Km \pip \pip $, several remarks are formulated.
\begin{itemize}
\item A $\pip \pip $ component is included only in the CLEO-c
measurement and it corresponds to $(15\pm 3)\%$ of the decay rate.
\item  The relative importance of $I=1/2$
and $I=3/2$ components can be different in $K\pi$ scattering and
in a three-body decay. This is because, even if Watson's theorem
is expected to be valid, it applies separately for the
$I=1/2$ and $I=3/2$ components and concerns only  the corresponding
phases of these amplitudes.
In E791 and CLEO-c they measured the total $K\pi$
$S$-wave amplitude and compared their results with the $I=1/2$
component from LASS. FOCUS \cite{ref:focuskpipi}, using the phase 
of the $I=3/2$ amplitude
measured in scattering experiments, had fitted separately the two
components and found large effects from the $I=3/2$ part.
In Fig.~\ref{fig:e791}-a the phase of the total $S$-wave amplitude
which contains contributions from the two isospin components,
as measured by FOCUS \cite{ref:focuskpipi2},
is plotted.  
\item Measured phases in Dalitz plot analyses have a global shift as compared
to the scattering case (in which phases are expected to be zero at 
threshold). Having corrected for this effect (with some arbitrariness),
the variation measured for the phase in three-body decays and in $K\pi$
scattering is roughly similar, but quantitative comparison is difficult.
Differences between the two approaches 
as a function of $m_{K\pi}$
are much larger than the quoted uncertainties. 
They may arise from the comparison 
itself, which considers the total $K\pi$ $S$-wave in one case and only the 
$I=1/2$ component for scattering. They could be due also to the interaction
of the bachelor pion which invalidates the application of the 
Watson theorem.
\end{itemize}

It is thus difficult to draw quantitative conclusions from results
obtained with $\Dp \rightarrow \Km \pip \pip $ decays. Qualitatively,
one can say that the 
phase of the $S$-wave component depends roughly similarly on 
$m_{K\pi}$ as the phase measured by LASS.
Below the $K^*_0(1430)$, the 
$S$-wave amplitude magnitude has a smooth variation versus $m_{K\pi}$.
At the $K^*_0(1430)$ average mass value and above, this magnitude has a sharp
decrease with the mass.

\subsection{$D_{\ell 4}$ decays}

The dominant hadronic contribution in the $D_{\ell 4}$
%\footnote{$D_{\ell 4}$ stands for a semileptonic decay of a D meson with 2 pseudo-scalar mesons in the final state.} 
decay channel comes from the 
($J^{P}$ =$1^{-}$) $K^{*}(892)$ resonant state. E687 \cite{ref:e687}
gave the first suggestion for an additional component.
FOCUS \cite{ref:focus1}, a few years later, measured the $S$-wave contribution
from the asymmetry in the angular distribution of the $K$ in 
the $K\pi$ rest frame. They concluded that the phase
difference between $S$- and $P$-waves was compatible with a constant equal
to $\pi/4$, over the $K^*(892)$ mass region. 

In the second publication \cite{ref:focus2} they
found that the asymmetry could be explained if they used 
the variation of the $S$-wave component
versus the $K\pi$ mass 
 measured by the LASS collaboration \cite{ref:lass1}.
 They did not fit to their data the two parameters that governed 
this phase variation but
took LASS results: 
\begin{eqnarray}
\cot{(\delta_{BG})}&=&\frac{1}{a_{S,BG}~p^*}+\frac{b_{S,BG}~p^*}{2},\\
a_{S,BG}&=&(4.03\pm1.72\pm0.06)~\gev^{-1}, \nonumber \\
b_{S,BG}&=&(1.29\pm0.63\pm 0.67) ~\gev^{-1}.\nonumber
\label{eq:lass1}
\end{eqnarray}

These values corresponded to the total $S$-wave amplitude 
measured by LASS which
was the sum of $I=1/2$ and $I=3/2$ contributions whereas only the former
component was present in charm semileptonic decays.
For the $S$-wave amplitude they assumed that it was proportional to
the elastic amplitude (see Eq.~(\ref{eq:amplielast})).
For the $P$-wave, they used a relativistic Breit-Wigner with mass 
dependent width
\cite{ref:angles}.
They fitted the values of the pole mass, the width and the 
Blatt-Weisskopf damping
parameter for the $K^*(892)$. 
These values from FOCUS are given in Table \ref{tab:kstar1}
and compared with present world averages \cite{ref:pdg10}.
, dominated
by the $P$-wave measurements from LASS.

\begin{table}
\begin{center}
  \caption {{ Parameters of the $K^{*}(892)^0$ measured by FOCUS are compared
with world average or previous values.}
  \label{tab:kstar1}}
  \begin{tabular}{c c c }
    \hline \hline\noalign{\vskip1pt}
Parameter & FOCUS results \cite{ref:focus2} & previous results\\
\noalign{\vskip1pt}
\hline \noalign{\vskip1pt}
$m_{K^{*0}}~(\mevcc)$ & $895.41\pm0.32^{+0.35}_{-0.43}$ & $895.94\pm0.22$ \cite{ref:pdg10} \\
\noalign{\vskip1pt}
$\Gamma_{K^{*0}}^0~(\mevcc)$ & $~47.79\pm0.86^{+1.32}_{-1.06}$ & $~48.7\pm0.8$ \cite{ref:pdg10} \\
\noalign{\vskip1pt}
$r_{BW}~(\gevc)^{-1}$ & $~~3.96\pm0.54^{+1.31}_{-0.90}$ & $~3.40\pm0.67$ \cite{ref:lass1} \\
\noalign{\vskip1pt}
\hline\hline
  \end{tabular}
\end{center}
\end{table}

They also compared the measured angular asymmetry of the $K$ in the $K\pi$ rest frame versus the $K\pi$ mass with
expectations from a $\kappa$ resonance and conclude that the presence of a $\kappa$ could
be neglected. 
They used a Breit-Wigner distribution for the $\kappa$ amplitude using
values measured by the E791 collaboration \cite{ref:e791kappa} for the mass and width of this resonance
($m_{\kappa}=797\pm19\pm43~\mevcc,~\Gamma_{\kappa}=410\pm43\pm87~\mevcc$).
This approach to search for a $\kappa$ does not seem to be appropriate.
Adding a $\kappa$ in this way violates the Watson theorem as the phase  
of the fitted $K\pi$ amplitude would differ greatly from the one measured by LASS.
In addition, the interpretation of LASS measurements in Ref. \cite{ref:seb1}
concluded there was evidence for a $\kappa$. 
In addition to the $K^*(892)$ they measured the rate
for the non-resonant $S$-wave contribution and placed 
limits on other components
(Table \ref{tab:ratesfocus}).

\begin{table}[htb!]
\begin{center}
  \caption {{Measured fraction of the non-resonant $S$-wave component
and limits on contributions from $K^*_0(1430)$ and $K^*(1680)$
in the decay $\Dp \rightarrow \Km  \pip  \mu^+\nu_{\mu}$, obtained by FOCUS
\cite{ref:focus2}.}
  \label{tab:ratesfocus}}
  \begin{tabular}{ c c }
    \hline \hline\noalign{\vskip1pt}
Channel & FOCUS \cite{ref:focus2} ($\%$)\\
\noalign{\vskip1pt}
\hline\noalign{\vskip1pt}
$\frac{\Gamma(\Dp \rightarrow \Km  \pip  \mu^+\nu_{\mu})_{NR}}{\Gamma(\Dp \rightarrow \Km  \pip  \mu^+\nu_{\mu})}$ & $5.30 \pm0.74 ^{+0.99}_{-0.96}$\\
\noalign{\vskip1pt}
\hline\noalign{\vskip1pt}
$\frac{\Gamma(\Dp \rightarrow \Km  \pip  \mu^+\nu_{\mu})_{K^*_0(1430)}}{\Gamma(\Dp \rightarrow \Km  \pip  \mu^+\nu_{\mu})}$ & $<0.64\%~{\rm at}~90\%$ C.L.\\
\noalign{\vskip1pt}
\hline\noalign{\vskip1pt}
$\frac{\Gamma(\Dp \rightarrow \Km  \pip  \mu^+\nu_{\mu})_{K^*(1680)}}{\Gamma(\Dp \rightarrow \Km  \pip  \mu^+\nu_{\mu})}$ & $<4.0\%~{\rm at}~90\%$ C.L.\\
\noalign{\vskip1pt}
\hline \hline
  \end{tabular}
\end{center}
\end{table}

Analyzing $\Dp  \rightarrow \Km  \pip  \ep  \nue $ events from a sample
corresponding to $281~{\rm pb}^{-1}$ integrated luminosity, the CLEO-c collaboration 
had confirmed the FOCUS result for the $S$-wave contribution. They did
not provide an independent measurement of the $S$-wave phase \cite{ref:cleoc1}.

\section{$\Dp  \rightarrow \Km  \pip  \ep  \nue $ decay rate formalism}
\label{sec:decayslbr}

The invariant matrix element for the $\Dp  \rightarrow \Km  \pip  \ep  \nue $
semileptonic decay is the product of a hadronic and a leptonic current.
\begin{eqnarray}
{\cal M}_{fi} &=& \frac{G_F}{\sqrt{2}}\Vcs \left < \pi(p_{\pip })K(p_{\Km })|
\overline{s}\g_{\mu}(1-\g_5)c|D(p_{\Dp })\right > \nonumber\\
& &\times \overline{u}(p_{\nue })\g_{\mu}(1-\g_5)v(p_{\ep }). 
\end{eqnarray}
In this expression, $p_{\Km },~p_{\pip },~p_{\ep }~{\rm and}~p_{\nue }$ are 
the $\Km ,~\pip ,~\ep $, and $\nue $ four-momenta, respectively.

The leptonic current corresponds to the 
virtual $W^+$ which decays into $\ep \nue $. 
The matrix element of the hadronic current can be written in terms 
of four form factors, but
neglecting the electron mass, only three are contributing to the
decay rate: $h$ and $w_{\pm}$. 
Using the conventions of Ref. \cite{ref:wise1}, 
the vector and axial-vector components are, respectively:
\begin{eqnarray}
\left < \pi(p_{\pip })K(p_{\Km })|\overline{s}\g_{\mu}c|D(p_{\Dp })\right >~~~~~~~~~~~~~~~~~~~~~~ \nonumber\\
=h\epsilon_{\mu\alpha\beta\g}p_{\Dp }^{\alpha}\left(p_{\Km }+p_{\pip }\right )^{\beta}\left(p_{\Km }-p_{\pip }\right )^{\g};\label{eq:ff1}\\
\left < \pi(p_{\pip })K(p_{\Km })|\overline{s}\g_{\mu}(-\g_5)c|D(p_{\Dp })\right >~~~~~~~~~~~~~~~ \nonumber\\
=iw_+\left(p_{\Km }+p_{\pip }\right )_{\mu}+iw_-\left(p_{\Km }-p_{\pip }\right )_{\mu}.~\label{eq:ff2}
\end{eqnarray}  

As there are 4 particles in the final state, 
the differential decay rate has five degrees of freedom
that can be expressed in the following variables
\cite{ref:cab1,ref:pais1}:
\begin{itemize}
\item $m^2$, the mass squared of the $K\pi$ system;
\item $q^2$, the mass squared of the $\ep \nue $ system;
\item $\cos{(\theta_K)}$, where $\theta_K$ is the angle between  the 
$K$ three-momentum in the $K\pi$ rest frame and the line of flight
of the $K\pi$ in the $D$ rest frame;
\item $\cos{(\theta_e)}$, where $\theta_e$ is the angle between the 
charged lepton three-momentum  in the $e\nue $ rest frame and the line of flight
of the $e\nue $ in the $D$ rest frame; 
\item $\chi$, the angle between the normals to the planes defined in the $D$ 
rest frame by the $K\pi$ pair and the $e\nue $ pair. 
$\chi$ is defined between $-\pi$ and $+\pi$.
\end{itemize}
\begin{figure}
	\centering
		\includegraphics[width=9cm]{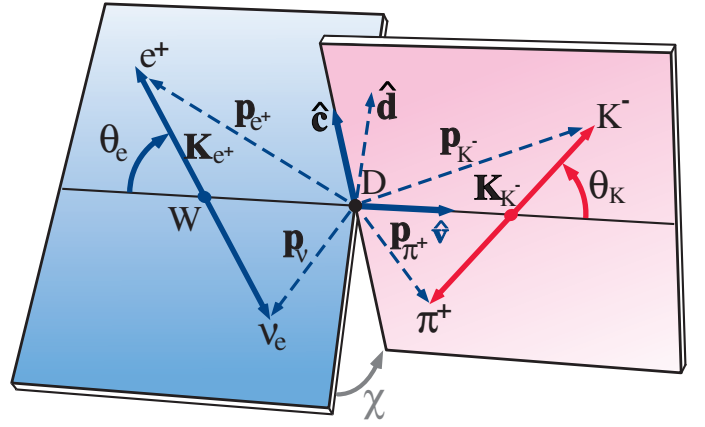}
	\caption{{(color online) Definition of angular variables.}
	\label{fig:angles}}
\end{figure}

The angular variables are shown in Fig.~\ref{fig:angles}, 
where $\mbox{\boldmath{$K$}}_{\Km }$ is the $\Km $ three-momentum in the $K\pi$ CM and $\mbox{\boldmath{$K$}}_{\ep }$ is the 
three-momentum of the positron in the virtual $W$ CM. 
Let $\hat{\mbox{\boldmath{$v$}}}$ be the unit vector along the $K\pi$ direction in the $D$ rest frame,
$\hat{\mbox{\boldmath{$c$}}}$ 
the unit vector along the projection of $\mbox{\boldmath{$K$}}_{\Km }$ perpendicular 
to $\hat{\mbox{\boldmath{$v$}}}$, and $\hat{\mbox{\boldmath{$d$}}}$ the unit vector along the projection of 
$\mbox{\boldmath{$K$}}_{\ep }$ perpendicular to $\hat{\mbox{\boldmath{$v$}}}$.
We have:
\begin{eqnarray}
m^{2}=(p_{\pip }+p_{\Km })^2,~  q^2=(p_{\ep }+p_{\nue })^2, \\
\cos{(\theta_{K})}=\frac{\hat{\mbox{\boldmath{$v$}}}\cdot\mbox{\boldmath{$K$}}_{\Km }}{|\mbox{\boldmath{$K$}}_{\Km }|},~\cos{(\theta_{e})}=-\frac{\hat{\mbox{\boldmath{$v$}}}\cdot\mbox{\boldmath{$K$}}_{\ep }}{|\mbox{\boldmath{$K$}}_{\ep }|},\nonumber \\
\cos{(\chi)}=\hat{\mbox{\boldmath{$c$}}}\cdot\hat{\mbox{\boldmath{$d$}}}; \; \;\sin{(\chi)}=(\hat{\mbox{\boldmath{$c$}}}\times\hat{\mbox{\boldmath{$v$}}})\cdot\hat{\mbox{\boldmath{$d$}}} \nonumber
\end{eqnarray}

The definition of $\chi$ is the same as proposed initially in 
Ref.~\cite{ref:cab1}.
When analyzing $\Dm $ decays, the sign of
$\chi$ has to be changed. This is because, if $\CP$ invariance is assumed
with the adopted definitions, $\chi$ changes sign through $\CP$ 
transformation of the final state \cite{ref:focus1}. 

For the differential decay partial width, 
we use the formalism given in Ref.~\cite{ref:wise1}, which generalizes 
to five variables
the decay rate given in Ref.~\cite{ref:rich1} in terms of 
$q^{2},~ \cos\theta_{K},~ \cos\theta_{e}$ and $\chi$ variables.
In addition, it provides a partial wave decomposition for the hadronic system.
Any dependence on the lepton mass is neglected as only electrons
or positrons are used in this analysis:
\begin{eqnarray}
{\rm d}^5\Gamma=\frac{G_F^2 \left | \Vcs \right |^2} 
{\left ( 4\pi\right )^6 m_D^3} X\beta {\cal I}(m^2,q^2,\theta_K,\theta_e,\chi) \nonumber \\
{\rm d}m^2 {\rm d}q^2 {\rm d}\cos{(\theta_K)} {\rm d}\cos{(\theta_e)}{\rm d}\chi.
\label{eq:decay1}
\end{eqnarray}
In this expression, 
$X=p_{K\pi}\, m_D$ where $p_{K\pi}$ is the momentum
of the $K\pi$ system in the $D$ rest frame, and $\beta=2 p^*/m$.
$p^*$ is the breakup momentum of the $K\pi$ system in its rest frame.
The form factors $h$ and $w_{\pm}$, introduced in 
Eq.~(\ref{eq:ff1}-\ref{eq:ff2}), 
are functions of $m^2$, $q^2$ and
$\cos{\theta_K}$.
In place of these form factors and to simplify the notations,
the quantities ${\cal F}_{1,2,3}$ are defined \cite{ref:wise1}:
\begin{eqnarray}
{\cal F}_1 &=& Xw_+ + \left [\beta (p_{\Km }+p_{\pip })(p_{\ep }+p_{\nue })\cos{\theta_K}\right .\nonumber\\
& +&\left .\frac{m_K^2-m_{\pi}^2}{m^2}X \right ]w_-,\nonumber\\
{\cal F}_2&=& \beta\, q\, m \,w_-,\\
{\cal F}_3 &=& \beta\, X\, q\, m \,h.\nonumber
\end{eqnarray}
The dependence of ${\cal I}$ on $\theta_e$ and $\chi$ is given by:
\begin{eqnarray}
{\cal I} &=& {\cal I}_1 +{\cal I}_2 \cos{2\theta_e}+{\cal I}_3\sin^2{\theta_e}\cos{2\chi}\\
& & +{\cal I}_4\sin{2\theta_e}\cos{\chi}+{\cal I}_5\sin{\theta_e}\cos{\chi}\nonumber \\
& &+{\cal I}_6\cos{\theta_e}+ {\cal I}_7\sin{\theta_e}\sin{\chi}\nonumber \\
& & +{\cal I}_8\sin{2\theta_e}\sin{\chi}+{\cal I}_9\sin^2{\theta_e}\sin{2\chi} \nonumber
\end{eqnarray}
where ${\cal I}_{1,...,9}$ depend on $m^2,~q^2$ and $\theta_{K}$. 
These quantities can be expressed in terms of the three form factors,
${\cal F}_{1,2,3}$.

\begin{eqnarray}
{\cal I}_1 &=&\frac{1}{4}\left \{ |{\cal F}_1|^2 
+\frac{3}{2}\sin^2{\theta_{K}} \left(|{\cal F}_2|^2 +|{\cal F}_3|^2 \right) \right \} \\
{\cal I}_2 &=&-\frac{1}{4}\left \{ |{\cal F}_1|^2 
-\frac{1}{2}\sin^2{\theta_{K}} \left(|{\cal F}_2|^2 +|{\cal F}_3|^2 \right) \right \} \nonumber \\
{\cal I}_3 &=&-\frac{1}{4}\left \{ |{\cal F}_2|^2 
-|{\cal F}_3|^2 \right \}\sin^2{\theta_{K}}  \nonumber \\
{\cal I}_4 &=&\frac{1}{2}{\rm Re} \left ({\cal F}_1^*{\cal F}_2 \right) \sin{\theta_{K}} \nonumber \\ 
{\cal I}_5 &=&{\rm Re} \left ({\cal F}_1^*{\cal F}_3 \right) \sin{\theta_{K}}  \nonumber \\
{\cal I}_6 &=&{\rm Re} \left ({\cal F}_2^*{\cal F}_3 \right) \sin^2{\theta_{K}} \nonumber \\ 
{\cal I}_7 &=&{\rm Im} \left ({\cal F}_1{\cal F}_2^* \right) \sin{\theta_{K}} \nonumber \\ 
{\cal I}_8 &=&\frac{1}{2}{\rm Im} \left ({\cal F}_1{\cal F}_3^* \right) \sin{\theta_{K}}  \nonumber\\ 
{\cal I}_9 &=&-\frac{1}{2}{\rm Im} \left ({\cal F}_2{\cal F}_3^* \right) \sin^2{\theta_{K}}  \nonumber
\end{eqnarray}

Form factors ${\cal F}_{1,2,3}$ can be expanded into partial waves
to show their explicit dependence on $\theta_{K}$. If only
$S$-, $P$- and $D$-waves are kept, this gives:

\begin{eqnarray}
{\cal F}_1 &=& {\cal F}_{10} + {\cal F}_{11} \cos{\theta_{K}}
+ {\cal F}_{12}\frac{3\cos^2{\theta_{K}}-1}{2};\nonumber\\
{\cal F}_2 &=& \frac{1}{\sqrt{2}}{\cal F}_{21} 
+\sqrt{\frac{3}{2}}{\cal F}_{22} \cos{\theta_{K}}; \\
{\cal F}_3 &=& \frac{1}{\sqrt{2}}{\cal F}_{31} 
+\sqrt{\frac{3}{2}}{\cal F}_{32} \cos{\theta_{K}}.\nonumber 
\label{eq:f123}
\end{eqnarray}
Form factors ${\cal F}_{ij}$ depend on $m^2$ and $q^2$.
${\cal F}_{10}$ characterizes the $S$-wave contribution whereas 
${\cal F}_{i1}$ 
and ${\cal F}_{i2}$ correspond to  the $P$- and $D$-wave, respectively.

\subsection{$P$-wave form factors}

By comparing expressions given in Ref. \cite{ref:wise1} and 
\cite{ref:rich1} it is possible to relate  ${\cal F}_{i1},~i=1,2,3$ 
with the helicity form factors $H_{0,\pm}$:
\begin{eqnarray}
{\cal F}_{11}  &=&  2 \sqrt{2}\alpha\, q \,H_0\nonumber\\
{\cal F}_{21}  &=& 2 \alpha \, q \left (H_+ + H_- \right )\label{eq:Fdeux} \\
{\cal F}_{31}  &=& 2 \alpha \, q \left (H_+ - H_- \right ) \nonumber
\end{eqnarray}
where $\alpha$ is a constant factor, its value is given in Eq.~(\ref{eq:alphadef}); 
it depends on the definition adopted for the mass distribution.
The helicity amplitudes can in turn be related to the two
axial-vector form factors $A_{1,2}(q^2)$, and to the vector form factor
$V(q^2)$:

\begin{eqnarray}
 H_0(q^2) &=&\frac{1}{2 m\, q} \left[ \left (
m_D^2-m^2-q^2 \right ) \left ( m_D+m \right ) A_1(q^2)\right .\nonumber\\ 
 &-& \left . 4 \frac{m_D^2\,p_{K\pi}^2}{m_D+m} A_2(q^2) \right ] \\
 H_{\pm}(q^2)& =&\left (m_D + m \right ) A_1(q^2)
\mp \frac{2m_D\, p_{K\pi}}{m_D+m} V(q^2).\nonumber
\label{eq:FF}
\end{eqnarray}

As we are considering resonances which have an extended
mass distribution, form factors 
can also have a mass dependence. We have assumed that the 
$q^2$ and $m$ dependence can be factorized: 
\begin{equation} 
(V,A_1,A_2)(q^2,m) = (V,A_1,A_2)(q^2)\times {\cal A}(m)
\label{eq:VAff}
\end{equation} 
where in case of a resonance ${\cal A}(m)$ is 
assumed to behave according to a Breit-Wigner distribution.

This factorized expression can be justified by the fact that the $q^2$ 
dependence of the form
factors is expected to be determined by the singularities 
which are nearest to the physical region:
$q^2\in[0,~q^2_{max}]$. These singularities are poles
or cuts situated at (or above) hadron masses $M_H \simeq 2.1$-$2.5~\gevcc$, depending on the form factor.
Because the $q^2$ variation range is limited
 to $q^2\simeq 1 ~\gev^2$, the proposed approach is equivalent
to an expansion in $q^2/M_H^2<0.2$.

For the $q^2$ dependence we use a single pole parameterization
and try to determine the effective pole mass.
\begin{eqnarray}
V(q^2)&=&\frac{V(0)}{1-\frac{q^2}{m_{V}^2}} \nonumber\\
A_1(q^2)&=&\frac{A_1(0)}{1-\frac{q^2}{m_A^2}} \label{eq:ffdef}\\
A_2(q^2)&=&\frac{A_2(0)}{1-\frac{q^2}{m_A^2}} \nonumber
\end{eqnarray}
where $m_V$ and $m_A$ are expected to be close to $m_{D^*_s}\simeq 2.1~\gevcc$ 
and  $m_{D_{s1}}\simeq 2.5~\gevcc$ respectively. Other parameterizations
involving a double pole in $V$ have been proposed 
\cite{ref:doublepole}, but as the present
analysis is not sensitive to $m_V$, the single pole ansatz is adequate.

Ratios of these form factors, evaluated at $q^{2}=0$, 
$r_{V}=\frac{V(0)}{A_{1}(0)}$ 
and $r_{2}=\frac{A_{2}(0)}{A_{1}(0)}$, are measured 
by studying the variation of the differential 
decay rate versus the kinematic variables. 
The value of $A_{1}(0)$ is determined by measuring 
the $\Dp \rightarrow \overline{K}^{*0}  \ep  \nue $
branching fraction.
For the mass dependence, in case
of the $K^*(892)$,  we use a Breit-Wigner distribution:
\begin{equation}
{\cal A}_{K^*(892)}=\frac{ m_{K^*(892)}\Gamma_{K^*(892)}^0 F_1(m) }{m_{K^*(892)}^2 -m^2 - im_{K^*(892)}\Gamma_{K^*(892)}(m) }.
\label{eq:kstar2}
\end{equation}
In this expression:
\begin{itemize}
\item $m_{K^*(892)}$ is the $K^*(892)$ pole mass;
\item $\Gamma_{K^*(892)}^0$ is the total width of the $K^*(892)$ for $m=m_{K^*(892)}$;
\item $\Gamma_{K^*(892)}(m)$ is the mass-dependent $K^*(892)$ width:
$\Gamma_{K^*(892)}(m)=\Gamma_{K^*(892)}^0\frac{p^*}{p^*_0}\frac{m_{K^*(892)}}{m}F^2_1(m)$;
\item $F_1(m)=\frac{p^*}{p^*_0}\frac{B(p^*)}{B(p^*_0)}$ where $B$ is
the Blatt-Weisskopf damping factor: $B=1/\sqrt{1+r_{BW}^2p^{*2}}$, $r_{BW}$ is the barrier factor, 
$p^*$ and $p^*_0$ are evaluated at the mass $m$ and $m_{K^*(892)}$
respectively and depend also on the masses of the $K^*(892)$ decay products.  
\end{itemize}
With the definition of the mass distribution given in Eq.~(\ref{eq:kstar2}),
the parameter $\alpha$ entering in  Eq.~(\ref{eq:Fdeux}) is equal to:
\begin{equation}
\alpha = \sqrt{\frac{3\pi B_{K^*}}{p^*_0 ~\Gamma_{K^*(892)}^0}}
\label{eq:alphadef}
\end{equation}
where $B_{K^*} = \BR \left (K^*(892) \rightarrow \Km  \pip  \right )= 2/3$.
\subsection{$S$-wave form factor}
In a similar way as for the $P$-wave, we need to have the 
correspondence between the $S$-wave amplitude ${\cal F}_{10}$ (Eq.~(\ref{eq:f123})) and the corresponding invariant form factor.
In an $S$-wave, only the helicity $H_0$ form factor can contribute
and we take:
\begin{equation}
{\cal F}_{10} = p_{K\pi} m_D \frac{1}{1-\frac{q^2}{m_{A}^{2}}} {\cal A}_{S}(m).
\label{eq:af10}
\end{equation}
The term ${\cal F}_{10}$ is proportional to $p_{K\pi}$ to ensure that the corresponding decay rate 
varies as $p_{K\pi}^{3}$ as expected from the $L=1$ angular momentum between the virtual $W$ and the $S$-wave 
$K\pi$ hadronic state.
Because the $q^{2}$ variation of the form factor is expected to be determined by the
contribution of $J^{P}=1^{+}~c\bar{s}$ states, we use the same $q^{2}$ dependence as for $A_{1}$ and $A_{2}$.
The term ${\cal A}_S(m)$ corresponds to the mass dependent $S$-wave amplitude.
Considering that previous charm Dalitz plot analyses have measured 
an $S$-wave amplitude magnitude which is essentially constant up to the 
$K_0^*(1430)$ mass and then drops sharply above this value, we have used the following ansatz:
\begin{eqnarray}
{\cal A}_S&=&r_S P(m) e^{i\delta_S(m)},~{\rm and}\label{eq:samp}\\
{\cal A}_S&=&r_S P(m_{K^{*}_0(1430)})  \nonumber \\
& \times& \sqrt{\frac{ (m_{K_0^*(1430)}\Gamma_{K_0^*(1430)})^2 }
{(m_{K_0^*(1430)}^2 -m^2)^2 +(m_{K_0^*(1430)}\Gamma_{K_0^*(1430)})^2 }}~e^{i\delta_S(m)},\nonumber
\end{eqnarray} 
respectively for $m$ below and above the $K^{*}_0(1430)$ pole mass value.
In these expressions, $\delta_S(m)$ is the $S$-wave phase, 
$P(m) = 1 + r_S^{(1)}\times x +  r_S^{(2)}\times x^{2} + ... $ and 
$x=\sqrt{ (\frac{m}{m_{K}+m_{\pi}})^{2} -1 }$. The coefficients $r^{(i)}_S$ have no dimension and their values are fitted, but in practice, the fit to data
is sensitive only to the linear term. 
We have introduced the constant
$r_{S}$
which measures the magnitude of the $S$-wave amplitude.
From the observed asymmetry of the $\cos{\theta_K}$ distribution
in our data, $r_S<0$. This relative sign between $S$ and $P$ waves
agrees with the FOCUS measurement \cite{ref:focus1}.

\subsection{$D$-wave form factors}
Expressions for the form factors ${\cal F}_{i,2}$ 
for the $D$-wave are \cite{ref:Seb_Dwave}:
\begin{eqnarray}
{\cal F}_{12}&=&\frac{m_D\,p_{K\pi}}{3}\left[ \left(m_D^2-m^2-q^2 \right)\left( m_D+m\right ) T_1(q^2) \right . \nonumber\\ 
 &-& \left . \frac{ m_D^2\,p_{K\pi}^2}{m_D+m} T_2(q^2) \right ],\nonumber\\
{\cal F}_{22}&=&\sqrt{\frac{2}{3}}m_D\, m\, q\, p_{K\pi}\left( m_D+m\right ) T_1(q^2),\\
{\cal F}_{32}&=&\sqrt{\frac{2}{3}}\frac{2 m_D^2\, m\, q\, p^2_{K\pi}}{\left(m_D+m \right )}T_V(q^2).\nonumber
\label{eq:ffd3}
\end {eqnarray}
These expressions are multiplied by a relativistic Breit-Wigner amplitude
which corresponds to the $K_2^*(1430)$:
\beq
{
{\cal A}_{K_2^*}=\frac{ r_D\,m_{K_2^*(1430)}\Gamma_{K_2^*(1430)}^0 F_2(m) }
{m_{K_2^*(1430)}^2 -m^2 - im_{K_2^*(1430)}\Gamma_{K_2^*(1430)}(m) }.}
\label{eq:FFD}
\eeq
$r_D$ measures the magnitude of the $D$-wave amplitude and
similar conventions as in Eq.~(\ref{eq:kstar2}) 
are used for the other variables
apart from the Blatt-Weisskopf term which is equal to: 
\beq
B_2=1/\sqrt{\left (r_{BW}^2p^{*2}-3\right )^2 +9 r_{BW}^2p^{*2}},
\label{eq:b2d}
\eeq
and enters into 
\beq
F_2(m)=\left ( \frac{p^*}{p^*_0}\right )^2\frac{B_2(p^*)}{B_2(p^*_0)}.
\label{eq:f2d}
\eeq

The form factors $T_{i}(q^2)$ ($i=1,~2,~V$) are parameterized assuming the 
single pole model with corresponding axial or vector poles. Values for these
pole masses are assumed to be the same as those considered before for the
$S$- or $P$-wave hadronic form factors. Ratios of $D$-wave hadronic form factors
evaluated at $q^2=0$, 
$r_{22}=T_{2}(0)/T_{1}(0)$ and  $r_{2V}=T_{V}(0)/T_{1}(0)$ are supposed
to be equal to one \cite{ref:dwave}.

\section{The \babar\ detector and dataset}
\label{sec:detector}
A detailed description of the \babar\ detector and of the algorithms used
for charged and neutral particle reconstruction and identification is 
provided elsewhere~\cite{ref:babar, ref:babardet}. 
Charged particles are reconstructed by matching hits in 
the five-layer double-sided silicon vertex tracker (SVT) 
with track elements in the 40 layer drift chamber (DCH), which is
filled with a gas mixture of helium and isobutane.
Slow particles which due to bending in the $1.5$ T magnetic field
do not have enough hits in the DCH, are reconstructed in the SVT only.
Charged hadron identification is performed combining the measurements of 
the energy deposition in the SVT and in the DCH with the information from the
Cherenkov detector (DIRC). Photons are detected and measured in the 
CsI(Tl) electro-magnetic calorimeter (EMC). 
Electrons are identified by the ratio of the track momentum to the
associated energy deposited in the EMC, the transverse profile of the shower,
the energy loss in the DCH, and the Cherenkov angle in the DIRC.
Muons are identified in the instrumented flux return, composed
of resistive plate chambers and limited streamer tubes interleaved with layers of steel and brass.

The results presented here are obtained using  
a total integrated luminosity of $347.5~\invfb$. 
Monte Carlo (MC) simulation samples of $\FourS $ decays,
charm, and light quark pairs from continuum, equivalent 
to  $3.3,~1.7,~{\rm and}~1.1$ times the data statistics, 
respectively, and have been
generated using  $\textsc{Geant4}$ 
\cite{ref:geant4}. These samples are used mainly
to evaluate background components. Quark fragmentation in continuum events is described 
using the JETSET package \cite{ref:jetset}. 
The MC distributions are rescaled to the 
data sample luminosity, using the expected cross sections 
of the different components : $1.3$ nb for $c\overline{c}$, $0.525$
nb for $\Bp\Bm$ and $B^0 \overline{B}^0$, and $2.09$ nb for light $u\bar{u}$, $d\bar{d}$, and $s\bar{s}$ quark events.
Dedicated samples of pure
signal events, equivalent to 4.5 times the data statistics,
are used to correct measurements for efficiency and 
finite resolution effects. 
Radiative decays $(\Dp  \rightarrow \Km  \pip  \ep  \nue  \g)$ are modeled by PHOTOS
\cite{ref:photos}. 
Events with  a $\Dp $ decaying into $\Km \pip  \pip $ 
are also reconstructed in data and simulation.
This control sample is  used to adjust the
$c$-quark fragmentation distribution and the kinematic
characteristics of particles accompanying the $\Dp $
 meson in order to better match the data.
It is used also to measure the reconstruction accuracy of the missing
neutrino momentum. Other samples with a $\Dz$, a $\Dstarp$, or a $\Ds$ meson
exclusively reconstructed are used to define corrections on production characteristics
of charm mesons and accompanying particles that contribute to the background.

\section{Analysis method}
\label{sec:analysis}

Candidate signal events are isolated from $\FourS$ and continuum events
using variables combined into two Fisher discriminants,
tuned to suppress $\FourS $ and continuum background events, respectively.
Several differences between distributions
of quantities entering in the analysis, in data and simulation, are
measured and corrected using dedicated event samples.

\subsection{Signal Selection}
\label{sec:sigsel}

The approach used to reconstruct $\Dp $ mesons decaying into $\Km \pip \ep \nue $ is similar to that
used in previous analyses studying $\Dz \rightarrow \Km \ep \nue $ 
\cite{ref:kenu} and $\Ds \rightarrow \Kp \Km  \ep \nue $ \cite{ref:kkenu}.
Charged and neutral particles are boosted to the CM system and the event thrust axis is determined. 
A plane perpendicular to this axis is used to define two hemispheres.

Signal candidates are extracted from a sample of events already enriched in charm semileptonic decays.
Criteria applied for first enriching selection are:
\begin{itemize}
\item an existence of a positron candidate with a momentum larger than $0.5~\gevc$ in the CM frame,
to eliminate most of light quark events. Positron candidates are
accepted based on a tight identification selection with a pion 
misidentified as an electron or a positron below one per
mill; 
\item a value of $R_2>0.2$, $R_2$ being the ratio between 
second- and zeroth-order Fox-Wolfram moments \cite{ref:r2}, to decrease
the contribution from $B$ decays; 
\item a minimum value 
for the invariant mass of the particles in the event hemisphere
opposite to the electron candidate, $m_{opp}>0.5~\gevcc$, to reject
lepton pairs and two-photon events;
\item the invariant mass of the system formed by the positron and the most energetic particle in the candidate hemisphere, $m_{tag}>0.13~\gevcc$,
to remove events where the lepton is the only particle in its hemisphere.
\end{itemize}

A candidate is a positron, a charged kaon, and a charged pion present in 
the same hemisphere.
A vertex is formed using these three tracks, and the
corresponding $\chi^{2}$ probability larger than $10^{-7}$ are kept.
The value of this probability is used in the following with other information 
to reject background events. 

All other tracks in the hemisphere are defined as ``spectators''. 
They most probably originate from
the beam interaction point and are emitted during hadronization
of the created $c$ and $\overline{c}$ quarks. The
``leading'' particle is the spectator particle having the highest momentum.
Information from the spectator system is used  to decrease the
contribution from the combinatorial background. As charm hadrons take a large
fraction of the charm quark energy, charm decay products
have, on average, higher energies than spectator particles.

To estimate the neutrino momentum, the $( \Km   \pip  \ep  \nue )$
system is constrained to the $\Dp $ mass.
In this fit, estimates of the $\Dp $ direction and of the neutrino energy are
included from measurements obtained from all tracks registered in the event.
The $\Dp $ direction estimate is taken as the direction of the vector opposite
to the momentum sum of all reconstructed particles but the kaon, the pion, and the positron.
 The neutrino energy is evaluated by subtracting from the hemisphere
energy the energy of reconstructed particles contained in that hemisphere.
The energy of each hemisphere is evaluated by considering that the
total CM energy is distributed between two objects of mass corresponding
to the measured hemisphere masses \cite{ref:hemass}.
 As a $\Dp $ is expected to be present in the analyzed hemisphere and as at least
a $D$ meson is produced in the opposite hemisphere, minimum values for hemisphere masses are imposed.

For a hemisphere $i$, with the index of the other hemisphere noted as
$j$, the energy $E^{(i)}_{hem}$ and the mass $m^{(i)}_{hem}$ are defined as: 
\begin{align}
E^{(i)}_{hem}=\frac{1}{2}\left [ \sqrt{s}+\frac{m^{2,(i)}_{hem}-m^{2,(j)}_{hem}}{\sqrt{s}} \right ]\\
m^{(i)}_{hem}={\rm max}(m^{(i)}_{hem}({\rm measured}),m_{D}).\nonumber
\end{align}

The missing energy in a hemisphere is the difference between the hemisphere energy and the sum 
of the energy of the particles contained in this hemisphere ($E_{hem}^{miss}=E_{hem} - \sum^{n_{hem}}_{i=1}E_i$).
In a given collision, some of the resulting particles might take a path close
 to the beam line, 
being therefore undetected. In such cases, as one uses all reconstructed 
particles in an event to estimate the $D$ meson direction,  
this direction is poorly determined. 
These events are removed by only accepting those in which 
the cosine of the angle between the thrust axis and 
the beam line, $\cos(\theta_{thrust})$, is smaller than 0.7.
In cases where there is a loss of a large fraction of the energy contained in the opposite hemisphere, 
the reconstruction of the $D$ is also damaged. To minimize the impact of these cases, events with a missing energy in the opposite 
hemisphere greater than 3 $\gev$ are rejected.

\begin{figure}
	\centering
		\includegraphics[width=9cm]{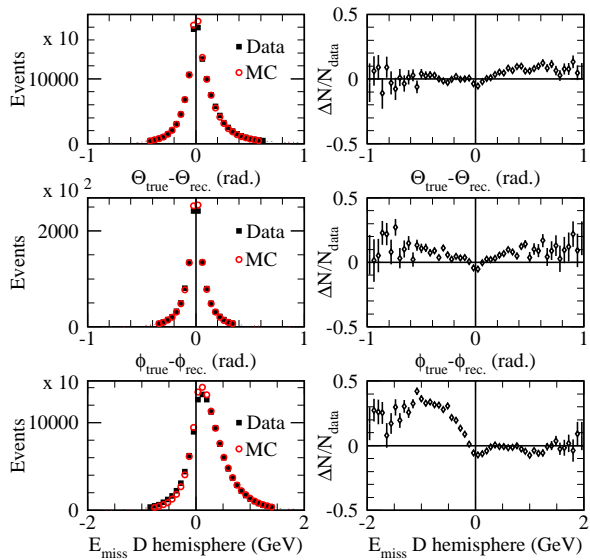}
	\caption{{(color online) Distributions of the difference (left)
between reconstructed and expected values, in the CM frame, 
for $\Dp $ direction angles 
($\theta,~\phi$) and 
for the missing energy in the candidate hemisphere. These distributions
are normalized to the same number of entries.
The $\Dp $ is reconstructed in the $\Km \pip \pip $ decay channel.
Distributions on the right display the relative difference between the 
histograms given on the left.}
	\label{fig:controldir}}
\end{figure}

The mass-constrained fit also requires estimates
of the uncertainties on the angles defining the $\Dp $ direction 
and on the missing energy must also be provided. These estimates are parameterized versus the 
missing energy in the opposite hemisphere which is used to quantify the quality of the reconstruction in a given event. 
Parameterizations of these uncertainties are obtained in data and in simulation using events with a reconstructed
$\Dp  \rightarrow \Km  \pip  \pip $, 
for which we can compare the measured $\Dp $ direction with its estimate using the algorithm employed  
for the analyzed semileptonic decay channel. 
$\Dp  \rightarrow \Km  \pip  \pip $ events also allow one to control the missing energy estimate and 
its uncertainty. Corresponding distributions obtained in data and with simulated events  are given in 
Fig.~\ref{fig:controldir}. These distributions are similar, and the
remaining differences are corrected as explained in 
Section \ref{sec:simultune}.

Typical values for the reconstruction accuracy of kinematic variables, obtained by 
fitting the sum of two Gaussian distributions for each variable, are given in 
Table~\ref{tab:resl}. These values are only indicative as the
matching of reconstructed-to-generated kinematic variables of events in five dimensions is included, 
event-by-event, in the fitting procedure.

\begin{table}[htbp]
\begin{center}
  \caption[]{{Expected resolutions for the five variables. They are obtained 
by fitting the distributions to the sum of two 
Gaussian functions. The fraction
of events fitted in the broad component is given in the last column.} 
  \label{tab:resl}}
  \begin{tabular}{c c c c }
    \hline\hline
variable & $\sigma_{1}$ & $\sigma_{2}$ & fraction of events  \\
     & & &in broadest Gaussian \\
\hline
$\cos\theta_{e}$ & 0.068 & 0.325  & 0.139\\
%\hline
$\cos\theta_{K}$ &  0.145 & 0.5  & 0.135\\
%\hline
$\chi~(\rad)$ & 0.223 & 1.174  & 0.135\\
%\hline
$q^{2}~(\gev^2)$ & 0.081 & 0.264 & 0.205  \\
%\hline
$m_{K\pi}~(\gevcc)$ & 0.0027 & 0.010 & 0.032\\
   \hline\hline	
  \end{tabular}
\end{center}
\end{table}

\subsection{Background rejection}
\label{sec:backgrej}
\begin{figure}
	\centering
\includegraphics[width=9.cm]{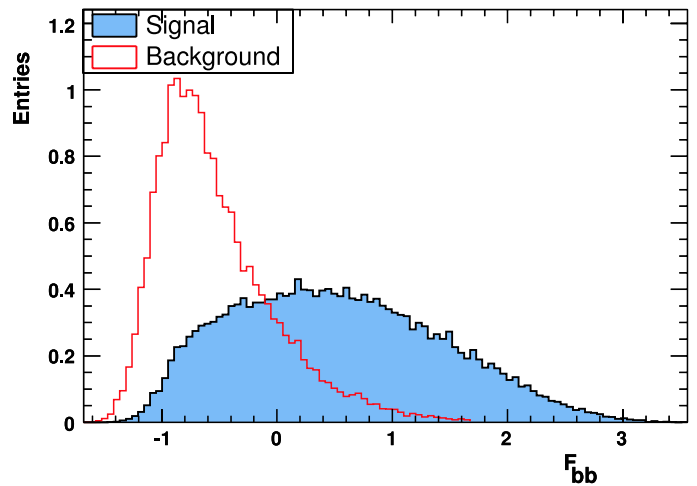}
\caption{{(color online) Distributions of $F_{bb}$ for signal and for $\FourS $ 
background events. The two distributions are normalized to the same number of entries.}
	\label{fig:bb_Fisher_dist}}
\end{figure}

Background events arise from $\FourS $ decays and hadronic 
events from the continuum. 
Three variables are used to decrease the contribution from $B\overline{B}$ events:
$R_2$, the total charged and neutral 
multiplicity, and the sphericity of the system of particles produced in the
event hemisphere opposite to the candidate. 
These variables use topological differences between events with $B$ decays and 
events with $c\bar{c}$ fragmentation. The particle distribution in $\FourS $ decay events 
tends to be isotropic as the $B$ mesons are heavy and produced near 
threshold, while the distribution in $c\bar{c}$ events 
is jet-like as the CM energy is well above the charm threshold. 
These variables are combined linearly in a Fisher discriminant
\cite{ref:fisher}, $F_{bb}$, and corresponding distributions
are given in Fig.~\ref{fig:bb_Fisher_dist}. 
The requirement $F_{bb}>0$ retains $70\%$ of signal and 15$\%$ of $B\overline{B}$-background events.

\begin{figure}
	\centering
\includegraphics[width=8cm]{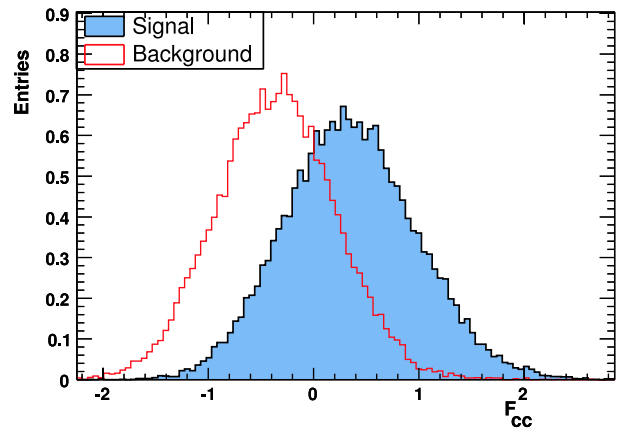}
	\caption{{(color online) Fisher discriminant variable $F_{cc}$ distribution for charm background and signal events. The two distributions are normalized to the same number of entries.}
	\label{fig:cc_Fisher_dist}}
\end{figure}

Background events from the continuum arise mainly from charm particles,
as requiring an electron and a kaon reduces the 
contribution from light-quark flavors
to a low level.
Because charm hadrons take a large
fraction of the charm quark energy, charm decay products
have higher average energies and different angular distributions  (relative to 
the thrust axis or to the $D$ direction) as compared to
other particles in the hemisphere, emitted from the hadronization
of the $c$ and $\overline{c}$ quarks. 
The $\Dp $ meson decays also at a measurable distance from the beam interaction
point, whereas background event candidates contain usually a pion 
from fragmentation.
Therefore, to decrease the amount of background from fragmentation
particles in $c \bar{c}$ events, the following variables are used: 
\begin{itemize}
\item the spectator system mass;
\item the momentum of the leading spectator track;
\item a quantity
derived from the $\chi^2$ probability of the $\Dp $
mass-constrained fit;
\item a quantity
derived from the $\chi^2$ vertex fit probability of the $K$, $\pi$
and $e$ trajectories;
\item the value of the $\Dp $ momentum
after the $\Dp $
mass-constrained fit; 
\item the significance of the flight length of the $\Dp $ from the beam interaction point until its decay point;
\item the ratio between the significances of the distance of the pion trajectory
to the $\Dp $ decay position and to the beam interaction point.
\end{itemize}
Several of these variables are transformed such that distributions of 
resulting (derived)
quantities have a bell-like shape.
These seven variables are combined linearly into a Fisher discriminant 
variable ($F_{cc}$) and the corresponding distribution is given in Fig.~\ref{fig:cc_Fisher_dist};
events are kept for values above 0.5.
This selection retains $40\%$ of signal events that were kept by the previous selection requirement 
on $F_{bb}$ and
rejects $94\%$ of the remaining background.
About $244\times10^3$ signal events are selected with a ratio $S/B=2.3$. In the
mass region of the $\akst$ this ratio increases to 4.6.
The average efficiency for signal is $2.9\%$ and is uniform when
projected onto individual kinematic variables.
A loss of efficiency, induced mainly
by the requirement of a minimal energy for the positron, is observed
for negative values of $\cos{\theta_e}$ and at low $q^2$.

\subsection{Simulation tuning}
\label{sec:alltuning}
Several event samples are used to correct differences between data 
and simulation. For the remaining $\FourS $ decays, the simulation
is compared to data as explained in Section \ref{sec:BB_section}.
For $\epem \rightarrow \ccbar$ events, 
corrections to the signal sample are different from those to the background 
sample. For signal,
events with a reconstructed $\Dp  \rightarrow \Km \pip \pip $ in data
and MC are used.
These samples allow us to compare the different distributions of the quantities
entering in the definition of the $F_{bb}$ and $F_{cc}$ discriminant
variables. Measured differences are then corrected, as explained below  (Section \ref{sec:simultune}). These samples
are used also to measure the reconstruction accuracy on the direction
and missing energy estimates for $\Dp  \rightarrow \Km \pip  \ep  \nue $.
For background events (Section \ref{sec:simulbackg}), the control of the simulation has to be extended
to $\Dz $, $D^{*+}$ and $\Ds$ production and to their accompanying
charged mesons. 
Additional samples with a reconstructed exclusive decay
of the corresponding charm mesons are used. Corrections are applied also
on the semileptonic decay models such that they agree with recent
measurements. Effects of these corrections are verified using wrong
sign events (Section \ref{sec:wsevt}), which are used also to correct 
for the production fractions
of charged and neutral $D$-mesons. Finally, absolute mass
measurement capabilities of the detector and the mass resolution
are verified (Section \ref{sec:massscale}) using 
$\Dz \rightarrow \Km \pip $ and 
$\Dp  \rightarrow \Km \pip \pip $ decay channels.

\subsubsection{Background from $\FourS$ decays}
\label{sec:BB_section}
\begin{figure}[htbp]
	\centering
\includegraphics[width=9.cm]
{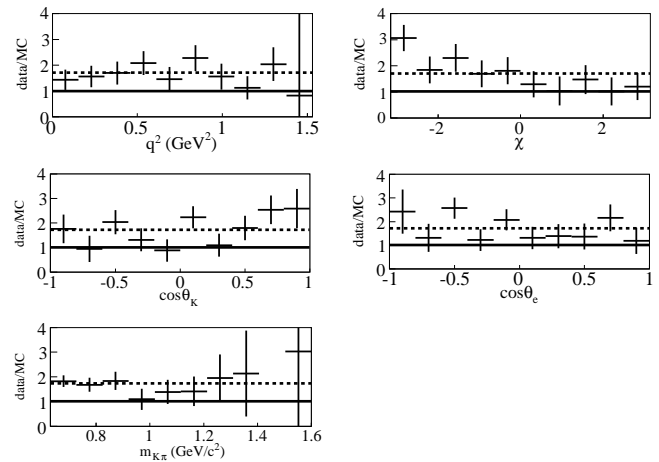}
	\caption{{Ratio (data/MC) distribution  for $\FourS$ decays
versus each of the five kinematic variables.  The dotted line corresponds 
to data/MC = 1.7. }
	\label{fig:bbratio}}
\end{figure}

The distribution of a given variable for events 
from the remaining
$\FourS \rightarrow \BB$ background is obtained by 
comparing corresponding distributions for events registered at the $\FourS$ resonance 
and 40 $\mev$ below. Compared with expectations from simulated events 
in Fig.~\ref{fig:bbratio}, distributions versus
the kinematic variables agree reasonably well in shape, within statistics, but the simulation needs to be scaled 
by $1.7 \pm 0.2 $. A similar effect was measured also in a previous
analysis of the $\Ds\rightarrow \Km  \Kp  \ep  \nue $ decay channel 
\cite{ref:kkenu}.

\subsubsection{Simulation tuning of signal events}
\label{sec:simultune}
Events with a reconstructed $\Dp  \rightarrow \Km \pip \pip $ candidate
are used to correct the simulation
of several quantities which contribute to the $\Km \pip  \ep  \nue $ 
event reconstruction.

Using the $\Km \pip \pip $ mass distribution, a signal region, 
between 1.849 and 1.889 $\gevcc$, and two sidebands 
($[1.798,~1.838]~{\rm and}~ [1.900,~1.940]~\gevcc$), are defined. 
A distribution of a given variable is obtained by subtracting 
from the corresponding distribution of events
in the signal region half the content
of those from sidebands. This approach is 
referred to as sideband 
subtraction in the following.
It is verified with simulated events that 
distributions obtained in this way agree
with those expected from true signal events.

\paragraph{control of the $c\rightarrow \Dp $ production mechanism:}
the Fisher discriminants $F_{bb}$ and $F_{cc}$ are functions 
of several variables, listed in Section \ref{sec:backgrej},
which have distributions that may differ between data and simulation.
For a given variable, weights are computed
from the ratio
of normalized distributions measured in data and simulation.
This procedure is repeated,
iteratively, considering the various 
variables, until 
corresponding projected distributions
are similar to those obtained in data.
There are remaining differences between data and simulation coming
from correlations between variables. To minimize their contribution,
the energy spectrum of $\Dp \rightarrow \Km \pip \pip  $ is weighted
in data and simulation to be similar to the spectrum
of semileptonic signal events.

We have performed another determination of the 
corrections without requiring that these 
two energy spectra are similar.
Differences between the fitted parameters
obtained using the two sets of corrections are taken as systematic 
uncertainties.

\paragraph{control of the $\Dp $ direction and missing energy measurements:}
the direction of a fully reconstructed $\Dp \rightarrow \Km  \pip  \pip $ 
decay is accurately measured and one can therefore compare 
the values of the two angles, defining its direction, with those obtained
when using all particles present in the event except those attributed to the
decay signal candidate. The latter procedure is used to estimate
of the $\Dp $ direction for 
the decay $\Dp \rightarrow \Km   \pip  \ep  \nue $.
Distributions of the difference between angles measured with the
two methods give the corresponding angular resolutions. 
This event sample allows also one to compare the missing energy measured in 
the $\Dp $ hemisphere and in the opposite hemisphere for data and
simulated events. These estimates for the $\Dp $ direction and momentum, 
and their corresponding
uncertainties are used in a mass-constrained fit.

For this study, differences between data and simulation in 
the $c\to \Dp $ fragmentation
characteristics are corrected as explained in the previous paragraph.
Global cuts similar to those applied for the
$\Dp \rightarrow \Km  \pip  \ep  \nue $ analysis are used such that
the topology of $\Dp \rightarrow \Km  \pip  \pip $ selected events is as
close as possible to that of semileptonic events.
Comparisons between angular resolutions measured in data and simulation
indicate that the ratio data/MC is 1.1
in the tails of the distributions
(Fig.~\ref{fig:controldir}). 
Corresponding distributions for the missing energy
measured in the signal hemisphere ($E^{same}_{miss.}$), 
in data and simulation, show that these distributions have an
offset of about 100 $\mevcc$ (Fig.~\ref{fig:controldir}) 
which corresponds to energy escaping detection
even in absence of neutrinos. To evaluate the neutrino energy
in $\Dp $ semileptonic decays this bias is corrected on average.

The difference between the exact and estimated values of the two angles and
missing energy is measured versus the value of the missing
energy in the opposite event hemisphere ($E_{miss.}^{opp.}$). 
This last quantity provides an estimate of the quality of the
energy reconstruction for a given event.
In each
slice of $E_{miss.}^{opp.}$, a Gaussian distribution is fitted and 
corresponding values of the average and standard deviation are measured.
As expected, the resolution gets worse when 
$E_{miss.}^{opp.}$ increases. 
These values are used as estimates for the bias and resolution 
for the considered
variable.
Fitted uncertainties are slightly higher in data than in the
simulation. From these measurements, a correction and a smearing are defined
 as a function of $E_{miss.}^{opp.}$. They are
applied to simulated event estimates of $\theta,~\phi$ and $E_{miss.}^{same}$.
This additional smearing is very small for the $\Dp $ direction
determination and is typically $\simeq 100~\mev$ on the missing energy estimate.

After applying corrections, the resolution on simulated events becomes slightly 
worse than in data.
When evaluating systematic uncertainties we have used the total deviation of 
fitted parameters obtained when applying or not applying the corrections.

\subsubsection{Simulation tuning of charm background events from continuum}
\label{sec:simulbackg}
As the main source of background originates from track combinations in which 
particles are from a charm meson decay, and others from hadronization, it is necessary to verify 
that the fragmentation of a charm quark into a charm meson and that  
the production characteristics of charged particles 
accompanying the charm meson are similar
in data and in simulation. 

In addition, most background events contain a lepton from a charm hadron semileptonic decay. The simulation
of these decays is done using the ISGW2 model \cite{ref:isgw2},
which does not agree with recent measurements \cite{ref:kenu}, 
therefore all simulated decay 
distributions are corrected.

\paragraph{Corrections on charm quark hadronization:}
\label{sec:corrcchad}
for this purpose,
distributions obtained in data and MC are compared. We study the event shape 
variables that enter in 
the Fisher discriminant $F_{bb}$ and for variables entering into $F_{cc}$,
apart from $\chi^2$ probability of the mass-constrained fit
 which is
peculiar  to the analyzed $\Dp $ semileptonic decay channel.
Production characteristics of charged pions and kaons 
emitted during the charm quark fragmentation, are also measured, 
and their rate, momentum,
and angle distribution relative to the simulated $D$ direction
are corrected. These corrections
are obtained separately for particles having the same or the opposite charge relative to the charm
quark forming the $D$ hadron. Corrections consist of a weight
applied to each simulated event. This weight is obtained iteratively,
correcting in turn each of the considered distributions. 
Measurements are done for $D^{*+}$, $\Dz $ (vetoing $\Dz $ from $D^{*+}$ decays) and for $\Dp $. 
For $D_s^+$ mesons, only the corresponding $c$-quark fragmentation distribution 
is corrected.

\begin{figure*}[htbp!]
	\center
\includegraphics[width=18cm]{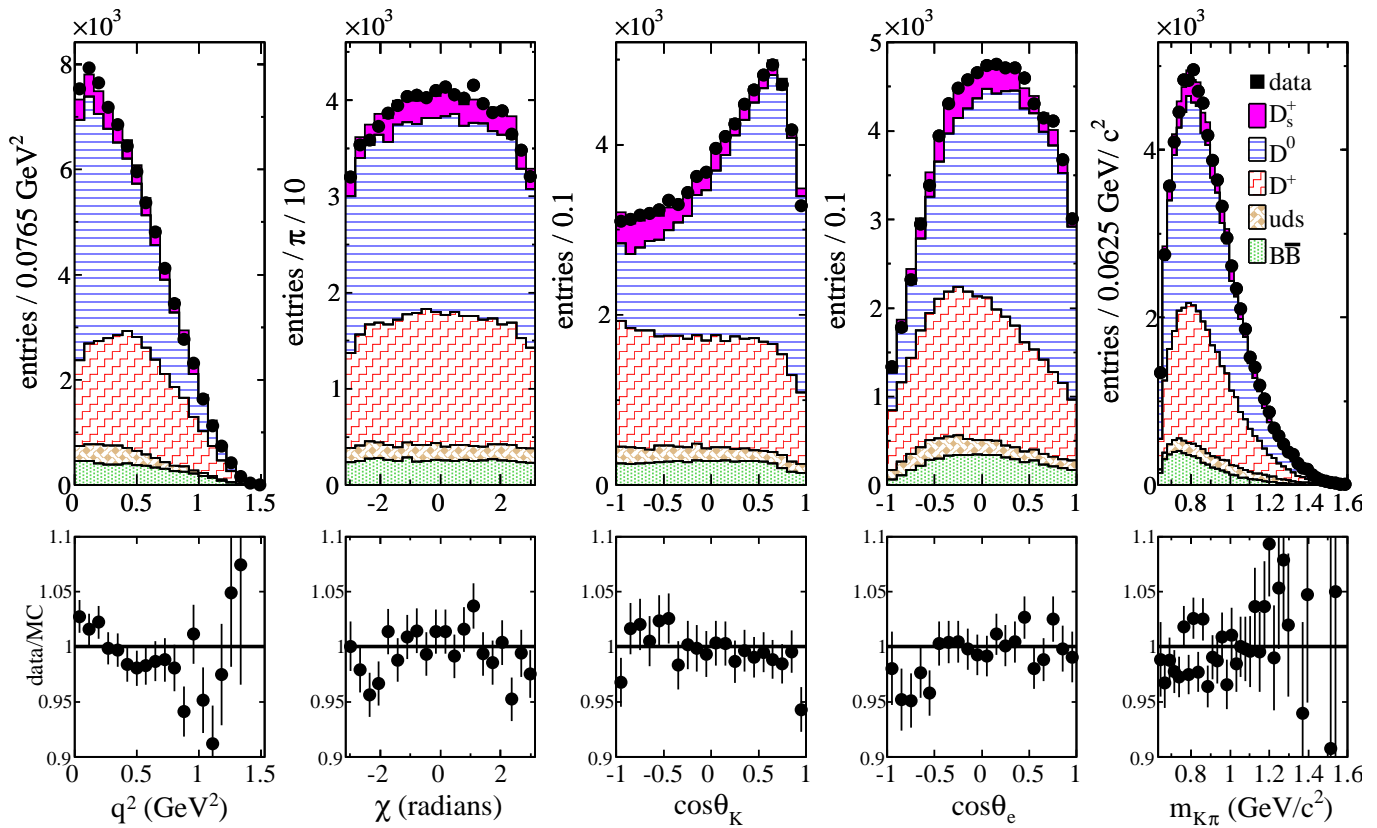}
	\caption{(color online) Distributions of the five dynamical variables for 
wrong-sign events in data (black dots) and MC (histograms), 
after all corrections. From top to bottom the background components displayed in the stacked
histograms are: $c\bar{c}~(D_{s},D^{0},D^{+}),~uds,~{\rm and}~B\overline{B}$ events respectively.
In the lower row, distributions of the ratio data/MC for upper row plots are given.}
	\label{fig:5VARS_WS_R4_step2}
\end{figure*}

\paragraph{Correction of D semileptonic decay form factors:}
\label{sec:corrccbr}
by default, $D$ semileptonic decays are generated in EvtGen \cite{ref:evtgen} using the ISGW2 decay model 
which does not reproduce present measurements (this was shown for instance in the \babar\ analysis 
of $D^{*+}\rightarrow \Dz \pip ,\Dz \rightarrow \Km \ep \nue $ \cite{ref:kenu}).
Events are weighted such that
they correspond to hadronic form factors behaving according to the 
single pole parameterization as in Eq.~(\ref{eq:ffdef}). 

For decay processes of the type $D\rightarrow P e \nue $,
where $P$ is a pseudoscalar meson,
 the weight is 
proportional  to the square of the ratio between
the corresponding hadronic form factors, and  
the total decay branching fraction
remains unchanged after the transformation. 
For all Cabibbo-favored decays a pole mass value equal to
$1.893~\gevcc$ \cite{ref:kenu} is used
whereas for Cabibbo-suppressed decays $1.9~\gevcc$ \cite{ref:cleo_pienu}
is taken. This value of the pole mass is used
also for $D_s$ semileptonic decays into a pseudoscalar meson.
For decay processes of the type $D\rightarrow V e \nue ,~(V\rightarrow P_{1}P_{2})$, where $P$ and $V$ are respectively pseudoscalar and vector mesons, 
corrections depend on the mass of the hadronic system, and on $q^{2},~\cos \theta_{e},~\cos \theta_{K}$ and $\chi$.
They are evaluated iteratively using projections of the differential
decay rate versus these variables, as obtained in EvtGen and in a 
simulation which contains the expected distribution. 
To account for correlations between these variables,
once distributions agree in projection, binned distributions over the
five dimensional space are compared and a weight is measured in each bin.
For Cabibbo-allowed decays, events are distributed over 2800 bins,
similar to those defined in Section \ref{sec:fitting}; 243 bins
are used for Cabibbo-suppressed decays.
Apart for the resonance mass and width which are different
for each decay channel, the same values, given in Table
\ref{tab:paramsyst}, are used for the other parameters
which determine the differential decay rate.

For decay channels $D \rightarrow K \pi \ep \nue $ an $S$-wave component
is added with the same characteristics as in the present measurements.
Other decay channels included in EvtGen \cite{ref:evtgen} 
and contributing to this same final 
state, 
such as a constant amplitude and 
the $\akstd$ components, are removed as they are
not observed in data.

All branching fractions used in the simulation agree within uncertainties
with the current measurements \cite{ref:pdg10} 
(apart for $D\rightarrow \pi \ep \nue $,  which is then rescaled). 
Only the shapes of charm semileptonic decay distributions are corrected.

Systematic uncertainties related to these corrections are estimated
by varying separately each parameter according to its expected uncertainty,
given in Table \ref{tab:paramsyst}.

\begin{table}[htbp]
\begin{center}
  \caption[]{{Central values and variation range for the various
parameters which determine the differential decay rate
in $D\rightarrow P/V\ep \nue $ decays, used to correct the simulation
and to evaluate corresponding systematic uncertainties. The
form factors $A_1(q^2),~A_2(q^2)~{\rm and}~V(q^2)$ 
and the mass parameters $m_A$ and $m_V$ are defined in 
Eq.~(\ref{eq:ffdef}).} 
  \label{tab:paramsyst}}
  \begin{tabular}{c c c }
    \hline \hline
parameter & central & variation   \\
     &value &interval \\
\hline\noalign{\vskip1pt}
$m_{pole}(D^{0,+}\rightarrow K\ep \nue )$ & $1.893~\gevcc$ & $\pm 30~\mevcc$ \\
$m_{pole}(D^{0,+}\rightarrow \pi \ep \nue )$ & $1.9~\gevcc$ & $\pm 100~\mevcc$ \\
$m_{pole}(D_s^+\rightarrow \eta/\eta^{\prime}\ep \nue )$ & $1.9~\gevcc$ & $\pm 100~\mevcc$ \\
\noalign{\vskip1pt}
\hline\noalign{\vskip1pt}
$r_2=|A_2(0)|/|A_1(0)|$ & $0.80$ & $\pm0.05$\\
$r_V=|V(0)|/|A_1(0)|$ & $1.50$ & $\pm0.05$\\
$m_A$ & $2.5~\gevcc$ & $\pm0.3~\gevcc$\\
$m_V$ & $2.1~\gevcc$ & $\pm0.2~\gevcc$\\
$r_{BW}$ & $3.0~(\gevc)^{-1}$ & $\pm0.3~(\gevc^{-1})$\\
\noalign{\vskip1pt}
\hline \hline
  \end{tabular}
\end{center}
\end{table}

\subsubsection{Wrong sign event analysis}
\label{sec:wsevt}
 Wrong-sign (WS) events of the type $\Km \pim \ep $ are used
to verify if corrections applied to the simulation improve the 
agreement with data,
because the origin of these events is quite similar to that of the background
contributing in right-sign $\Km \pip \ep $ (RS) events.  
The ratio between the measured and expected number of WS events is $0.950 \pm 0.005$. In RS events the
number of background candidates is a free parameter in the fit.  

At this point corrections have been evaluated separately for charged and
neutral $D$ mesons. As the two charged states correspond to background
distributions having different shapes, it is also possible to correct for
their relative contributions. We improve the agreement with data by increasing 
the fraction of events with a $\Dz $ meson in MC by 4$\%$
and correspondingly decreasing the fraction of $\Dp $ by $5\%$. 
After corrections, projected distributions of the five kinematic variables obtained in data and simulation
are given in Fig.~\ref{fig:5VARS_WS_R4_step2}.

\subsubsection{Absolute mass scale.}
\label{sec:massscale}
The absolute mass measurement is verified using exclusive
reconstruction of charm mesons in data and simulation.
For candidate events
$D^{*+}\rightarrow \Dz \pip ,~\Dz \rightarrow \Km \pip $,   
the mean and RMS values of the
$\Dz $ mass distribution are measured from a fit of 
the sum to a Gaussian distribution for the signal and a first 
order polynomial for the background. 
The $\Dz $ mass reconstructed in simulation is very close
to expectation, $\Delta_m^{MC}=(-0.07\pm 0.01)~\mevcc$, whereas
in data it differs by $\Delta_m^{data}=(-1.07\pm 0.17)~\mevcc$.
Here $\Delta_m$ is the difference between  the reconstructed and the
exact or the world average mass values when analyzing MC or data respectively.
The uncertainty quoted for $\Delta_m^{data}$ is from Ref.~\cite{ref:pdg10}.
To correct for this effect the momentum ($p$) of each track in data, measured
in the laboratory frame, is increased by an amount:
$\Delta_p^{data}=0.7\times 10^{-3}p$. The standard deviation
of the Gaussian fitted on the $\Dz $ signal is slightly
smaller in simulation, $(7.25\pm0.01)~\mevcc$, than in data,
$(7.39\pm0.01)~\mevcc$. The difference 
between the widths of reconstructed $\Dz $ signals in the two samples,
is measured versus the
transverse momentum of the tracks emitted in the decay. 
In simulation, the measured transverse momenta of the 
tracks
are smeared to correct for this difference.

Having applied these corrections, $\Dp $ mass distributions,
for the decay $\Dp \rightarrow \Km \pip \pip $ obtained in data and
simulation are compared. The standard deviation of the fitted
Gaussian distribution on signal is now similar in data and simulation.
The reconstructed $\Dp $ mass is higher by $0.23~\mevcc$ in simulation
(on which no correction was applied) and by $0.32~\mevcc$ in data.
These remaining differences are not corrected and included as 
uncertainties.

\subsection{Fitting procedure} 
\label{sec:fitting}
A binned distribution of data events is analyzed.
The expected number of events in each bin depends on signal and background estimates and 
the former is a function of the values of the fitted parameters.

We perform a minimization of a negative log-likelihood
distribution. This distribution has two parts. One corresponds
to the comparison between measured and expected number of events in
bins which span the five dimensional 
space of the differential decay rate. The other part uses the distribution of the values
of the Fisher discriminant variable $F_{cc}$ to measure the fraction of background events.

There are respectively 5, 5 and 4 equal size bins  for the variables $\chi$, $\cos \theta_{K}$
and $\cos \theta_{e}$. For $q^{2}$ and $m_{K\pi}$ we use respectively 4 and 7 bins of different size
such that they contain approximately the same number of signal events.
There are 2800 bins ($N_{\mathrm{bins}}$) in total.

The likelihood expression is:

\begin{eqnarray}
{\cal L}&=&\prod^{N_{\mathrm{bins}}}_{i=0}{P(n^{i}_{\mathrm{data}}|n^{i}_{\mathrm{MC}})} \nonumber\\
&\times& \prod^{N_{\mathrm{data}}}_{j=1}
\left [ \frac{N_{\mathrm{sig}}}{N_{\mathrm{sig}} + N_{\mathrm{bkg}}} \times pdf_{\mathrm{sig}}^{j} \right.  \nonumber\\
&+& \left .\frac{N_{\mathrm{bkg}}}{N_{\mathrm{sig}} + N_{\mathrm{bkg}}}\times{pdf_{\mathrm{bkg}}^{j}}\right ],
\label{eq:likelihood}
\end{eqnarray}

\noindent where $n_{\mathrm{data}}^{i}$ is the number of data events in bin 
$i$ and $n_{\mathrm{MC}}^{i}$ is the sum of MC estimates for 
signal and background events in the same bin.
$P(n^{i}_{\mathrm{data}}|n^{i}_{\mathrm{MC}})$ is the Poisson probability 
for having
 $n_{\mathrm{data}}^{i}$ events in bin $i$ where $n_{\mathrm{MC}}^{i}$ 
events are expected,
on average, where:

\begin{eqnarray}
n_{\mathrm{MC}}^{i}& =& \sum^{N_{\mathrm{events}}^{ \mathrm{bin}~i}}_{j=0} \left [\frac{N_{\mathrm{sig}}}{W_{\mathrm{fit}}^{\mathrm{tot}}(\vec{\lambda_{0}},\vec{\lambda})}\frac{W_{j}(\vec{\lambda})}{W_{j}(\vec{\lambda_{0}})}C_{j}\right]\nonumber\\
& + & \frac{N_{\mathrm{bkg}}}{W_{\mathrm{bkg}}^{\mathrm{tot}}}\times{W_{\mathrm{bkg}}^{i}}; \\
W^{\mathrm{tot}}_{\mathrm{fit}}(\vec{\lambda_{0}},\vec{\lambda}) &=& \sum^{N_{\mathrm{events}}^{\mathrm{all~bins}}}_{j=0} \frac{W_{j}(\vec{\lambda})}{W_{j}(\vec{\lambda_{0}})}C_{j}. \nonumber
\end{eqnarray}

The summation to determine $n_{\mathrm{MC}}^{i}$ extends over all generated
signal events which are reconstructed in bin $i$.
The terms $\vec{\lambda} ~{\rm and}~\vec{\lambda_{0}}$ are, respectively, the values of parameters used in the fit and 
those used to produce simulated events. $W_{j}(\vec{\lambda})$ is the value of 
the expression for the decay rate (see Eq.~(\ref{eq:decay1})) for event $j$ 
using the set of parameters $(\vec{\lambda})$.
In these expressions, generated values of the kinematic variables are used.
   $C_{j}$ is the weight applied to each signal event to correct for
differences between data and simulation. It is left unchanged during the fit.
 $W_{\mathrm{bkg}}^{i}$ is the estimated number of background events in bin $i$
given by the simulation, corrected for measured differences
with data, as explained in Section \ref{sec:alltuning}. $W_{\mathrm{bkg}}^{\mathrm{tot}}$ 
is the estimated total number of background events. 

\begin{figure}[htbp!]
	\center
\includegraphics[width=8cm]{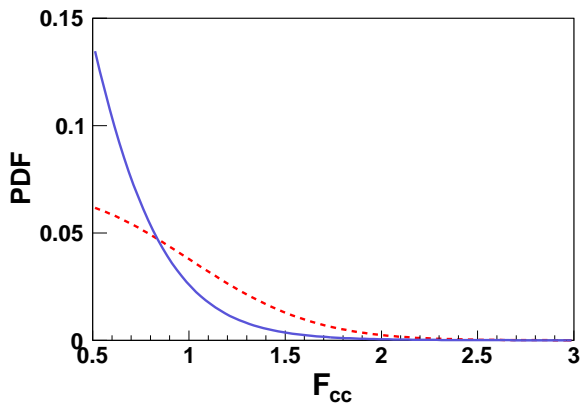} 
	\caption{{(color online) Probability density functions for signal (red dashed line) and 
background (blue full line) events versus the values of the discriminant variable $F_{cc}$.}
	\label{fig:fccpdf}}
\end{figure}

$N_{\mathrm{sig}}$ and $N_{\mathrm{bkg}}$ are respectively the total number of signal and 
background events fitted in the data sample which contains $N_{\mathrm{data}}$ events. 
$pdf_{\mathrm{sig}}^j$ and $pdf_{\mathrm{bkg}}^j$ are the probability density functions  for signal and background, respectively,
evaluated at the value of the $F_{cc}$ variable for event $j$.
The following expressions are used :
\begin{eqnarray}
\displaystyle
pdf_{\mathrm{sig}}(F_{cc}) &=& {\cal N}_{\mathrm{sig}}\left \{ c_{2}\times \exp\left [ \frac{-(F_{cc}-c_{0})^{2}}{2c_{1}^{2}}\right ] \right . \nonumber\\
&+&\left . c_{5}\times \exp \left [ \frac{-(F_{cc}-c_{3})^{2}}{2c_{4}^{2}} \right ]\right \},\nonumber\\
pdf_{\mathrm{bkg}}(F_{cc}) &=&  {\cal N}_{\mathrm{bkg}}\left \{\exp \left [\sum_{i=0}^{4} d_{i}(F_{cc})^{i}\right ]\right \}
\end{eqnarray}
and values of the corresponding parameters 
$c_{0-5}$ and $d_{0-4}$ 
are determined from fits to binned distributions of $F_{cc}$
in simulated signal and background samples. 
${\cal N}_{\mathrm{sig}}$ and ${\cal N}_{\mathrm{bkg}}$ are normalization 
factors.
In Fig.~\ref{fig:fccpdf} these two distributions are drawn to illustrate
their different behavior versus the values of $F_{cc}$ for signal and 
background events. As expected, the $pdf_{\mathrm{bkg}}$ 
distribution has higher
values at low $F_{cc}$ than the corresponding distribution for signal.

\subsubsection{Background smoothing}
\label{sec:smooth}
As the statistics of simulated background events for the charm continuum is only $1.6$ times the data,
biases appear in the determination of the fit parameters if we use simply, as estimates for background in each bin,
the actual values obtained from the MC. 
Using a parameterized event generator, this effect is measured using 
distributions of the difference between the fitted and exact values of a 
parameter divided by its fitted uncertainty (pull distributions).
To reduce these biases, a smoothing \cite{ref:Cranmer} of 
the background distribution is performed.
It consists of distributing the contribution of 
each event, in each dimension, according to a Gaussian distribution. 
In this procedure correlations between variables are neglected.
To account for boundary effects, the dataset is reflected about each boundary.
$\chi$ is essentially  uncorrelated with all other variables and
in particular with $\cos \theta_{l}$. Therefore,
for each bin in ($m,~q^{2}~{\rm and}~ \cos \theta_{K}$), a smoothing
of the $\chi$ and $\cos \theta_{l}$ distributions is done in the hypothesis
that these two variables are independent.

\section{$\Dp  \rightarrow \overline{K}^{*0} \ep  \nue $ hadronic form factor measurements}
\label{sec:LASSAMP_sec}
We first consider a signal made of the $\akst$ and $S$-wave components. 
Using the LASS parameterization of the $S$-wave phase versus the $K\pi$ mass (Eq.~(\ref{eq:phaselass})),
values of the following quantities 
(quoted in Table \ref{tab:fit_data_ampphase_2} second column) are obtained from a fit to data :
\begin{itemize}
\item parameters of the $\kst$ Breit-Wigner distribution: $m_{K^*(892)}$, $\Gamma^0_{K^*(892)}$, and $r_{BW}$, the Blatt-Weisskopf
parameter;
\item parameters of the $\Dp  \rightarrow \overline{K}^{*0} \ep  \nue $ hadronic form factors: $r_2,~r_V$, and $m_A$. The parameter
$m_V$ which determines the $q^2$ variation of the vector form factor is fixed to $2.0~\gevcc$;
\item parameters which define the $S$-wave component: $r_S$ and $r_S^{(1)}$ for the amplitude
(Eq.~(\ref{eq:samp})), $a_{S,BG}^{1/2}$ and $b_{S,BG}^{1/2}$ for the phase (Eq.~(\ref{eq:phasebg}));
\item and finally the total numbers of signal and background events,
$N_{sig}$ and $N_{bkg}$. 
\end{itemize}

\begin{table*}[!htb]
\begin{center}
 \caption[]{{ {Values of fitted parameters assuming that the final state consists of a
sum of $S$-wave and $\akst$ components (second column), includes the $\akstp$
in the $P$-wave (third column) and a $D$-wave (last column). 
The variation of the $S$-wave phase versus the $K\pi$ mass is parameterized according to
Eq.~(\ref{eq:phaselass}) whereas the $S$-wave amplitude is parameterized as in Eq.~(\ref{eq:samp}).
Fit results including the $\akstp$ are discussed in Section \ref{sec:rwithpprim}. 
Values given in the third column of this Table are the central results
of this analysis.}
  \label{tab:fit_data_ampphase_2}}}
{
\begin{tabular}{c c c c}
\hline\hline\noalign{\vskip1pt}

variable & $S+\akst$ & $S+\akst $ & $S+\akst $  \\
         &         & $\akstp$ & $\akstp+D$ \\
\hline\noalign{\vskip1pt}
{ $m_{K^{*}(892)}(\mevcc)$} & {$894.77\pm{0.08}$}& $895.43\pm{0.21}$ & $895.27\pm{0.15}$ \\
%\hline
{ $\Gamma^0_{K^{*}(892)}(\mevcc)$} & {$45.78\pm{0.23}$} & $46.48\pm{0.31}$ & $46.38\pm{0.26}$\\
%\hline
{$r_{BW}(\gevc)^{-1}$ }& {$3.71\pm{0.22}$} & $2.13\pm{0.48}$ & $2.31\pm{0.20}$ \\
%\hline
{ $m_{A} (\gevcc)$} & {$2.65\pm 0.10$} & $2.63 \pm 0.10$ & $2.58 \pm 0.09$ \\
%\hline
{$r_{V}$} & {$1.458\pm{0.016}$}& $1.463\pm{0.017}$& $1.471\pm{0.016}$ \\
%\hline
{$r_{2}$} & {$0.804\pm{0.020}$} & $0.801\pm{0.020}$ & $0.786\pm{0.020}$   \\
%\hline
{$r_{S}(\gev)^{-1}$} & {$-0.470\pm{0.032}$} & $-0.497\pm{0.029}$  & $-0.548\pm{0.027}$  \\
%\hline
{$r^{(1)}_{S}$} & {$0.17\pm{0.08}$} & $0.14\pm{0.06}$ & $0.03\pm{0.06}$  \\
%\hline
{$a_{S,BG}^{1/2}(\gevc)^{-1}$} & {$1.82\pm{0.14}$} & $2.18\pm{0.14}$ & $2.10\pm{0.10}$ \\
%\hline
{$b_{S,BG}^{1/2}(\gevc)^{-1}$ }& {$-1.66\pm{0.65}$} & $1.76$ fixed  & $1.76$ fixed  \\
% \hline
 $r_{\kstp}$ & & $0.074\pm{0.016}$  & $0.052\pm{0.013}$   \\
 %\hline
 $\delta_{\kstp}({\rm degree})$ & & $8.3\pm{13.0}$ & $0$ fixed    \\
 %\hline
 $r_{D}(\gev)^{-4}$ & &  & $0.78\pm0.18$   \\
 %\hline
 $\delta_{D}({\rm degree})$ & & & $0$ fixed    \\
 %\hline
{$N_{sig}$} & {$243850\pm{699}$}& $243219\pm{713}$& $243521\pm{688}$\\
%\hline
{$N_{bkg}$} & {$107370\pm{593}$}& $108001\pm{613}$ & $107699\pm{583}$\\
Fit probability & $4.6\%$ & $6.4\%$ & $8.8\%$ \\
\hline\hline
\end{tabular}}
\end{center}
\end{table*}

Apart from the effective range parameter, $b_{S,BG}^{1/2}$, all other quantities are accurately measured.
Values for the $S$-wave parameters depend on the parameterization used for the $P$-wave and as the LASS
experiment includes a $K^*(1410)$ and other components one cannot directly compare our results on $a_{S,BG}^{1/2}$ and $b_{S,BG}^{1/2}$
with those of LASS.
We have obtained the first measurement for $m_A$ which gives the $q^2$
variation of the axial vector hadronic form factors.
Using the values of fitted  parameters and integrating
the corresponding differential decay rates, fractions of $S$- and $P$-wave 
are given in the second column of Table \ref{tab:fr_SPPpr_SP}.

\begin{table*}[htbp]
\begin{center} 
 \caption{{Fractions for signal components assuming that the final state consists of a
sum of $S$-wave and $\akst$ components (second column), including the $\akstp$
in the $P$-wave (third column) and a $D$-wave (last column).
In the second and third cases, the sum of the fractions for the two  
$\overline{K}^{*}$ does not correspond exactly 
to the total $P$-wave fraction because of interference.}
  \label{tab:fr_SPPpr_SP}}
\begin{tabular}{c c c c}
\hline\hline\noalign{\vskip1pt}
Component & $S+\akst$ & $S+\akst$& $S+\akst$ \\
   & $(\%)$& $+\akstp(\%)$& $\akstp+D(\%)$\\
\noalign{\vskip1pt}
\hline\noalign{\vskip1pt}
 $S$-wave &  $5.62\pm0.14\pm0.13$ & $5.79\pm 0.16\pm0.15$& $5.69\pm 0.16\pm0.15$\\
\hline
 $P$-wave & $94.38$ & $94.21$& $94.12$\\
\hline\noalign{\vskip1pt}
$\akst$ & $94.38$& $94.11\pm0.74\pm0.75$ & $94.41\pm0.15\pm0.20$ \\
%\hline
 $\akstp$ &  $0$ & $0.33\pm0.13\pm0.19$ & $0.16\pm0.08\pm0.14$\\
\hline
$D$-wave &  $0$ & $0$& $0.19\pm0.09\pm0.09$\\\hline\hline
\end{tabular}
\end{center}
\end{table*}

\begin{figure*}
	\centering
\includegraphics[width=18cm]
{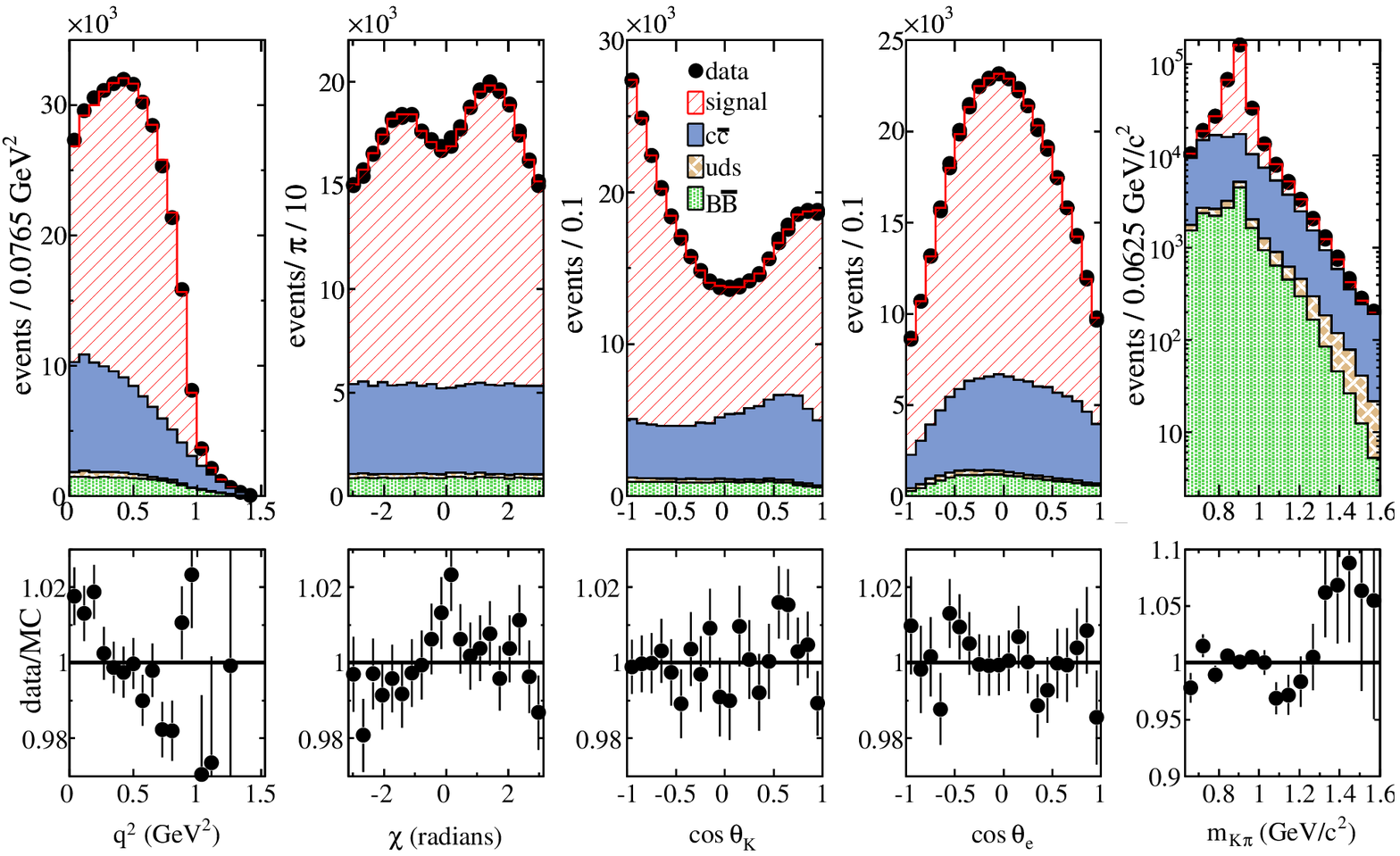}
	\caption{{(color online) Projections of data (black dots) and of the fitted MC 
distribution (histograms) 
versus each of the five kinematic variables. 
The signal contains an $S$-wave and a $\akst$ components.
From top to bottom the fitted background components displayed 
in the stacked
histograms are: $c\bar{c},~uds,~{\rm and}~B\overline{B}$ events respectively.
In the lower row, distributions of the ratio data/MC for upper row plots
are given.}
	\label{fig:main_result}}
\end{figure*}

Projected distributions, versus the five variables, obtained in data and from the $S$-wave + $\akst$ fit result are displayed
in Fig.~\ref{fig:main_result}.
The total $\chi^2$ of this fit is 2914 for 2787 degrees of freedom which corresponds to the probability of 4.6$\%$.
Fit results including the $\akstp$ and $D$-wave are discussed 
in Section \ref{sec:SPPrime_1}. 

\subsection{Systematic uncertainties}
\label{sec:systeval}

The systematic uncertainty on each fitted parameter ($x$) is defined
 as the difference between the fit results in nominal conditions $(x[0])$ 
and those obtained, $(x[i])$, after changing a variable or a condition ($i$)
by an amount which corresponds to an estimate of the uncertainty 
in the determination of this quantity: 
\begin{align}
\Delta x = x[0] - x[i].
\end{align}
Values are given in Table \ref{tab:syst_final}.
Some of the corrections induce a variation on the $F_{cc}$ 
distributions for signal or background which are therefore 
reevaluated.

\subsubsection{Signal production and decay}

\paragraph{Corrections of distributions of Fisher input variables (I):}
the signal control sample is corrected as explained in 
Section \ref{sec:simultune}. 
The corresponding systematic uncertainty 
is obtained by defining new event weights  without taking into account 
that the momentum distribution of reconstructed $D$ mesons is different in hadronic 
and in semileptonic samples. 

\normalfont

\paragraph{Simulation of radiative events (II):}
most of radiative events correspond to radiation from the charged lepton, although a non-negligible fraction comes 
from radiation of the $\kst$ decay products.
In $D\rightarrow P e \nue $, by comparing two generators (PHOTOS 
\cite{ref:photos} and KLOR \cite{ref:klor}), the CLEO-c collaboration has used a variation of $16\%$ 
to evaluate corresponding systematic uncertainties \cite{ref:cleocrad}.
We have increased the fraction of radiative events (simulated by PHOTOS) by 30$\%$ (keeping constant the total number of events) and obtained 
the corresponding variations on fitted parameters.
\begin{table*}[!htbp]
\begin{center}
\caption[]{{ Systematic uncertainties on parameters fitted using
the $S$-wave and $\akst$ model, expressed as $(x[0] - x[i])/\sigma_{stat}$: (I) uncertainty associated with the tuning of the signal control sample,
 (II) fraction of radiative signal events increased by $30\%$, (III) no PID
corrections on electron or kaon in MC signal events,
(IV)  no smearing applied on $\theta_{D},~\phi_{D}~{\rm and}~E_{miss}$ for simulated signal events, (V) $B\overline{B}$ background rate lowered by 
the statistical uncertainty
of its determination, (VI) uncertainty associated with the tuning of fragmentation in charm background events, (VII) remaining uncertainty on semileptonic
decay models in charm background events, (VIII) uncertainty associated with c-meson relative fractions, 
(IX) uncertainty remaining from the smoothing of the background distribution, (X) effects from limited statistics in simulation, (XI) variation  of parameters 
that were kept constant in the fit and, (XII) absolute mass scale uncertainties. }
  \label{tab:syst_final}}

\begin{tabular}{c c c c c c c c c c c c c }
\hline\hline\noalign{\vskip1pt}
variation & $ \Delta M_{\kst}$  &  $ \Delta \Gamma_{\kst}$  &  $\Delta r_{BW}$  &  $\Delta m_{A}$  &  $ \Delta r_{V} $  &  $\Delta r_{2} $  &  $\Delta r_{S}$  &  $\Delta r^{(1)}_{S}$ & $\Delta a_{S,BG}^{1/2}$  & $\Delta b_{S,BG}^{1/2}$  &  $\Delta N_{S}$  &  $ \Delta N_{B}$   \\
\noalign{\vskip1pt}
\hline
signal\\
\hline
I &  -0.13 & -0.16 & -0.10 & -0.18 & ~0.28 & ~0.18 & -0.40 & -0.43 & ~0.02 & ~0.00 & -0.36 & ~0.44 \\
%\hline
II & -0.36 & ~0.07 & ~0.02 & -0.11 &~0.34 & ~0.10 &  ~0.26 & ~0.20 & ~0.17 & ~0.21 & -0.21 & ~0.26 \\
%\hline
III & ~0.21 & ~0.13 & ~0.27 & ~0.69 & ~0.78 & ~0.51 & ~0.17 & ~0.16 & ~0.29 & ~0.17 & ~0.18 & ~0.22 \\
%\hline
IV &  ~0.29 & ~0.36 & ~0.20 & -0.18 & ~0.07 & -0.25 & ~0.15 & ~0.19 & -0.31 & -0.23 & ~0.57 & -0.70 \\
\hline\noalign{\vskip1pt}
$B\overline{B}$ bkg.\\
\hline
V & -0.06 & ~0.32 & ~0.09 & ~0.22 & -0.13 & ~0.03 & ~0.30 & ~0.31 & ~0.14 & ~0.30 & -0.09 & ~0.11 \\
\hline
$c\overline{c}$  bkg. \\
\hline
VI &  -0.04 & ~0.21 & -0.61 & ~0.10 & -0.08 & ~0.07 & ~0.33 & ~0.32 & ~0.13 & ~0.27 & ~0.06 & -0.08 \\
%\hline
VII & ~0.53 & ~0.19 & ~0.14 & ~0.16 & ~0.13 & ~0.07 & ~0.10 & ~0.10 & ~0.17 & ~0.19 & ~0.16 & ~0.22 \\
%\hline
VIII & ~0.24 & ~0.36 & ~0.11 & -0.49 & ~0.85 & ~0.04 & -0.76 & -0.68 & -0.77 & ~1.02 & ~0.76 & -0.91 \\
\hline
Fitting procedure\\
\hline
IX & ~0.13 & ~0.17 & ~0.25 & ~0.29 & ~0.30 & ~0.25 & ~0.25 & ~0.25 & ~0.32 & ~0.32 & ~0.13 &~0.13 \\
%\hline
X  & ~0.70  & ~0.70  & ~0.70  & ~0.70  & ~0.70  & ~0.70  & ~0.70  & ~0.70  & ~0.70  & ~0.70 & ~0.70 & ~0.70 \\
%\hline
XI  & ~0.00 & ~0.00 & ~0.07 & ~0.07 & ~1.15 & ~0.08 & ~0.05 & ~0.05 &  ~1.43 & ~0.46 & ~0.01 & ~0.01 \\
%\hline
XII & -0.93 &~ -0.06 & ~0.09 & ~0.09 & -0.05 & ~0.04 & ~0.03 & ~0.02 & ~0.07 & -0.05 & ~0.00 & ~0.00 \\
%\hline
%total\\ 
\hline
$\sigma_{syst}$ &~1.41  & ~1.00  &~1.06  & ~1.21 & ~1.87 & ~0.97 & ~1.27 & ~1.23  & ~1.87 & ~1.47 & ~1.29 & ~1.48 \\
\hline\hline
\end{tabular}
\end{center}
\end{table*}

\paragraph{Particle identification efficiencies (III):}
the systematic uncertainty is estimated by not correcting for 
remaining differences between data and MC on particle identification.

\paragraph{Estimates of the values and uncertainties for the $D$ direction and missing energy (IV):}
in Section \ref{sec:simultune} it is observed that estimates  of the $\Dp $
direction and energy are more accurate in the simulation than in data.
After applying smearing corrections, the result of this comparison is reversed.
The corresponding systematic uncertainty is equal to the difference on fitted
parameters obtained with and without smearing.

\subsubsection{$B\overline{B}$ background correction (V)}
The number of remaining $B\overline{B}$ background events 
expected from simulation is rescaled by $1.7\pm0.2$ 
(see Section \ref{sec:BB_section}).
The uncertainty on this quantity is used to evaluate corresponding systematic uncertainties. 

\subsubsection{Corrections to the $c\bar{c}$ background}

\paragraph{Fragmentation associated systematic uncertainties (VI):}
after applying corrections explained in Section \ref{sec:corrcchad}, the  
remaining differences between data and simulation for the considered distributions  are  five times smaller.
Therefore, 20$\%$ of the 
full difference measured before applying corrections
is used as the systematic uncertainty.

\paragraph{Form-factor correction systematics (VII):}
corresponding systematic uncertainties depend on uncertainties
on parameters used to model the differential semileptonic decay
rate of the various charm mesons (see Section \ref{sec:corrccbr}).

\paragraph{Hadronization-associated systematic uncertainties (VIII):}
using WS events, it is found in Section \ref{sec:wsevt} 
that the agreement between data and simulation 
improves by changing the hadronization fraction of the different charm mesons. 
Corresponding variations of relative hadronization fractions are 
compatible with current experimental uncertainties on these quantities. 
The corresponding systematic uncertainty is obtained by not applying 
these corrections.

\subsubsection{Fitting procedure}

\paragraph{Background Smoothing (IX):}
the MC background distribution  is smoothed as explained in Section \ref{sec:smooth}. 
 The evaluation of the associated systematic uncertainty is performed 
by measuring with  simulations based on parameterized distributions, 
the dispersion of displacements of the fitted quantities 
when the smoothing is or is not applied in a given experiment.
It is verified that uncertainties on the values of the two parameters
used in the smoothing have negligible contributions to the resulting 
uncertainty.

\paragraph{Limited statistics of simulated events (X):}
fluctuations of the number of MC events in each bin are 
not included in the likelihood expression, therefore one quantifies this 
effect using fits on distributions obtained with a parameterized event 
generator.
Pull distributions of fitted parameters, obtained in similar conditions as in data, have an RMS of 1.2. 
This increase is attributed to the limited MC statistics used  for the signal (4.5 times the data)
and, also, from the available statistics used to evaluate the 
background from $\epem \rightarrow \ccbar$ continuum events.
We have included this effect as a systematic uncertainty corresponding to 0.7 times the quoted
statistical uncertainty of the fit. It corresponds to the additional 
fluctuation needed to obtain a standard deviation of 1.2 of the pull 
distributions.

\subsubsection{Parameters kept constant in the fit (XI).}
The signal model has three fixed parameters, the vector pole mass $m_{V}$ 
and the mass and width of the $\overline{K}^{*}_{0}(1430)$ resonance. 
Corresponding systematic uncertainties are obtained by varying the values of these parameters.
For $m_V$ a $\pm 100~\mevcc$ variation is used, whereas for the other two quantities we take
respectively $\pm 50~\mevcc$ and $\pm 80~\mevcc$ \cite{ref:pdg10}.

\subsubsection{Absolute mass scale (XII).}

When corrections defined in Section \ref{sec:massscale} are applied, 
in data and simulation, for
the $\Dp  \rightarrow \Km \pip  \ep  \nue $ decay channel, the fitted
$\kst$ mass in data increases by $0.26~\mevcc$ and its
width decreases by $0.12~\mevcc$.
The uncertainty on the absolute mass measurement of the $\kst$
is obtained by noting that a mass variation, $\Delta_m^{\mathrm{data}}$,
of the $D$ reference signal
is reduced by a factor of four in the $K^*$ mass region; this gives:
\begin{eqnarray}
\sigma(m_{\kst})&=&\sqrt{0.17^2 +0.23^2}\frac{\Delta_m^{\mathrm{data}}(K\pi)}{\Delta_m^{\mathrm{data}}(D^{0,+})}\nonumber\\
&\simeq & 0.07~\mevcc.
\end{eqnarray}
In this expression, $0.17~\mevcc$ is the uncertainty on the 
$\Dz $ mass  \cite{ref:pdg10} and $0.23~\mevcc$ is the difference between 
the reconstructed
and exact values of the $\Dp $ mass in

\begin{table*}[htbp!]
\begin{center}
 \caption[]{{ Systematic uncertainties  on parameters fitted using
a model for the signal which contains $S$-wave, $\akst$ and $\akstp$ components, expressed as $(x[0] - x[i])/\sigma_{stat}$: (I) uncertainty associated with the tuning of the signal control sample,
 (II) fraction of radiative signal events increased by $30\%$, 
(III) no PID corrections on electron or kaon in MC signal events,
(IV)  no smearing applied on $\theta_{D},~\phi_{D}~{\rm and}~E_{miss}$ for simulated signal events, (V) $B\overline{B}$ background rate lowered by 
the statistical uncertainty
of its determination, (VI) uncertainty associated with the tuning of 
fragmentation in charm background events, (VII) remaining uncertainty 
on semileptonic
decay models for background events, (VIII) uncertainty associated with c-meson relative fractions, 
(IX) uncertainty remaining from the smoothing of the background distribution, (X) effects from limited statistics in simulation, (XI) variation of parameters 
that were kept constant in the fit, (XII) uncertainties on absolute mass scale.}
  \label{tab:syst_final_SPPprime}}
{\scriptsize
\begin{tabular}{c c c c c c c c c c c c c c }
\hline\hline\noalign{\vskip1pt}
variation & $ \Delta M_{\kst}$  &  $ \Delta \Gamma_{\kst}$  &  $\Delta r_{BW}$  &  $\Delta m_{A}$  &  $ \Delta r_{V} $  &  $\Delta r_{2} $  &  $\Delta r_{S}$  & $\Delta r^{(1)}_{S}$  & $\Delta a_{S,BG}^{1/2}$  &  $\Delta r_{\kstp}$ &  $\Delta \delta_{\kstp}$ &  $\Delta N_{S}$  &  $ \Delta N_{B}$   \\
\noalign{\vskip1pt}
\hline
\normalsize{signal}\\
\hline\noalign{\vskip1pt}
\small (I) &   ~0.17 &  ~0.05 & -0.23 & -0.22 & -0.31 & ~0.18 & ~0.14 & -0.14 & -0.13  & ~0.23 &  -0.19 & -0.39 & ~0.45 \\
%\hline
\noalign{\vskip1pt}
\small (II) &  -0.18 & ~0.06 & -0.01 & -0.14 & -0.36 & ~0.09 & -0.10 & ~0.08 & ~0.05 & -0.08 & -0.08 & -0.23 & ~0.26 \\
%\hline
\noalign{\vskip1pt}
\small (III) &  ~0.02 & ~0.10 & ~0.06 & ~0.70 & ~0.73 & ~0.53 & ~0.14 & ~0.07 & ~0.41 & ~0.08 & ~0.22 & ~0.10 & ~0.12 \\
%\hline
\noalign{\vskip1pt}
\small (IV) &  -0.13 & ~0.03 & ~0.29 & -0.18 & -0.04 & -0.27 & -0.02 & ~0.02 & -0.17 & -0.18 & ~0.32 & ~0.61 & -0.70 \\ 
\noalign{\vskip1pt}
\hline\noalign{\vskip1pt}
\normalsize{$B\overline{B}$ bkg.}\\
\noalign{\vskip1pt}
\hline\noalign{\vskip1pt}
\small (V) & -0.41 & -0.04 & ~0.34 & ~0.26 & ~0.16 & ~0.05 & -0.12 & ~0.12 & ~0.08 & -0.46 & ~0.22 & -0.01 & ~0.02 \\
\noalign{\vskip1pt}
\hline\noalign{\vskip1pt}
\normalsize{$c\overline{c}$ bkg. }\\
\noalign{\vskip1pt}
\hline\noalign{\vskip1pt}
\small (VI) & -0.14 & ~0.07 & -0.08 & ~0.13 & ~0.09 & ~0.09 & -0.16 & ~0.14 & -0.01 & -0.24 & -0.03 & ~0.08 & -0.09 \\  
%\hline
\noalign{\vskip1pt}
\small (VII) & ~0.09 & ~0.08 & ~0.14 & ~0.19 & ~0.14 & ~0.08 & ~0.18 & ~0.15 & ~0.06 & ~0.10 & ~0.11 & ~0.14 & ~0.18 \\
%\hline
\noalign{\vskip1pt}
\small (VIII) & -0.44 & -0.19 & ~0.59 & -0.48 & -0.75 & ~0.04 & ~0.98 & -0.94 & ~0.28 & -0.42 & -0.21 & ~1.03 & -1.23 \\
\noalign{\vskip1pt}
\hline\noalign{\vskip1pt}
\normalsize{Fitting procedure}\\
\noalign{\vskip1pt}
\hline\noalign{\vskip1pt}
\small(IX) & ~0.13 & ~0.17 & ~0.25 & ~0.29 & ~0.30 & ~0.25 & ~0.25 & ~0.25 & ~0.32 & ~0.30 & ~0.30 & ~0.13 & ~0.13 \\
\noalign{\vskip1pt}
%\hline
\small(X)  & ~0.70  & ~0.70  & ~0.70  & ~0.70  & ~0.70  & ~0.70  & ~0.70  & ~0.70  & ~0.70 & ~0.70 & ~0.70 & ~0.70 & ~0.70\\
\noalign{\vskip1pt}
%\hline
\small(XI) & ~0.27 & ~0.12 & ~0.29 & ~0.07 & ~1.15 & ~0.08 & ~0.57 & ~0.55 & ~3.25 & ~0.89 & ~0.40 & ~0.09 & ~0.10 \\
\noalign{\vskip1pt}
%\hline
\small (XII) & -0.33 & -0.05 &~ 0.03 & ~0.09 & ~0.05 & ~0.04 & -0.05 & ~0.03 & ~0.06 & -0.02 & -0.01 & -0.02 & ~0.02 \\
\noalign{\vskip1pt}
%\hline
%\normalsize{total}\\ 
\hline
\small $\sigma_{syst}$&  ~1.08 & ~0.78 & ~1.13 & ~1.24 & ~1.81 & ~0.99 & ~1.40 & ~1.35 & ~3.39 & ~1.39 & ~1.02 & ~1.48 & ~1.69 \\
\hline\hline
\end{tabular}
}
\end{center}
\end{table*}

simulation (see Section 
\ref{sec:massscale}).
Uncertainty on the $K^*$ width measurement from track resolution effects
is negligible.

\subsubsection{Comments on systematic uncertainties}

The total systematic uncertainty is obtained by summing in quadrature
the various contributions.
The main systematic uncertainty on $r_V$ comes from the assumed
variation for the parameter $m_V$ because these two parameters
are correlated. Values of the parameters $a_{S,BG}^{1/2}$ and
$b_{S,BG}^{1/2}$ depend on the mass
and width of the $\overline{K}^*_0(1430)$ because the measured $S$-wave phase
is the sum of two components: a background term and the $\overline{K}^*_0(1430)$.

\begin{figure*}[htbp!]
	\centering
\includegraphics[width=18cm]
{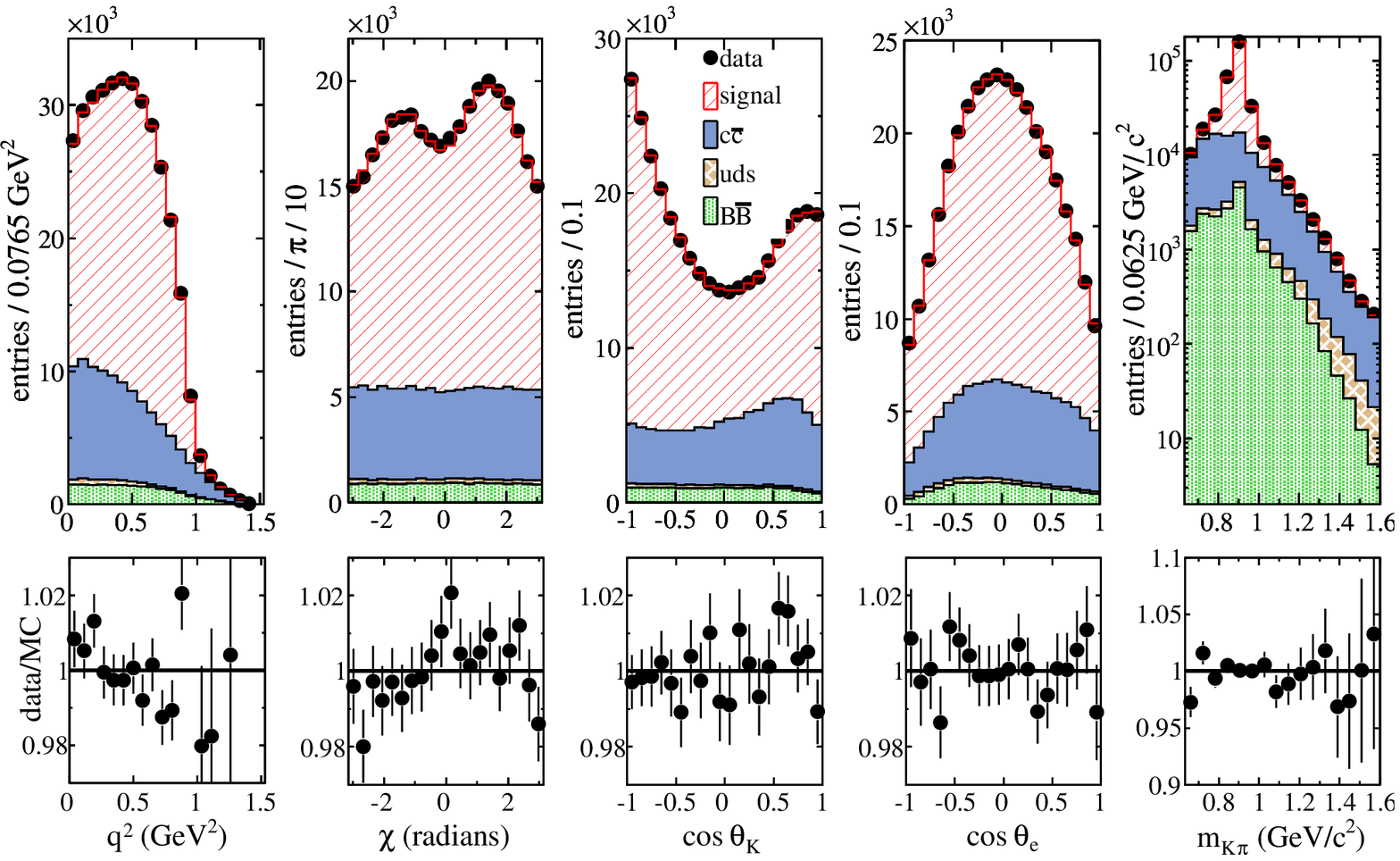}
	\caption{{(color online) Projections of data (black dots)  and of the fitted 
MC distribution (histograms) 
versus each of the five kinematic variables. 
The signal contains $S$-wave, $\akst$ and $\akstp$ components.
From top to bottom the background components displayed in the stacked
histograms are: $c\bar{c},~uds,~{\rm and}~B\overline{B}$ events respectively.
In the lower row, distributions of the ratio data/MC for upper row plots
are given.}
	\label{fig:spprim5d}}
\end{figure*}

\section{Including other components}
\label{sec:SPPrime_1}
A contribution to the $P$-wave from the $\akstp$ radial excitation 
was measured by LASS \cite{ref:lass1} in $Kp$ interactions
at small transfer and in $\tau$ decays \cite{ref:taubelle}.
As is discussed in the following, even if the statistical significance
of a signal at high mass does not reach the level to claim an observation,
data favor such a contribution and a signal containing the
$\akst$, the $\akstp$ and an $S$-wave components is considered as our
nominal fit to data.

To compare present results for the $S$-wave with LASS measurements 
a possible contribution from the $\akstp$ is included in the signal model.
It is parameterized using a similar Breit-Wigner expression as for the 
$\akst$ resonance.
The L=1 form factor components are in this case written as:

\begin{eqnarray}
{\cal F}_{11}  &\propto& (BW + r_{\kstp} e^{i\delta_{\kstp}}BW^{\prime})   2 \sqrt{2} q H_0\\
{\cal F}_{21}  &\propto& (BW + r_{\kstp} e^{i\delta_{\kstp}}BW^{\prime}) 2 q \left (H_+ + H_- \right )\nonumber \\
{\cal F}_{31}  &\propto& (BW + r_{\kstp} e^{i\delta_{\kstp}}BW^{\prime}) 2 q \left (H_+ - H_- \right ) \nonumber
\label{eq:FF2_prime}
\end{eqnarray}
where $BW$ stands for the $\akst$ Breit-Wigner distribution (Eq.~(\ref{eq:kstar2})) and $BW^{\prime}$ for that of the $\akstp$.
As the phase space region where this last component contributes is scarcely populated (high $K\pi$ mass), 
this analysis is not highly sensitive to the exact shape of the resonance. 
Therefore the Breit-Wigner parameters of the $\akstp$ 
(given in Table \ref{tab:kpistates}) are fixed and 
only the relative strength ($r_{\kstp}$) and phase ($\delta_{\kstp}$) are fitted. 
For the same reason, the value of $b_{S,BG}^{1/2}=1.76~\gev^{-1}$ 
is fixed to the LASS result (given in Table~\ref{tab:lassfit}).

\begin{figure}[htbp!]
	\centering
\includegraphics[width=8cm]
{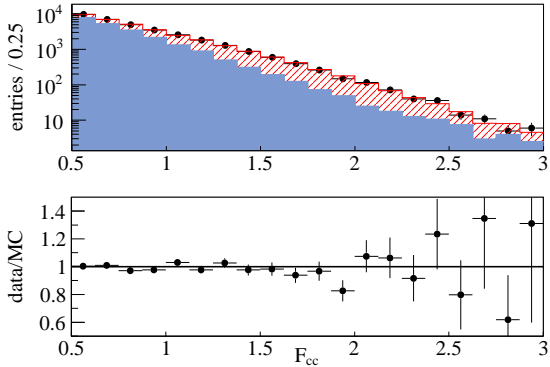}
\caption{{(color online) Comparison between measured and fitted distributions
of the values of the $F_{cc}$ discriminant variable. Points with
error bars correspond to data. The histogram is the fitted distribution.
It is the sum of a background (blue, filled histogram) and signal (hatched)
components.}
	\label{fig:fccfit}}
\end{figure}

\begin{figure}[htbp!]
	\centering
\includegraphics[width=8cm]
{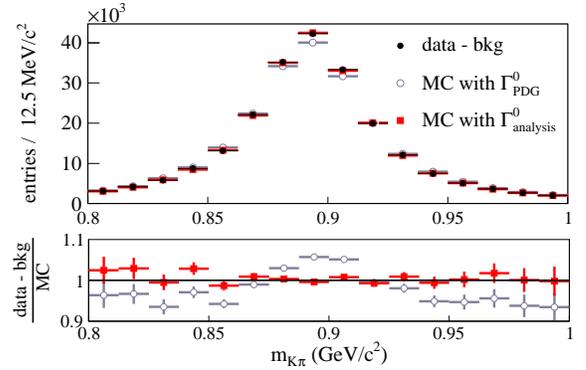}
\caption{{(color online) Comparison between measured and fitted $K\pi$ mass distributions
in the $\akst$ region. Results of a fit in which the width of the
$\akst$ meson is fixed to $50.3~\mevcc$ (value quoted in 2008 by the Particle Data Group) are also given.}
	\label{fig:kstarmass}}
\end{figure}

\subsection{ Results with a  $\akstp$ contribution included}
\label{sec:rwithpprim}
Results are presented in Table \ref{tab:fit_data_ampphase_2} (third column) 
using the same $S$-wave parameterization
as in Section \ref{sec:LASSAMP_sec}. They correspond to the central results
of this analysis.

\begin{figure*}[htbp!]
  \begin{center}
\mbox{\epsfig{file=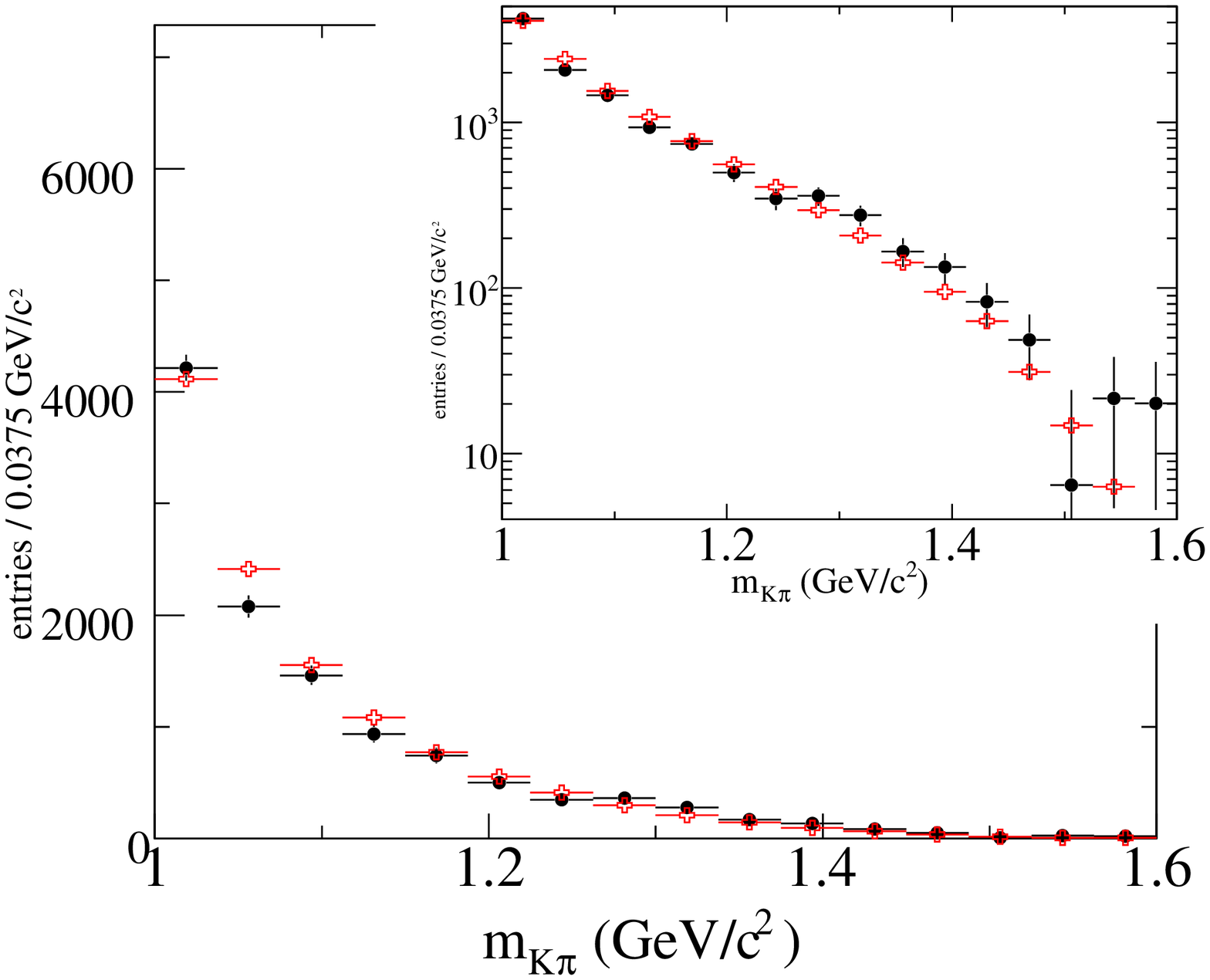,width=.5\textwidth}
\epsfig{file=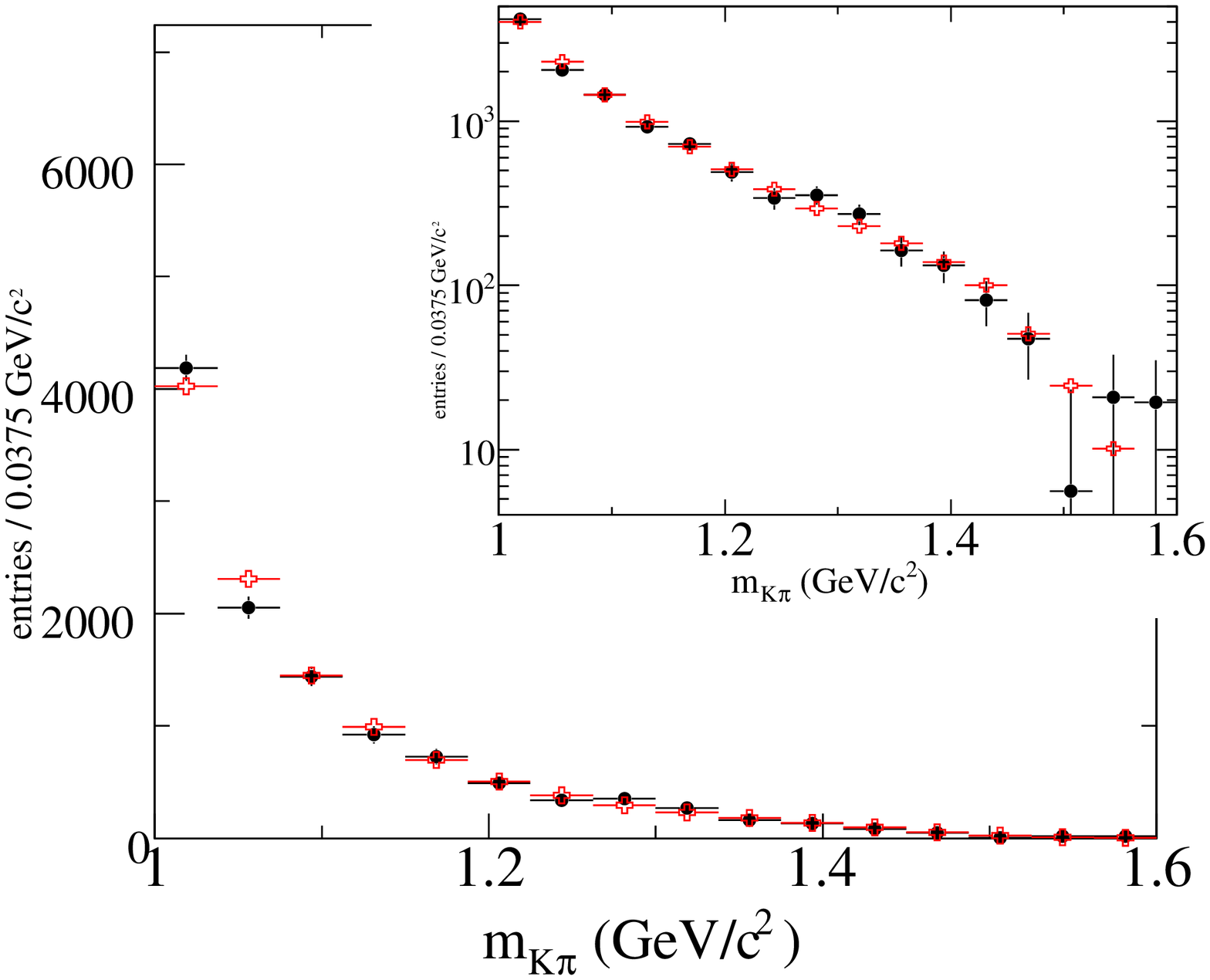,width=.5\textwidth}
}
  \end{center}
  \caption[]{{ (color online) Background subtracted data mass distribution (black full dots) 
and fit result (red open crosses) for the $S$-wave and $\akst$ model (left) 
and the $S$-wave, $\akst$ and $\akstp$ model (right), in the high mass region. 
Error bars correspond to statistical uncertainties only.}
   \label{fig:zoom_highmass_SPP}}
\end{figure*}

The total $\chi^{2}$ value is 2901 and the number of degrees of freedom is 
2786. This corresponds to a probability of $6.4\%$. 
Systematic uncertainties, evaluated as in Section \ref{sec:systeval}, are  given in Table \ref{tab:syst_final_SPPprime}.
The statistical error matrix of fitted parameters,
a table showing individual contribution of sources of systematic
uncertainties, which were grouped in the entries of Tab. 
\ref{tab:syst_final_SPPprime} labelled III, VII and XI, and the
full error matrix of systematic uncertainties are given in 
\ref{sec:appendixa}.
Projected distributions versus the five variables obtained in data and from 
the fit result are displayed in Fig.~\ref{fig:spprim5d}. 
Measured and fitted distributions of the values of the 
$F_{cc}$ discriminant variable are compared in Fig.~\ref{fig:fccfit}.

The comparison between measured and fitted, background subtracted, mass
distributions is given in Fig.~\ref{fig:kstarmass}. Results of a fit in which
the width of the $\kst$ resonance is fixed to $50.3~\mevcc$ (the value
quoted in 2008 by the Particle Data Group) are also 
given.  

Background subtracted projected distributions versus $m_{K\pi}$
for values higher than 1 $\gevcc$, obtained in data and using the fit results 
with and without including the $\akstp$, are displayed in Fig.~\ref{fig:zoom_highmass_SPP}.

 The measured fraction
of the $\akstp$ is compatible with the value obtained in $\tau$
decays \cite{ref:taubelle}. The relative phase between the 
$\akst$ and $\akstp$ is compatible with zero, as expected.
Values of the hadronic form factor parameters for the decay
$\Dp  \rightarrow \overline{K}^{*0} \ep  \nue $
are almost identical with those obtained without including the $\akstp$. 
The fitted value for  $a_{S,BG}^{1/2}$ is compatible with the result from LASS
reported in Table \ref{tab:lassfit}.

\begin{figure*}[htb!]
	\centering
\includegraphics[width=12cm]
{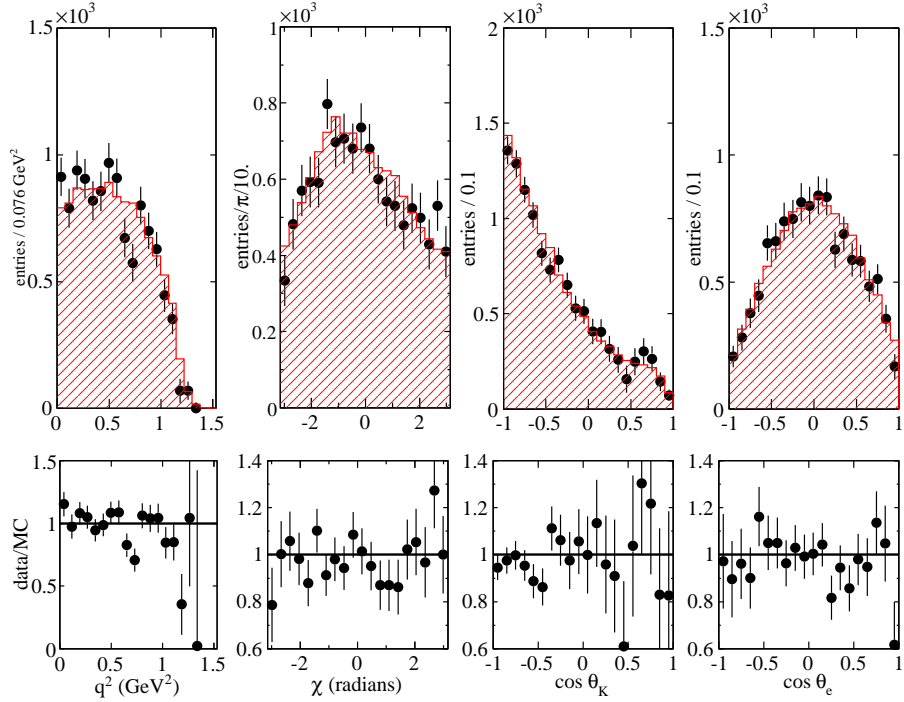}
	\caption{{ (color online) Projections of background subtracted data (black dots)  and  fitted MC signal distributions (hashed histogram)
versus the four  kinematic variables in the mass region between threshold and 800 $\mevcc$. Error bars correspond to statistical uncertainties only.
The signal contains $S$-wave, $\akst$ and $\akstp$ components.
Lower plots are the ratio between data and the fitted signal.}
	\label{fig:C4VARS_630_800MeV_SPP}}
\end{figure*}

\begin{figure*}[htb!]
	\centering
\includegraphics[width=12cm]
{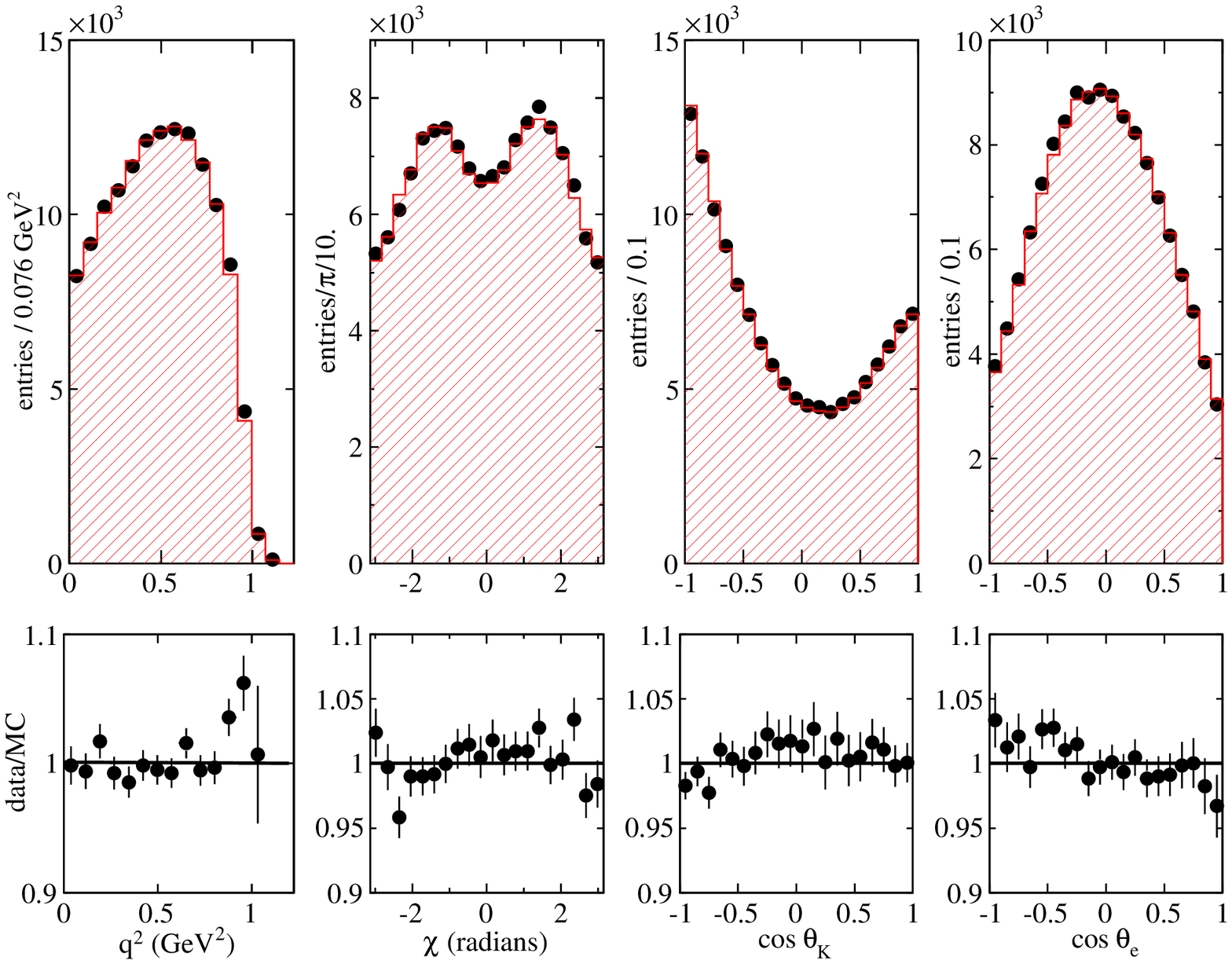}
	\caption{{ (color online) Projections of background subtracted data (black dots) and  fitted MC signal distributions (hashed histogram) 
versus the four kinematic variables in the mass region between 800 and 900 $\mevcc$. Error bars correspond to statistical uncertainties only.
The signal contains $S$-wave, $\akst$ and $\akstp$ components.
Lower plots are the ratio between data and the fitted signal.}
	\label{fig:C4VARS_800_900MeV_SPP}}
\end{figure*}

\begin{figure*}[htb!]
	\centering
\includegraphics[width=12cm]
{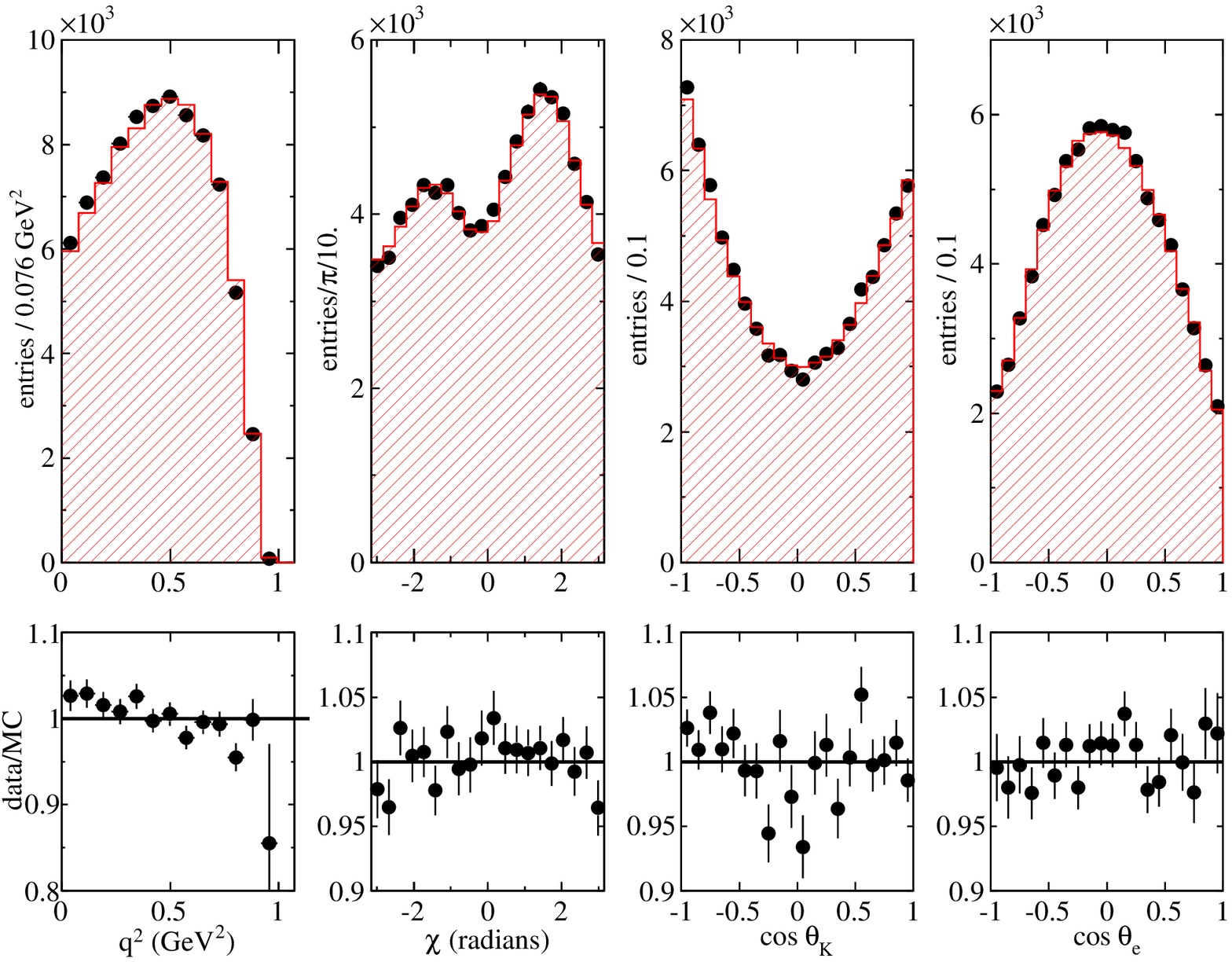}
\caption{{ (color online) Projections of background subtracted data (black dots) and  fitted MC signal distributions (hashed histogram)
versus  the four kinematic variables in the mass region between 900 and 1000 $\mevcc$. Error bars correspond to statistical uncertainties only.
The signal contains $S$-wave, $\akst$ and $\akstp$ components.
Lower plots are the ratio between data and the fitted signal.}
	\label{fig:C4VARS_900_1000MeV_SPP}}
\end{figure*}

\begin{figure*}[htb!]
	\centering
\includegraphics[width=12cm]
{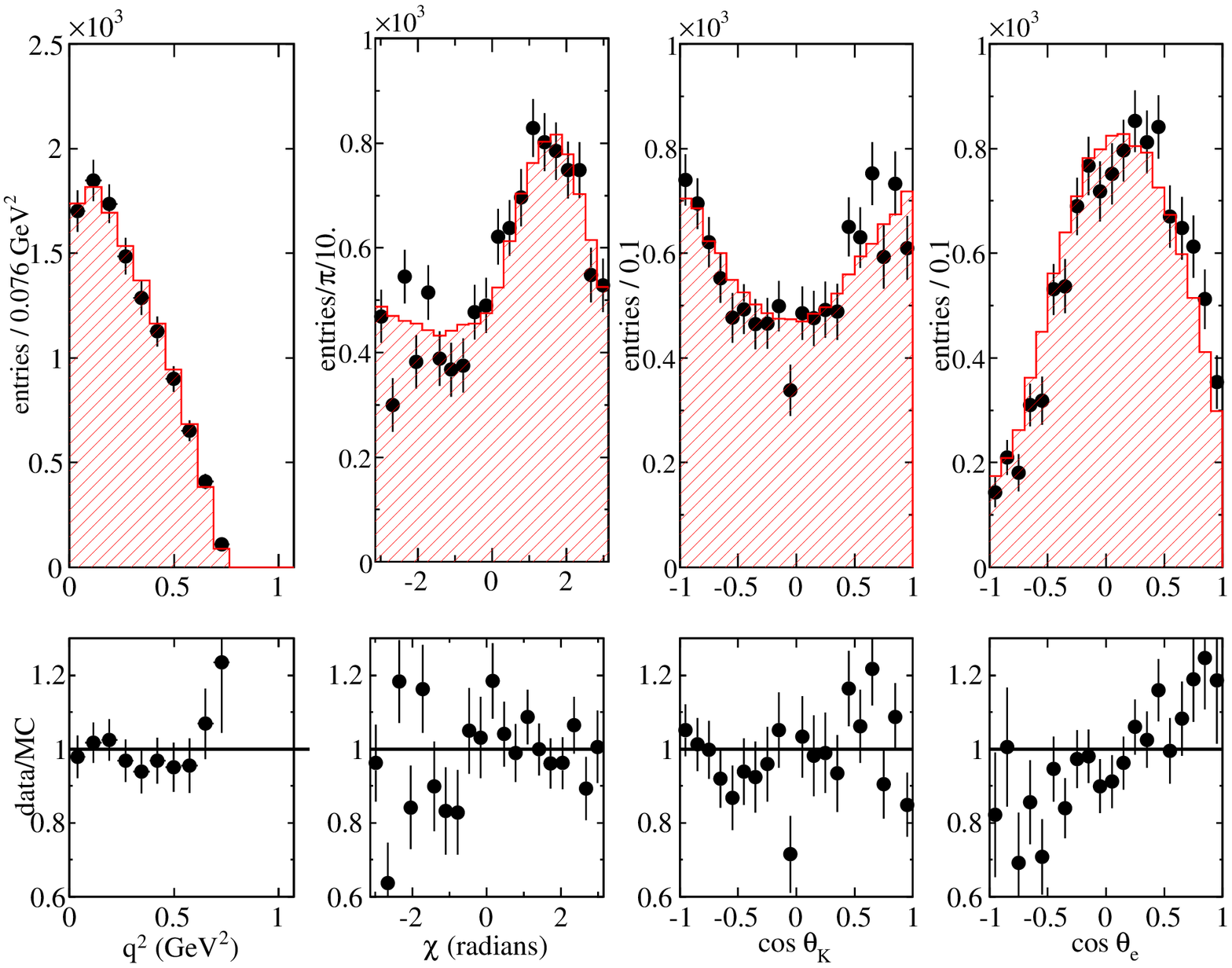}
	\caption{{ (color online) Projections of background subtracted data (black dots) and  fitted MC signal distributions (hashed histogram) 
versus the four kinematic variables in the mass region between 1000 and 1600 $\mevcc$. Error bars correspond to statistical uncertainties only.
The signal contains $S$-wave, $\akst$ and $\akstp$ components.
Lower plots are the ratio between data and the fitted signal.}
	\label{fig:C4VARS_highMass_SPP}}
\end{figure*}

The total fraction of the $S$-wave is 
compatible with the previous value. Fractions for each component are given in 
the third column of Table \ref{tab:fr_SPPpr_SP}.

Considering several mass intervals, background subtracted projected 
distributions versus the four other variables, obtained in data and 
from the fit results, are displayed in 
Fig.~\ref{fig:C4VARS_630_800MeV_SPP} to \ref{fig:C4VARS_highMass_SPP}.

\begin{table*}[!htbp]
\begin{center}
 \caption[]{{Systematic uncertainties  on parameters fitted using
a model for the signal which contains  $S$-wave, $\akst$ and $\akstp$ components in which the $\akst$ parameters are fixed,
expressed as $(x[0] - x[i])/\sigma_{stat}$:: (I) uncertainty associated with the tuning of the signal control sample,
 (II) fraction of radiative signal events increased by $30\%$, 
(III) no PID correction on electron or kaon in MC signal events,
(IV)  no smearing applied on $\theta_{D},~\phi_{D}~{\rm and}~E_{miss}$ for 
simulated signal events, (V) $B\overline{B}$ background rate lowered by 
the statistical uncertainty
of its determination, (VI) uncertainty associated with the tuning of 
fragmentation in charm background events, (VII) remaining uncertainty on 
semileptonic
decay models for background events, (VIII) uncertainty associated with 
c-meson relative fractions, 
(IX) uncertainty remaining from the smoothing of the background distribution, (X) effects from limited statistics in simulation, (XI) variation of parameters 
that were kept constant in the fit, (XII) uncertainties on absolute mass scale. }
\label{tab:syst_final_MIAMP_Pprime}}
{\scriptsize
\begin{tabular}{c c c c c c c c c c c c c c }
\hline\hline\noalign{\vskip1pt}
variation & $\Delta r_{\kstp}$ & $\Delta \delta_{\kstp}$ & $\Delta r_{S}$ &  $\Delta r^{(1)}_{S}$ & $\Delta \delta_{1}$  & $\Delta \delta_{2}$  & $\Delta \delta_{3}$  & $\Delta \delta_{4}$  & $\Delta \delta_{5}$  & $\Delta \delta_{6}$  & $\Delta \delta_{7}$  & $\Delta \delta_{8}$  & $\Delta \delta_{9}$   \\
\noalign{\vskip1pt}
\hline\noalign{\vskip1pt}
\small (I) &  ~0.23  & -0.08  & -0.13  & -0.16  &  ~0.02  & -0.07  & -0.09  & ~0.08  & ~0.05  & -0.06  & ~0.01  & ~0.01  & -0.09 \\
%\hline
\noalign{\vskip1pt}
\small (II) &  -0.34  & ~0.02  & 0  & -0.03  & ~0.04  & ~0.01  & -0.01  & ~0.01  & ~0.21  & -0.14  & -0.03  & ~0.04  & ~0.17 \\ 
%\hline
\noalign{\vskip1pt}
\small (III) & -0.01  & -0.05  & -0.11  & -0.11    & ~0.05  & -0.03  & -0.11  & ~0.29  & ~0.55  & ~0.05  & ~0.10  & ~0.04  & ~0.03 \\
%\hline
\noalign{\vskip1pt}
\small (IV) & -0.92  & ~0.26  & -0.12  & -0.14 & -0.08  & ~0.12  & ~0.02  & -0.11  & 0  & -0.21  & -0.22  & -0.03  & ~0.50 \\
\noalign{\vskip1pt}
\hline\noalign{\vskip1pt}
\normalsize{$B\overline{B}$ bkg.}\\
\noalign{\vskip1pt}
\hline\noalign{\vskip1pt}
\small (V) &  -1.05  & ~0.17  & -0.03  & -0.08 & ~0.17  & ~0.21  & ~0.27  & ~0.14  & -0.12  & -0.36  & -0.36  & -0.19  & ~0.59\\
\noalign{\vskip1pt}
\hline\noalign{\vskip1pt}
\normalsize{$c\overline{c}$ bkg. }\\
\noalign{\vskip1pt}
\hline\noalign{\vskip1pt}
\small (VI) &  -0.17  & -0.01  & ~0.16  & ~0.12  & -0.02  & ~0.01  & -0.01  & ~0.01  & ~0.01  & -0.03  & -0.05  & -0.05  & -0.08 \\
%\hline
\noalign{\vskip1pt}
\small (VII) & ~0.15 & ~0.09 & ~0.11 & ~0.13  & ~0.20 & ~0.11 & ~0.08 & ~0.03 & ~0.09 & ~0.11 & ~0.13 & ~0.10 & ~0.06\\
%\hline
\noalign{\vskip1pt}
\small (VIII) &-2.85  & -0.36  & -0.22& -0.19& -0.12  & -0.37  & -0.1  & ~0.59  & ~0.82  & ~0.27  & ~0.15  & ~0.14  & ~1.29\\ 
\noalign{\vskip1pt}
\hline\noalign{\vskip1pt}
\normalsize{Fitting procedure}\\
\noalign{\vskip1pt}
\hline\noalign{\vskip1pt}
\small(IX) &  ~0.60 & ~0.60 & ~0.60 & ~0.60 & ~1.06 & ~0.64 & ~0.47 & ~0.42 & ~0.40 & ~0.49 & ~0.54 & ~0.63 & ~0.82 \\
%\hline
\noalign{\vskip1pt}
\small(X)  & ~0.70 & ~0.70 & ~0.60  & ~0.61  & ~0.53 & ~0.54 & ~0.53 & ~0.53 & ~0.78 & ~0.54 & ~0.53 & ~0.54 & ~0.98\\
%\hline
\noalign{\vskip1pt}
\small(XI)  &  ~1.07  & ~0.27  & ~0.23  & ~0.26  & ~0.16  & ~0.07  & ~0.09  & ~0.16  & ~0.28  & ~0.16  & ~0.14  & ~0.14  & ~0.53\\
%\hline
\noalign{\vskip1pt}
\small(XII) &  -0.49  & ~0.01  & -0.01  & -0.04 & -0.03  & ~0.12  & 0  & ~0.33  & ~0.50  & ~0.10  & ~0.12  & ~0.14  & ~0.27  \\
\noalign{\vskip1pt}
\hline
%\normalsize{total}\\ 
%\hline
\small $\sigma_{syst}$  & ~3.70  & ~1.09  & ~0.94  & ~0.96 & ~1.24  & ~0.99  & ~0.83  & ~1.04  & ~1.47  & ~0.99  & ~0.98  & ~0.91  & ~2.07 \\
\hline\hline
\end{tabular}
} 
\end{center}
\end{table*}

\subsection{Fit of the $\akstp$ contribution and of the $S$-wave amplitude and phase}
\label{sec:MIAMP_SPP}

Fixing the parameters which determine the $\kst$ 
contribution to the values obtained in the previous fit,
we measure the $S$-wave parameters entering in Eq.~(\ref{eq:samp}) in which the $S$-wave phase
is assumed to be a constant within each of the considered $K\pi$ mass intervals.Values of $m_{K\pi}$ which correspond to the center and to half the width
of each mass interval are given in Table \ref{tab:massbins}.
The two parameters which define the $\akstp$ are also fitted.
Numbers of signal and background events are fixed to their previously determined values.
Values of fitted parameters are given in Table \ref{tab:fit_data_ampphase_3}.

\begin{table}
\begin{center}
 \caption[]{{Positions of the center and values of 
half the mass intervals used in the phase measurement.}
  \label{tab:massbins}}
\begin{tabular}{c c c  }
\hline\hline\noalign{\vskip1pt}
mass bin &  $m_{K\pi}~(\gevcc)$ & $\Delta m_{K\pi}~(\gevcc)$\\
\noalign{\vskip1pt}
\hline
$1$ & 0.707 & 0.019\\
%\hline
$2$ &  0.761 & 0.035\\
%\hline
$3$ &  0.828 & 0.032 \\
%\hline
$4$ &  0.880 &  0.020\\
%\hline
$5$ & 0.955 & 0.055\\
%\hline
$6$ & 1.047 & 0.037\\
%\hline
$7$ & 1.125 & 0.041  \\
%\hline
$8$ & 1.205 & 0.039 \\
%\hline
$9$ & 1.422 & 0.178\\
\hline\hline
\end{tabular}
\end{center}
\end{table}

\begin{table}
\begin{center}
 \caption[]{{Fit results for a signal made of $S$-wave, $\akst$ and $\akstp$ 
components. The $S$-wave phase 
is measured for several values of the $K\pi$ mass and its amplitude is parameterized
according to Eq.~(\ref{eq:samp}). The two last columns give the values of the 
$P$-wave phase, which includes $\akst$ and $\akstp$ components,
and the values of the difference between the $S$- and $P$-wave phases.
Quoted uncertainties are statistical only, systematic uncertainties are
given in Table \ref{tab:syst_final_MIAMP_Pprime}. The same 
uncertainties apply to $\delta_S$ and $\delta_S-\delta_P$.}
  \label{tab:fit_data_ampphase_3}}
\begin{tabular}{c c c c}
\hline\hline
variable & result & &\\
\hline
$r_{\kstp}$ & $0.079\pm0.004$ & &\\
%\hline
$\delta_{\kstp}(^{\circ})$ & $-8.9 \pm 21.5 $ & & \\
%\hline
$r_{S}$ & $0.463 \pm 0.068 $ & & \\
%\hline
$r^{(1)}_{S}$ & $0.21 \pm 0.18$ & &\\
\hline\noalign{\vskip1pt}
&$\delta_S(^{\circ})$ & $\delta_P(^{\circ})$ & $\delta_S-\delta_P(^{\circ})$\\
\noalign{\vskip1pt}
\hline
$\delta_{1}$ & $16.8 \pm 11.7$ & 2.0 & 14.8\\
%\hline
$\delta_{2}$ & $31.3 \pm 5.5$ &  4.4 & 26.9\\
%\hline
$\delta_{3}$ & $30.4 \pm 3.1$ &  13.6& 16.9\\
%\hline
$\delta_{4}$ & $34.7 \pm 2.6$ &  54.0 & -19.3\\
%\hline
$\delta_{5}$ & $47.7 \pm 1.4$ & 152.2 & -104.4\\
%\hline
$\delta_{6}$ & $55.0 \pm 4.2$ & 161.4 & -106.4\\
%\hline
$\delta_{7}$ & $71.2 \pm 6.9$ & 159.1 & -87.9  \\
%\hline
$\delta_{8}$ & $60.6 \pm 12.8$ & 148.1 & -87.5 \\
%\hline
$\delta_{9}$ & $85.3 \pm 8.8$ & 130.9 & -45.6\\
\hline\hline
\end{tabular}
\end{center}
\end{table}

\begin{figure*}[!htb]
	\centering
\includegraphics[width=14cm]
{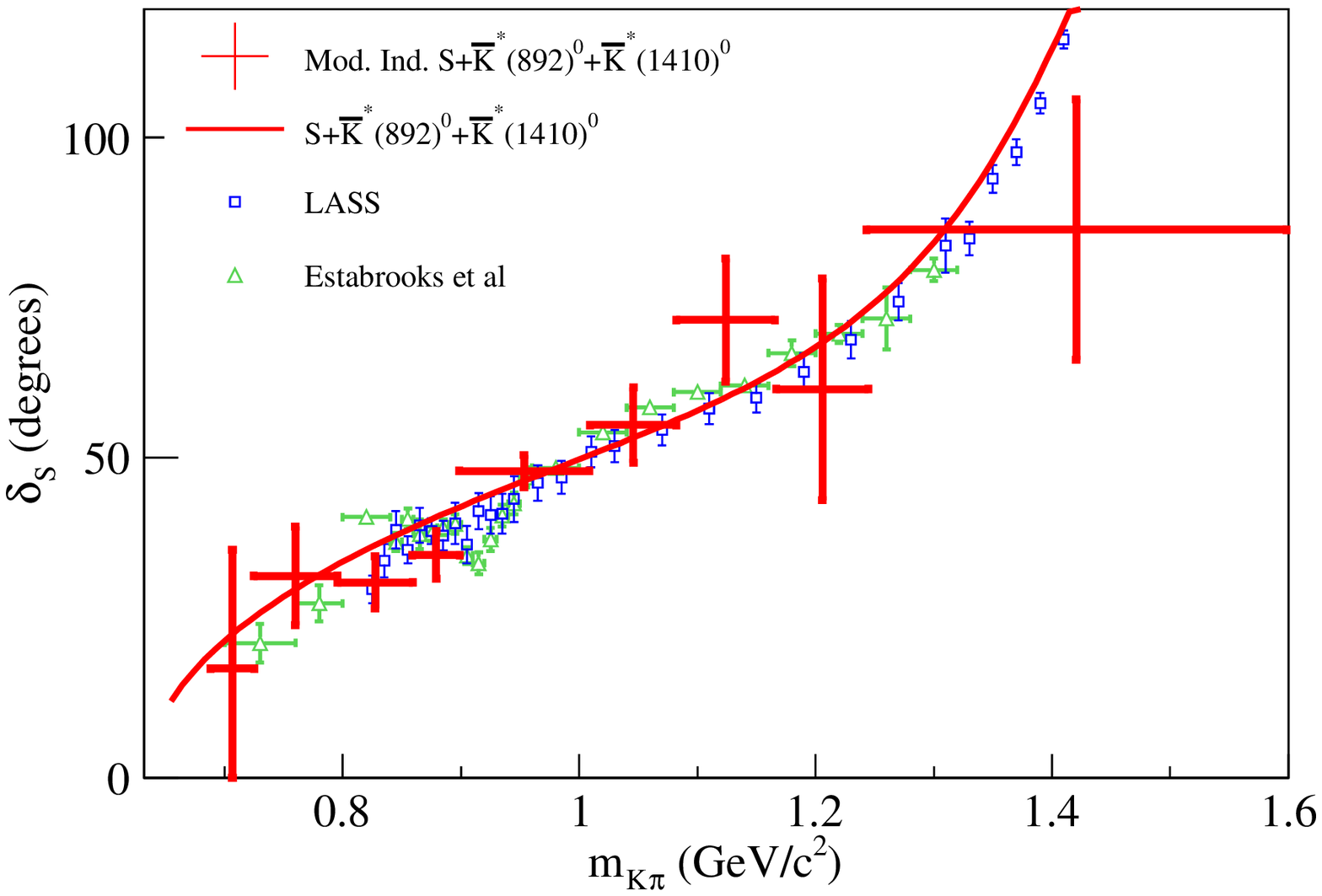}
\caption{{ (color online) Points (full circles) give the $S$-wave phase variation 
assuming a signal containing $S$-wave, $\akst$ and $\akstp$ components. 
The $S$-wave phase 
is assumed to be constant within each considered mass interval and
parameters of the $\akst$ are fixed to the values given in 
the third column of Tab. \ref{tab:fit_data_ampphase_2}. Error bars include systematic uncertainties.
The full line corresponds to the parameterized $S$-wave phase variation 
obtained from the values of the parameters
quoted in the same column of Table \ref{tab:fit_data_ampphase_2}.  
The phase variation measured in $K\pi$ scattering
by Ref. \cite{ref:easta1} (triangles) and LASS \cite{ref:lass1} (squares), 
after correcting 
for $\delta^{3/2}$, are given.}
	\label{fig:Swave_phase_with_Pprime_neat}}
\end{figure*}

The variation of the $S$-wave phase is given in Fig.~\ref{fig:Swave_phase_with_Pprime_neat} 
and compared with LASS results and with the result found in 
Section \ref{sec:SPPrime_1} where the $S$-wave phase variation was parameterized versus the $K\pi$ mass.
Systematic uncertainties are given  
in Table \ref{tab:syst_final_MIAMP_Pprime}.

\begin{figure*}[htbp!]
  \begin{center}
\mbox{\epsfig{file=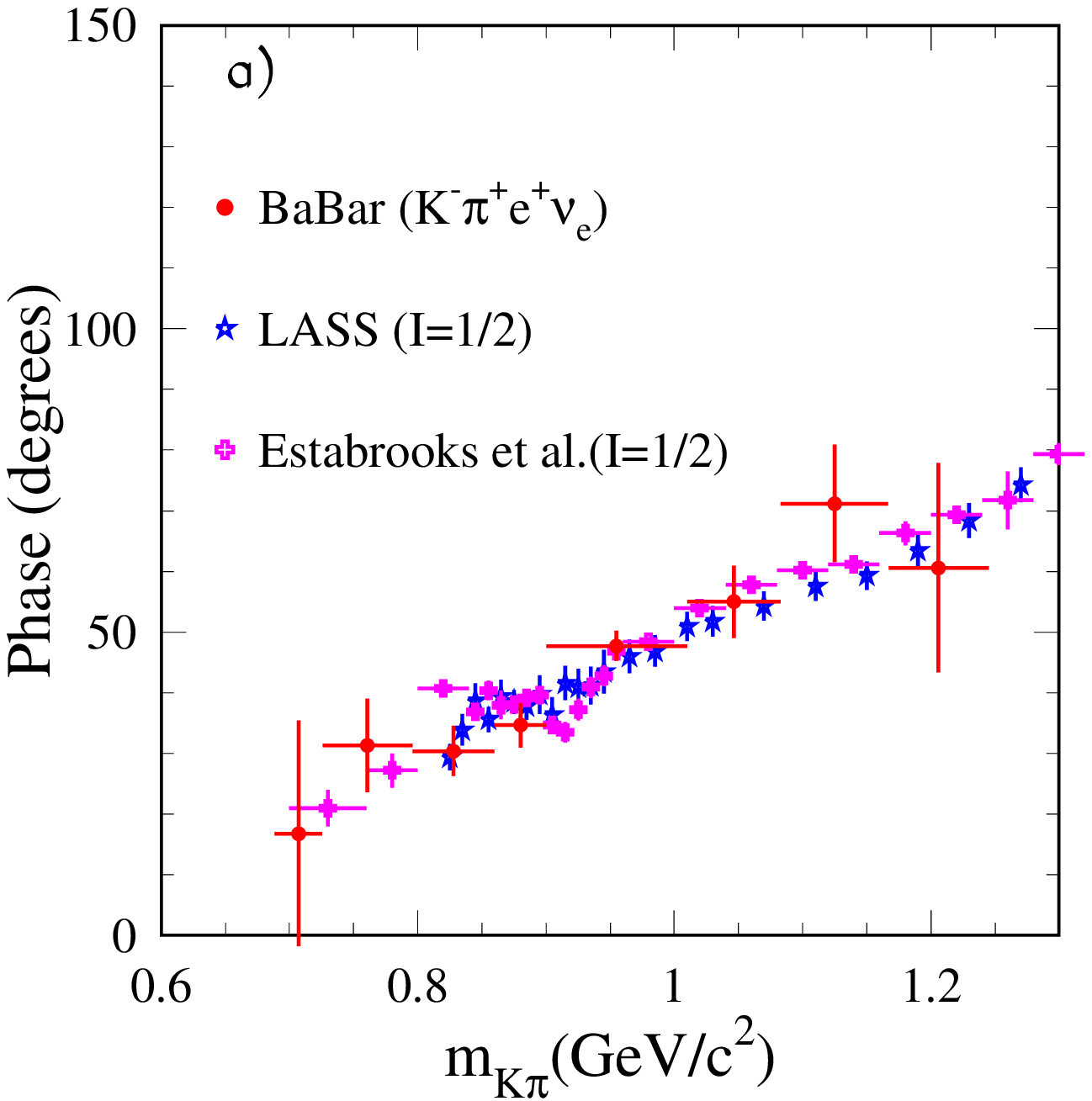,width=.4\textwidth}
\epsfig{file=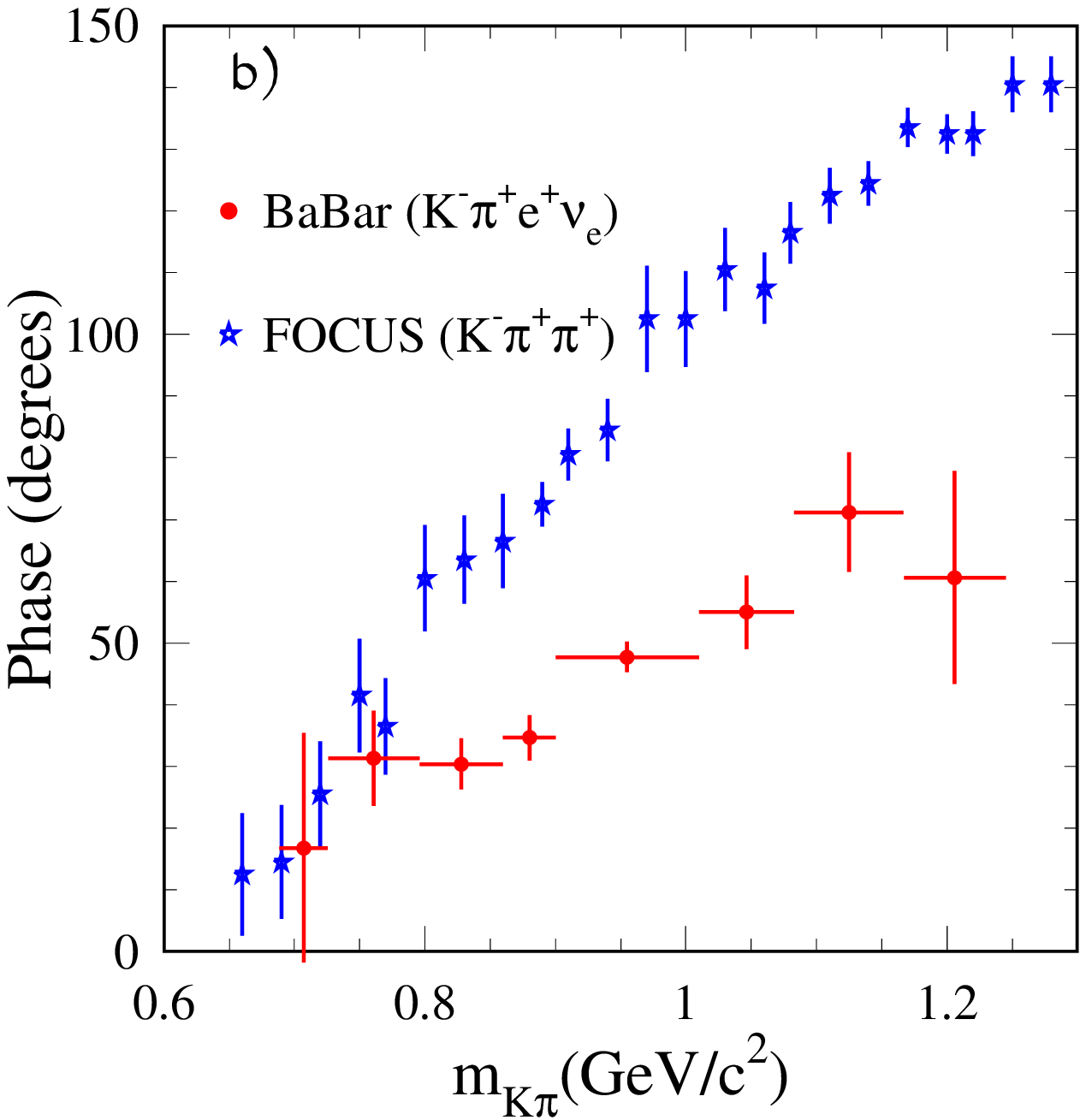,width=.4\textwidth}
}
\mbox{\epsfig{file=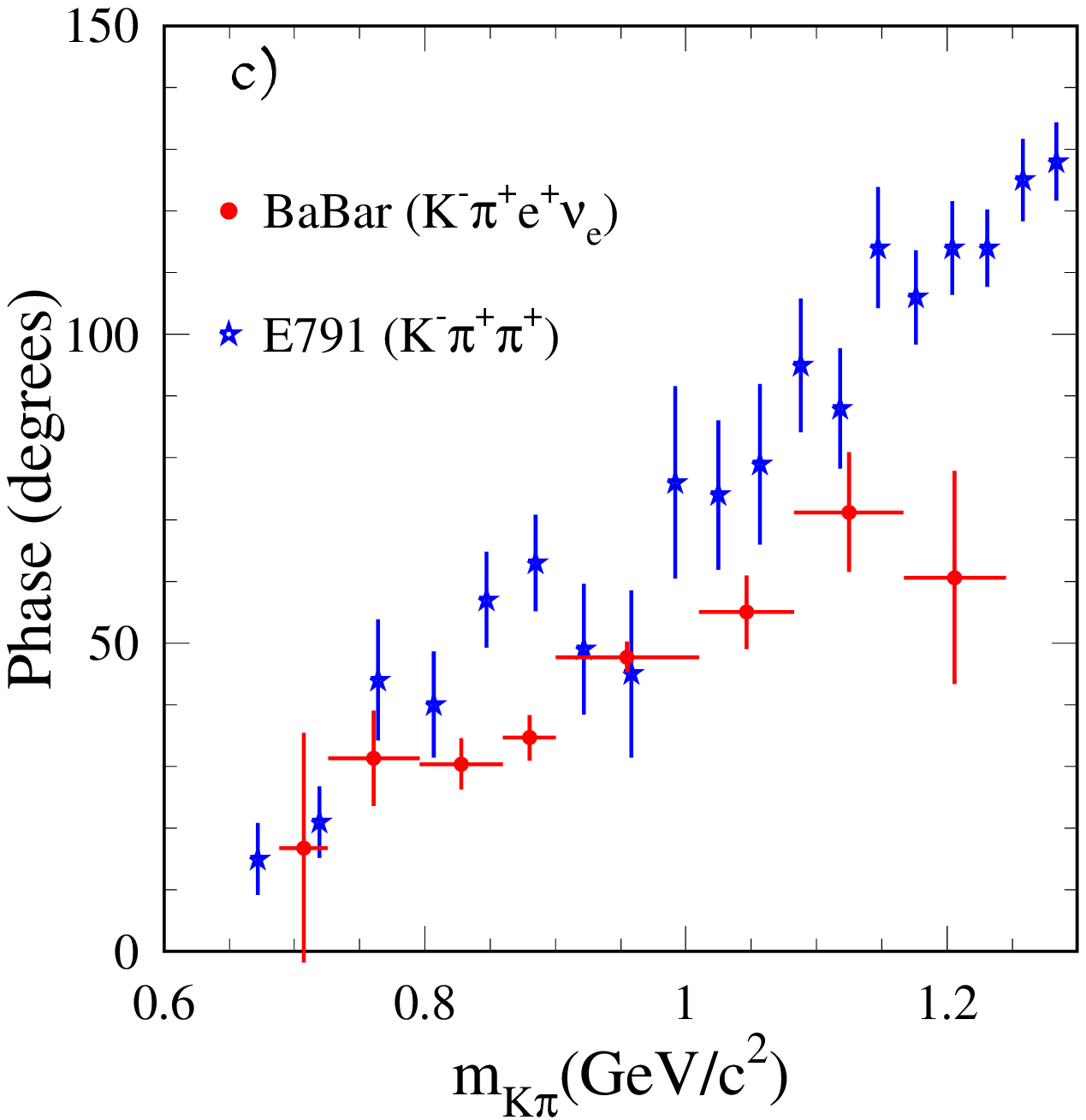,width=.4\textwidth}
\epsfig{file=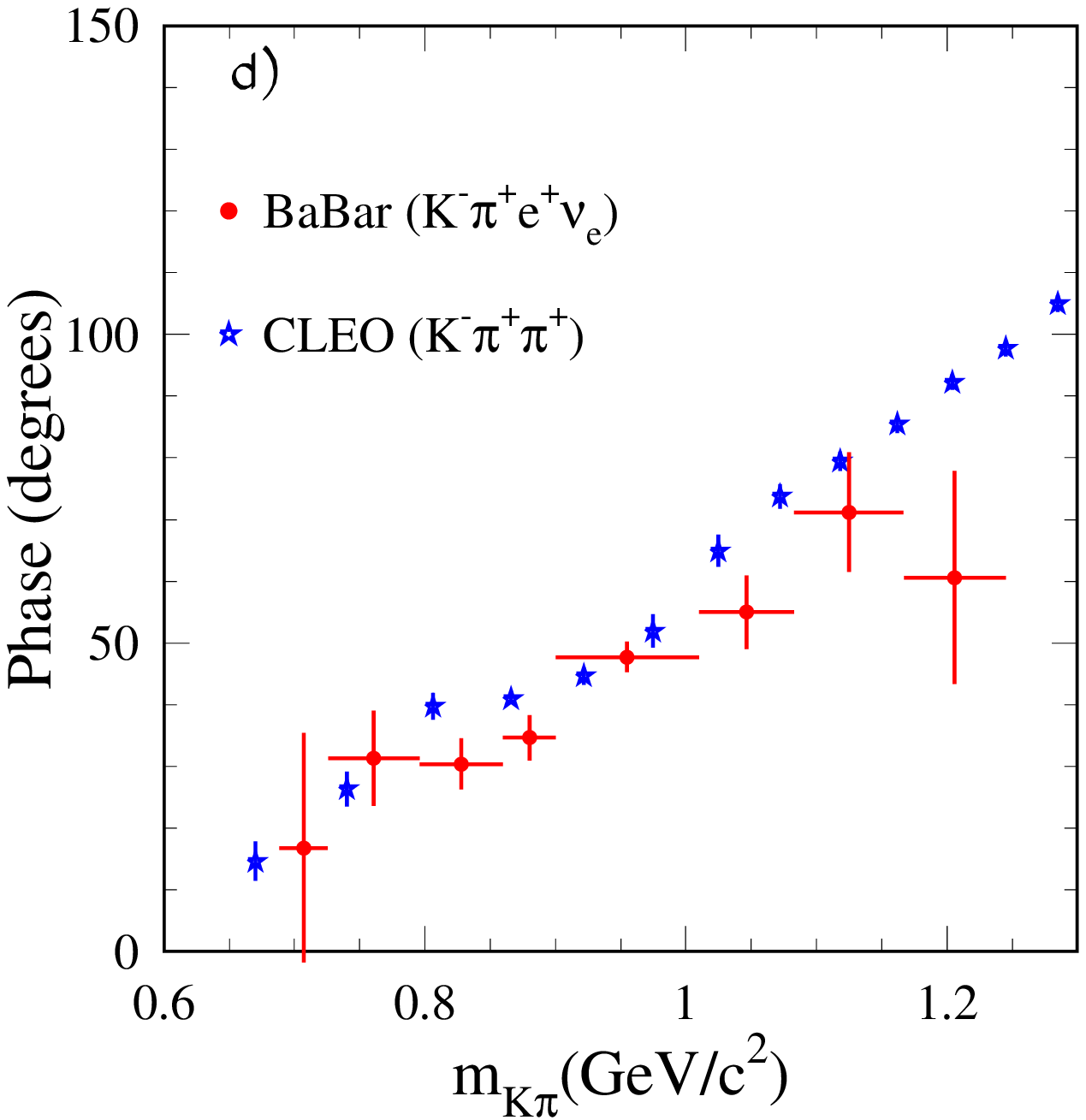,width=.4\textwidth}
}
  \end{center}
  \caption[]{{(color online) Comparison between present measurements of the $I=1/2$ 
$S-$wave phase variation with the $K\pi$ mass and previous results
from Estabrooks et al. \cite{ref:easta1}, LASS \cite{ref:lass1}, 
E791 \cite{ref:e791kpipi}, FOCUS \cite{ref:focuskpipi,ref:focuskpipi2}, 
and CLEO \cite{ref:kpipi_cleoc}.}
   \label{fig:sphase_babar_expts}}
\end{figure*}

In Fig.~\ref{fig:sphase_babar_expts} measured values
of the $S$-wave phase obtained by various experiments in the elastic
region are compared. Fig.~\ref{fig:sphase_babar_expts}-a is a zoom of Fig.
\ref{fig:Swave_phase_with_Pprime_neat}. 
Fig.~\ref{fig:sphase_babar_expts}-b to -d compare present measurements
with those obtained in Dalitz plot analyses of the decay
$\Dp  \rightarrow \Km  \pip \pip $. 
For the latter, the $S$-wave phase is obtained by reference
to the phase of the amplitude of one of the contributing channels
in this decay. To draw
the different figures it is assumed that the phase of the $S$-wave is
equal at $m_{K\pi}=0.67~\gevcc$ to the value given by the fitted
parameterization on LASS data. It is difficult to draw clear conclusions
from these comparisons as Dalitz plot analyses do not provide usually
the phase of the $I=1/2$ amplitude alone but the phase for the
total $S$-wave amplitude.

\subsection{$\delta_S-\delta_P$ measurement}
\begin{figure*}[!htbp]
	\centering
\includegraphics[width=14cm]{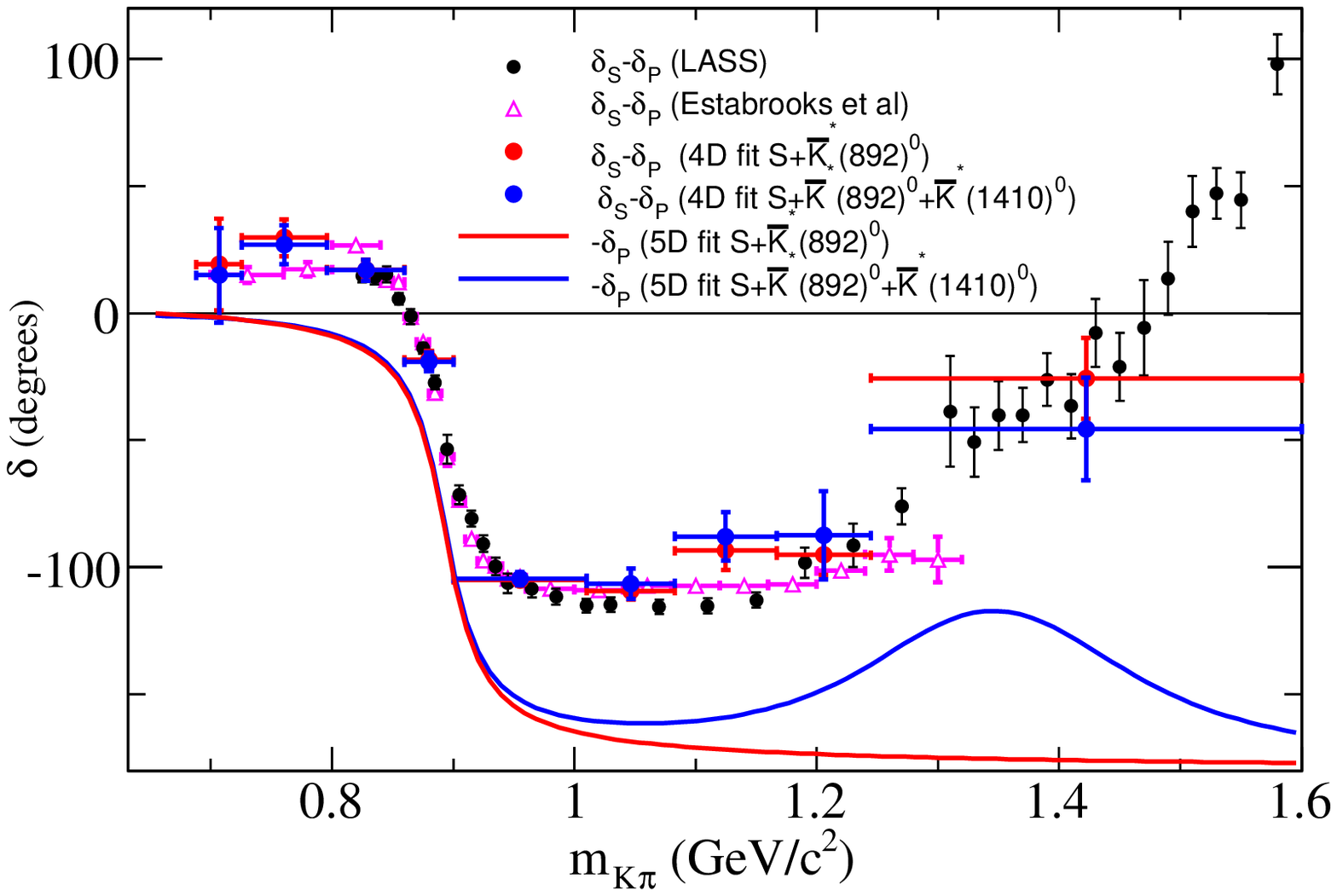}
\caption{{(color online) Difference between the $I=1/2$ $S$- and $P$-wave phase versus the $K\pi$ mass. Measurements 
are similar whether or not the $\akstp$ is included in the $P$-wave parameterization. Results are compared
with measurements from $K\pi$ scattering \cite{ref:easta1,ref:lass1}.
The continuous and dashed lines give the phase variation 
with a minus sign ($-\delta_P$)
for the $\akst$ and $\akst+\akstp$, respectively. The difference between these
curves and the measured points corresponds to the $S$-wave contribution.}
	\label{fig:dsmdp}}
\end{figure*}

As explained in previous sections, measurements are sensitive to the phase difference between $S$- and $P$-waves.
This quantity is given in Fig.~\ref{fig:dsmdp} for different values of
the $K\pi$ mass using results from the fit explained in Section 
\ref{sec:MIAMP_SPP}. Similar values are obtained
if the $\akstp$ is not included in the $P$-wave.

\subsection{Search for a $D$-wave component}
A $D$-wave component, assumed to correspond to the $\akstd$, is added
in the signal model using expressions given in Eq.~(\ref{eq:f123}) and 
(\ref{eq:ffd3}-\ref{eq:f2d}). As the phase of the $\akstp$, relative to the 
$\akst$, is compatible with zero this value is imposed in
the fit. For the $D$-wave, its phase $(\delta_D)$ is allowed to be 
zero or $\pi$. Fit results are given in the last column of Table
\ref{tab:fit_data_ampphase_2}. The total $\chi^{2}$ value is 2888 and 
the number of degrees of freedom is 
2786. This corresponds to a probability of $8.8\%$. 
The value zero is favored
for $\delta_D$. 
The fraction of the decay rate which corresponds to the $D$ wave
is given in Table \ref{tab:fr_SPPpr_SP} and is similar to 
the $\akstp$ fraction.

\section{Decay rate measurement}
\label{sec:decayrate}
The $\Dp  \rightarrow \Km  \pip  \ep  \nue (\g)$
branching fraction is measured relative to the reference decay
channel,  
$\Dp  \rightarrow \Km  \pip  \pip (\g)$. Specifically, in Eq.~(\ref{eq:rd1}) we
compare the ratio of rates for the decays 
$\Dp  \rightarrow \Km  \pip  \ep  \nue (\g)$ and $\Dp  \rightarrow \Km  \pip  \pip (\g)$
in data and simulated events: this way, many systematic uncertainties cancel:

{\small
\begin{eqnarray}
R_{D} &=& \frac{\BR(\Dp  \rightarrow \Km \pip  \ep  \nue )_{data}}{\BR(\Dp  \rightarrow \Km  \pip  \pip )_{data}}\\
 & =&  \frac{N(\Dp  \rightarrow \Km \pip  \ep  \nue )_{data}} {\epsilon(\Km  \pip  \ep  \nue )_{data}}\nonumber\\
&\times&\frac{\epsilon(\Km  \pip  \pi)_{data} }{N(\Dp  \rightarrow \Km \pip  \pip  )_{data}}
\times \frac{ {\cal L} (K\pi \pi)_{data} }{{\cal L}(K\pi e \nu)_{data}}. \nonumber 
\label{eq:rd1}
\end{eqnarray}
}
Introducing the reconstruction efficiency measured for the two channels with simulated 
events, this expression can be written:

{\small
\begin{eqnarray}
R_{D} &=& \frac{N(\Dp  \rightarrow \Km  \pip  \ep  \nue )_{data}}{N(\Dp  \rightarrow \Km \pip  \pip )_{data}} \times   \frac{ {\cal L} (K\pi \pi)_{data} }{{\cal L}(K\pi e \nue )_{data}} \nonumber\\
&\times& \frac{\epsilon( \Km \pip  \ep  \nue )_{MC}}{\epsilon( \Km  \pip  \ep  \nue )_{data}} \times
\frac{\epsilon(\Km \pip \pip )_{data}}{\epsilon( \Km \pip \pip )_{MC}}\nonumber\\
& \times& \frac{\epsilon( \Km \pip  \pip )_{MC}}{\epsilon( \Km  \pip  \ep  \nue )_{MC}} 
\label{eq:rd2}
\end{eqnarray}
}
The first line in this expression is the product between the ratio of measured number of signal events in data for the semileptonic 
and hadronic channels, and the ratio of the corresponding integrated luminosities analyzed for the two channels: 

{\small
\begin{eqnarray}
   \frac{ {\cal L} (K\pi \pi)_{data} }{{\cal L}(K\pi e \nue )_{data}} =  \frac{98.7~ \mbox{fb$^{-1}$}}{100.5 ~\mbox{fb$^{-1}$}}
\end{eqnarray}
}
The second line of Eq.~(\ref{eq:rd2})  corresponds to the ratio between efficiencies in data and in simulation, for the two channels. 
The last line is the ratio between efficiencies for the two channels measured using simulated events.
Considering that a special event sample is generated for the semileptonic decay channel, in which each event
contains a decay $\Dp  \rightarrow \overline{K}^{*0} \ep  \nue ,~\overline{K}^{*0}\rightarrow \Km  \pip $, whereas
the $\Dp  \rightarrow \Km \pip  \pip $ is reconstructed using the 
$\epem \rightarrow \ccbar$ generic simulation, the last term
in Eq.~(\ref{eq:rd2}) is written:

{\small
\begin{eqnarray}
&&\frac{\epsilon(\Km \pip \pip )_{MC}}{\epsilon(\overline{K}^{*0} \ep  \nue )_{MC}}=\frac{N(\Dp  \rightarrow \Km  \pip  \pip )_{MC}}{N(\Dp  \rightarrow \overline{K}^{*0} \ep  \nue )_{MC}} \\
& &\times
\frac{N(\Dp  \rightarrow \overline{K}^{*0} \ep  \nue )_{MC}^{gen}}{2N(c\bar{c})_{K\pi\pi}{\cal P}(c \rightarrow \Dp )\BR(\Dp  \rightarrow \Km  \pip  \pip )_{MC} }\nonumber
\end{eqnarray}
}

where:
\begin{itemize}
\item $N(\Dp  \rightarrow \overline{K}^{*0} \ep  \nue )_{MC}^{gen} =1.17 \times 10^{7} $ is the number of generated signal events;\\
\item $N(c\bar{c})_{K\pi\pi}= 1.517 \times 10^{8}$ is the number of $\epem \rightarrow \ccbar$ events analyzed to reconstruct the
$\Dp  \rightarrow \Km  \pip  \pip $ channel;\\
\item ${\cal P}(c \rightarrow \Dp )=26.0\%$ is the probability that a $c$-quark hadronizes into a $\Dp $ in simulated events. The $\Dp $ is prompt
or is cascading from a higher mass charm resonance;\\
\item $ \BR(\Dp  \rightarrow \Km  \pip  \pip )_{MC}= 0.0923$ is the branching fraction used in the simulation.
\end{itemize}

\subsection{Selection of candidate signal events}
To minimize systematic uncertainties, common selection criteria
are used, as much as possible, to reconstruct the two decay channels.

\subsubsection{The $\Dp  \rightarrow \Km  \pip  \pip $ decay channel}
\begin{figure}[!htbp]
	\centering
\includegraphics[width=8cm]{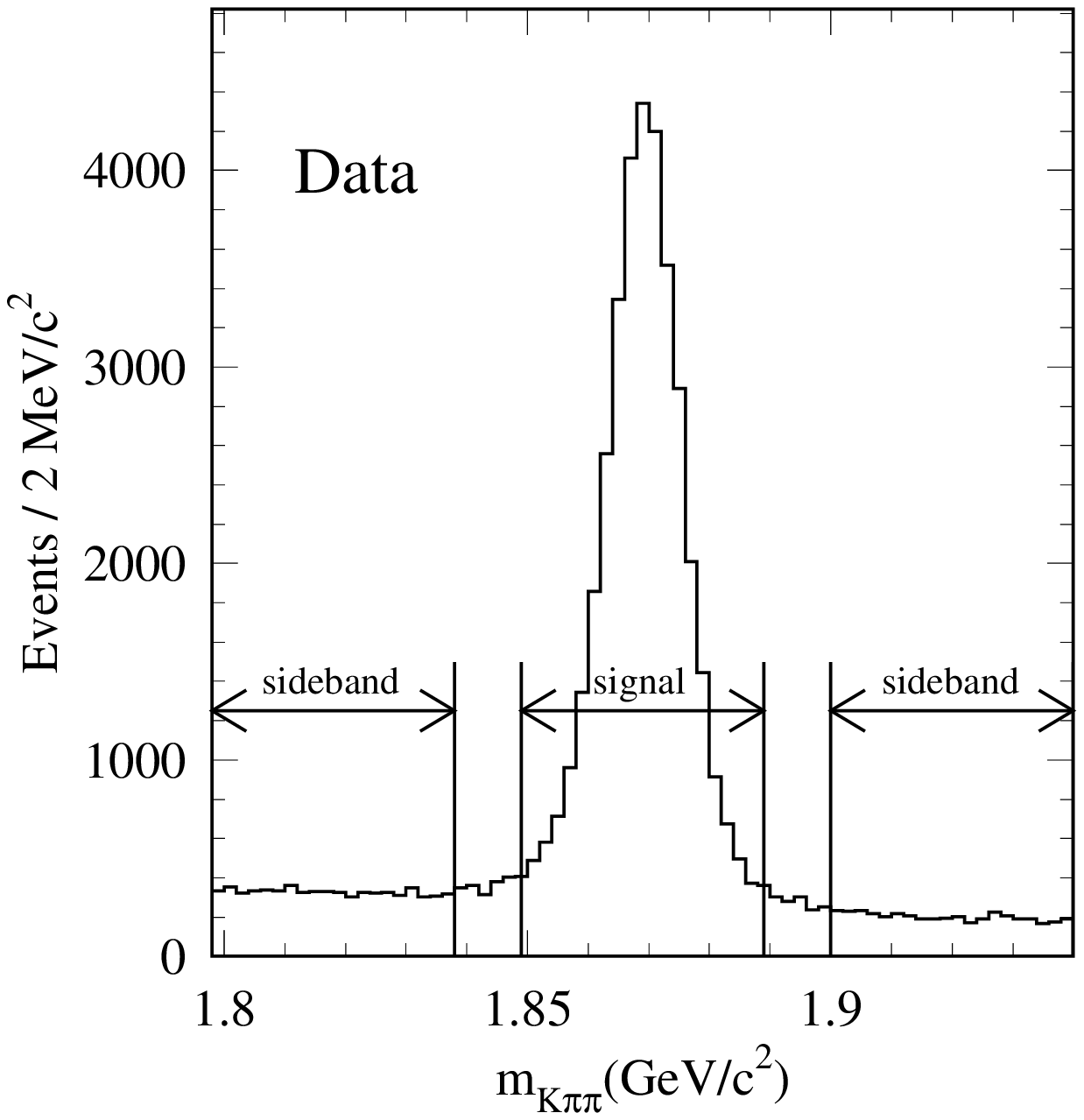}
\caption{{$\Km \pip \pip $ mass distribution measured in data. The signal
and sideband regions are indicated.}
	\label{fig:mkpipi}}
\end{figure}

As compared to the semileptonic decay channel, the selection criteria 
described in Section \ref{sec:sigsel} are used, apart for those involving the lepton.
The number of signal candidates is measured from the
$\Km \pip \pip $ mass distribution, after subtraction of events
situated in sidebands.
The signal region corresponds to the mass interval
$[1.849,~1.889]~\gevcc$ whereas sidebands are selected
within $[1.798,~1.838]$ and $[1.900,~1.940]~\gevcc$.
Results are given in Table \ref{tab:rate} and an example of
the $\Km \pip \pip $ mass distribution measured on data is 
displayed in Fig.~\ref{fig:mkpipi}.

The following differences between data and simulation are considered:

\begin{itemize}
\item {\it the signal mass interval.} 

\begin{table}[!htbp]
\begin{center}
 \caption[]{{{Measured numbers of signal events in data and simulation satisfying $F_{cc}>0.5$.}}
  \label{tab:rate}}
\begin{tabular}{c c c }
\hline\hline
Channel & Data & Simulation   \\
\hline\noalign{\vskip1pt}
$\Km  \pip  \ep  \nue $& $70549 \pm 363$ & $330969$ \\
$\Km  \pip  \pip  $& $52595\pm251$ & $68468\pm283$\\
 \hline\hline	
\end{tabular}
\end{center}
\end{table}

Procedures have been defined in Section \ref{sec:massscale} such that
the average mass and width of the $\Dp  \rightarrow \Km \pip \pip $
reconstructed signal in data and simulated events are similar. 

\item {\it the Dalitz plot model.}
Simulated events are generated using a model which differs from
present measurements of the event distribution over the Dalitz plane.
Measurements from CLEO-c \cite{ref:kpipi_cleoc} are used to reweight simulated events and
we measure that the number of reconstructed signal events changes by a factor
$1.0017\pm 0.0038$. 
This small variation is due to the approximately uniform acceptance of the analysis 
for this channel.

\item {\it the pion track.}
As compared with the $\Km \pip \ep \nue $ final state, there is a $\pip $
in place of the $\ep $ in the reference channel. As there is no requirement
on the PID for this pion we have considered that possible differences between
data and simulation on tracking efficiency cancel when considering
the simultaneous reconstruction of the pion and the electron. What remains
is the difference between data and simulation for electron identification
which is included in the evaluation of systematic uncertainties.

\end{itemize}

\subsubsection{The $\Dp  \rightarrow \Km  \pip  \ep  \nue $ decay channel}

The same data sample as used to measure the $\Dp  \rightarrow \Km  \pip  \pip $
is analyzed. Signal events are fitted as in Section \ref{sec:LASSAMP_sec}.
The stability of the measurement is verified versus the value of the cut
on $F_{cc}$ which is varied between 0.4 and 0.7.
Over this range the number of signal and background events change by factors
0.62 and 0.36 respectively. The variation of the ratio between the number of selected events, 
\begin{equation}
R_N = \frac{N(\Dp  \rightarrow \Km  \pip  \pip )_{MC}}{N(\Dp  \rightarrow \overline{K}^{*0} \ep  \nue )_{MC}}
\frac{N(\Dp  \rightarrow \Km \pip  \ep  \nue )_{data}} {N(\Dp  \rightarrow \Km \pip  \pip  )_{data}}
\end{equation}
in data and simulation is given in Table \ref{tab:stabrn}.

\begin{table*}[!htbp]
\begin{center}
 \caption[]{{{Variation of the ratio between the numbers of selected events
in data and simulation for different values of the cut on $F_{cc}$.}}
  \label{tab:stabrn}}
\begin{tabular}{c c c c}
\hline\hline
   &$F_{cc}>0.4$ &$F_{cc}>0.5$ &$F_{cc}>0.7$ \\           
\hline\noalign{\vskip1pt}
$N(\Dp  \rightarrow \Km  \pip  \pip )_{MC}$ & $72206\pm292$ &$ 60468\pm283$ & $59259\pm259$\\
$N(\Dp  \rightarrow \Km \pip  \pip  )_{data}$ & $55361\pm260$ & $52595\pm251$ & $45627\pm230$ \\
$N(\Dp  \rightarrow \overline{K}^{*0} \ep  \nue )_{MC}$ & $381707$ & $330969$ & $237104$ \\
$N(\Dp  \rightarrow \Km \pip  \ep  \nue )_{data}$ & $81322 \pm 383$ & $70549 \pm 363$ & $50989 \pm 303$\\
\noalign{\vskip1pt}
\hline
$R_N$ & $0.2779\pm0.022$ & $0.2775\pm0.0023$ & $0.2793\pm0.0026$ \\
 \hline	\hline
\end{tabular}
\end{center}
\end{table*}
Relative to the value for the nominal cut ($F_{cc}>0.5$), the value of $R_N$
for $F_{cc}>0.4$ is higher by $0.00038\pm 0.00063$ and for $F_{cc}>0.7$
it is higher by $0.0018\pm 0.0011$. Quoted uncertainties take into account
events that are common when comparing the samples. These variations are compatible with
statistical fluctuations and no additional systematic uncertainty is included.

To select semileptonic decay candidates a cut is applied on the probability of the
$\Dp $ mass-constrained fit at $0.01$. In a previous analysis 
of the decay $\Dz  \rightarrow \Km \ep  \nue $ \cite{ref:kenu} we measured
a value of $1.0062\pm0.0006$ for the ratio between the efficiency of this cut
in simulation and data. We use the same value in the present analysis because
this probability depends on the capability to reconstruct the $D$ direction and momentum
and to estimate corresponding uncertainties on these quantities which
are obtained, not from the studied decay channel, but from the rest of the event.

\subsection{Decay rate measurement}
Measurement of the 
$\Dp  \rightarrow \Km \pip  \ep  \nue $ branching fraction and of the
contributing $S-$wave, $\akst$, $\akstp$ and $\akstd$ components
is important to verify if the sum of exclusive channels in $D$-meson
semileptonic decays agrees with the inclusive value.
From the measurement of $\BR(\Dp  \rightarrow \akst \ep  \nue )$
the value of $|A_1(0)|$ is obtained and provides, with $r_2$ 
and $r_V$, the absolute
normalization for the corresponding hadronic form factors. These values can
be compared with Lattice QCD determinations.

Combining all measured quantities in Eq.~(\ref{eq:rd2}), the relative
decay rate is:
\begin{equation}
 R_D= 0.4380\pm0.0036\pm0.0042
\end{equation}
where uncertainties are statistical and systematic respectively.
Using the CLEO-c value for the branching fraction
$\BR(\Dp  \rightarrow \Km  \pip  \pip )=(9.14 \pm 0.20)\%$ \cite{ref:cleockpipi}, gives:
\begin{equation} 
\BR(\Dp  \rightarrow \Km \pip  \ep  \nue ) = ( 4.00 \pm 0.03 \pm 0.04 \pm 0.09)\times 10^{-2}
\end{equation}
where the last quoted uncertainty comes from the 
accuracy of $\BR(\Dp  \rightarrow \Km  \pip \pip )$.
To evaluate the contribution from the $\overline{K}^{*0}$, results obtained
with the $S+\akst+\akstp$ signal model are used.
The branching fraction for $\Dp \rightarrow \overline{K}^{*0} \ep  \nue $ is obtained after subtracting the $S$- and $\akstp$-wave contributions:
\begin{eqnarray}
~&\BR(\Dp \rightarrow \overline{K}^{*0} \ep  \nue )\times \BR(\overline{K}^{*0}\rightarrow \Km  \pip ) =\\
~& (3.77\pm0.04 \pm0.05 \pm0.09)\times10^{-2}. \nonumber
\end{eqnarray}
 The last uncertainty corresponds to external inputs.

The corresponding value of $A_{1}(0)$ is obtained by integrating Eq.~(\ref{eq:decay1}), restricted to the $\overline{K}^{*0}$ contribution,
over the three angles:
\begin{eqnarray}
\frac{{\rm d}\Gamma}{{\rm d}q^{2}{\rm d}m^{2}}& =& \frac{1}{3}\frac{G_F^2 \left | \Vcs \right |^2}
{\left ( 4\pi\right )^5 m_D^2} \beta \, p_{K\pi}\nonumber\\
&\times&\left [\frac{2}{3} \left \{|{\cal F}_{11}|^2 + |{\cal F}_{21}|^2 + |{\cal F}_{31}|^2 \right \}\right ]
\end{eqnarray}

Assuming that the $\kst$ meson is infinitely narrow,
integrating over the remaining variables gives:

\begin{eqnarray}
\Gamma &=& \frac{\hbar \BR(\Dp  \rightarrow \overline{K}^{*0} \ep  \nue )\BR(\overline{K}^{*0}\rightarrow \Km  \pip ) }{\tau_{\Dp }}\\
&=&\frac{G_F^2 \left | \Vcs \right |^2}
{96\pi^3} \frac{2}{3}|A_{1}(0)|^{2} \cal I, \nonumber
\end{eqnarray}

with 
\begin{eqnarray}
{\cal I }&=& \int^{q^{2}max}_{0} \frac{p_{K\pi}\,q^2}{|A_{1}(0)|^{2}m_D^2} \nonumber\\
& \times & \left [|H_0|^2 + |H_+|^2 + |H_-|^2 \right ]{\rm d}q^{2}
\end{eqnarray}

and:
\begin{equation}
A_1(0) = 0.6200 \pm 0.0056 \pm 0.0065 \pm 0.0071.
\label{eq:fplus}
\end{equation}
For this last evaluation, the values 
$\tau_{\Dp }= (10.40\pm0.07) \times 10^{-13}s$ for the $\Dp $ lifetime \cite{ref:pdg10}
and $\Vcs =0.9729\pm0.0003$ are used. Corresponding uncertainties are
included in the last quoted error in Eq.~(\ref{eq:fplus}).

If instead of considering a $\kst$ with zero width, the fitted mass 
distribution of the resonance is used in the integral of
the differential decay rate versus $q^2$ and $m^2$, the form
factor normalization becomes:
\begin{equation}
A_1(0)|_{q^2,m^2} = 0.9174 \pm 0.0084 \pm 0.0097 \pm 0.0105.  
\end{equation}
This value depends also on the normalization adopted for the
mass distribution which is given in Eq.~(\ref{eq:alphadef}).

\section{Summary}
\label{sec:summary}
We have studied the decay $\Dp  \rightarrow \Km \pip  \ep  \nue $
with a sample of approximately $244\times10^3$ 
signal events, which greatly exceeds 
any previous
measurement. The hadronic system in this decay is dominated by the $\overline{K}^{*0}$ meson. 
In addition to the $\overline{K}^{*0}$ meson we measure a contribution of the $\Km \pip $
$S$-wave component of $(5.79 \pm 0.16 \pm 0.15)\%$. We find a small 
contribution
from the $\akstp$ equal to $(0.33 \pm 0.13 \pm 0.19)\%$. 
This value agrees with the naive expectation based on corresponding 
measurements
in $\tau$ decays. The relative phase
between the $\akst$ and $\akstp$ components is compatible with zero
whereas there is a negative sign between the $S$- and 
$P$-wave amplitudes.
A fit to data of similar probability is obtained including
a $D$-wave component with a fraction equal to $(0.19 \pm 0.09 \pm 0.09)\%$.
In this case the $\akstp$ fraction becomes $(0.16 \pm 0.08 \pm 0.14)\%$. 
As these two components do not exceed a significance of three standard 
deviations,  upper limits at the 90$\%$ C.L. are quoted in
Table \ref{tab:comparison}.

Using a model for signal which includes $S$-wave, $\akst$- and $\akstp$ contributions,
hadronic form factor parameters of the $\overline{K}^{*0}$ component are 
obtained
from a fit to the five-dimensional decay distribution, 
assuming single pole
dominance: $r_V=V(0)/A_1(0)=1.463\pm0.017\pm0.032$,
$r_2=A_2(0)/A_1(0)=0.801\pm0.020\pm0.020$ and the pole mass of the axial vector form
factors  $m_A=\left ( 2.63 \pm 0.10 \pm 0.13\right ) ~\gevcc$.
For comparison with previous measurements we also perform a fit to data 
with fixed
pole mass $m_A=2.5~\gevcc$ and $m_V=2.1~\gevcc$ and including only
the $S$ and $\akst$ signal components; it gives 
$r_V=1.493\pm0.014\pm0.021$ and $r_2=0.775\pm0.011\pm0.011$. 

We have measured the phase of the $S$-wave component for several values of the $\Km  \pip $ mass.
Contrary to similar analyses using charm meson decays, as in  $\Dp  \rightarrow \Km  \pip  \pip $,
we find agreement 
with corresponding $S$-wave phase measurements done  in $\Km p$
interactions producing  $\Km \pip $ at small transfer. This is a confirmation of these
 results and illustrates the importance of final state interactions in $D$-meson hadronic
decays. As compared with elastic $\Km \pip $ scattering there is an additional
negative
sign between the $S$- and $P$-wave, in the $\Dp $ semileptonic decay channel.
This observation does not contradict the Watson theorem.
We have determined the parameters of the $\kst$ meson and found, in particular, a width
smaller than the value quoted in \cite{ref:pdg10}. Our result agrees 
with recent measurements
from FOCUS \cite{ref:focus1}, CLEO \cite{ref:kpipi_cleoc} and, $\tau$ decays 
(for the charged mode)
\cite{ref:taubelle}.
Comparison between these measurements and present world average values is illustrated
in Tab. \ref{tab:comparison}. Our measurements of the $S$-wave phase
have large uncertainties in the threshold region and it remains to evaluate
how they can improve the determination of chiral parameters using, 
for instance,
the framework explained in ref. \cite{ref:seb1}.

\begin{table*}[!htb]
\begin{center}
 \caption[]{{Comparison between these measurements and present world average results. Values for $\BR(\Dp \rightarrow \akstp / \akstd \ep \nue )$ are corrected for their respective
branching fractions into $\Km \pip $.}
  \label{tab:comparison}}
\begin{tabular}{c c c }
\hline\noalign{\vskip1pt}
Measured quantity & This analysis & World average \cite{ref:pdg10}\\
\noalign{\vskip1pt}
\hline\hline\noalign{\vskip1pt}
{ $m_{\kst}(\mevcc)$} & {$895.4\pm{0.2}\pm0.2$}& $895.94\pm{0.22}$ \\
{ $\Gamma^0_{\kst}(\mevcc)$} & {$46.5\pm{0.3}\pm0.2$} & $48.7\pm{0.8}$\\
{$r_{BW}(\gevc)^{-1}$ }& {$2.1\pm{0.5}\pm 0.5$} & $3.4\pm{0.7}$ \cite{ref:lass1}  \\
\noalign{\vskip1pt}
\hline
{$r_{V}$} & {$1.463\pm{0.017}\pm 0.031$}& $1.62\pm{0.08}$ \\
{$r_{2}$} & {$0.801\pm{0.020}\pm 0.020$} & $0.83\pm{0.05}$   \\
{ $m_{A} (\gevcc)$} & {$2.63\pm 0.10 \pm 0.13$} & no result \\
\hline\noalign{\vskip1pt}
$\BR(\Dp  \rightarrow \Km \pip  \ep  \nue )(\%)$ &$4.00 \pm 0.03 \pm 0.04 \pm 0.09$ & $4.1 \pm 0.6$\\
$\BR(\Dp \rightarrow \Km  \pip  \ep  \nue )_{\akst}(\%)$ & $3.77\pm0.04\pm0.05 \pm0.09 $ &$3.68 \pm 0.21$ \\ 
$\BR(\Dp  \rightarrow \Km \pip  \ep  \nue )_{S-wave}(\%)$ &$0.232 \pm0.007  \pm0.007  \pm0.005  $ & $0.21 \pm 0.06$\\
$\BR(\Dp \rightarrow \akstp \ep  \nue )(\%)$ & $0.30\pm0.12\pm0.18\pm0.06$ ($<0.6$ at 90$\%$ C.L.) & $$ \\ 
$\BR(\Dp \rightarrow \akstd \ep  \nue )(\%)$ & $0.023\pm0.011\pm0.011\pm0.001$ ($<0.05$ at 90$\%$ C.L.) & $$ \\ 
\hline\hline
\end{tabular}
\end{center}
\end{table*}

\section{Acknowledgments}
\label{sec:Acknowledgements}
The authors would like to thank S. Descotes-Genon and A. Le Yaouanc for fruitful discussions especially
on the charm meson semileptonic decay rate formalism. We also
thank V. Bernard, B. Moussallam and E. Passemar for discussions
on chiral perturbation theory and different aspects of the 
$K\pi$ system.  

\input acknowledgements.tex

%--------------------------------------------------------------------------------------

\appendix 

\section{Error matrices for the nominal fit}
\label{sec:appendixa}

The correlation matrix between statistical uncertainties is given
in Tab. \ref{tab:statcorrel} for parameters fitted using the nominal model.
Statistical uncertainties are quoted on the diagonal.
The elements of the statistical error matrix
are equal to $\rho_{ij}\,\sigma_i\,\sigma_j$, where $\rho_{ij}$
is an off diagonal element
or is equal to 1 on the diagonal.
\begin{table*}[htbp!]
\begin{center}
\caption[]{{ Correlation matrix for the $S+\akst +\kstp$ nominal fit.
On the diagonal, statistical uncertainties ($\sigma_i^{stat.}$) of fitted 
quantities ($i$) are given.}
\label{tab:statcorrel}}
{\small
\begin{tabular}{c c c c c c c c c c c c c }
\hline\hline\noalign{\vskip1pt}
$ \Delta M_{\kst}$  &  $ \Delta \Gamma_{\kst}$  &  $\Delta r_{BW}$  &  $\Delta m_{A}$  &  $ \Delta r_{V} $  &  $\Delta r_{2} $  &  $\Delta r_{S}$  & $\Delta r^{(1)}_{S}$  & $\Delta a_{S,BG}^{1/2}$  &  $\Delta r_{\kstp}$ &  $\Delta \delta_{\kstp}$ &  $\Delta N_{S}$  &  $ \Delta N_{B}$   \\
\hline
\textbf{0.211} & ~0.656 & -0.842 & -0.158 & ~0.142 & -0.131 & ~0.116 & -0.043 & -0.673 & ~0.899 & -0.774 & -0.254 & ~0.304 \\
%\hline
 & \textbf{0.315} & -0.614 & -0.002 & ~0.020 & ~0.007 & ~0.007 & ~0.024 & -0.470 & ~0.624 & -0.632 & ~0.027 & -0.021 \\
%\hline
  &  & \textbf{0.476} & ~0.163 & -0.165 & ~0.141 & -0.347 & ~0.270 & ~0.657 & -0.907 & ~0.846 & ~0.334 & -0.394 \\
%\hline
 &  &  & \textbf{0.0972} & -0.548 & ~0.840 & -0.045 & -0.070 & ~0.065 & -0.080 & ~0.087 & ~0.099 & -0.118 \\
%\hline
&  &  & & \textbf{0.0166} & -0.518 & ~0.048 & ~0.034 & -0.060 & ~0.101 & -0.126 & -0.124 & ~0.136 \\
%\hline
 &  &  & &  & \textbf{0.0201} & -0.016 & -0.080 & ~0.051 & -0.058 & ~0.080 & ~0.116 & -0.133 \\
%\hline
 &  &  &  &  & & \textbf{0.0286} & -0.968 & -0.157 & ~0.191 & -0.048 & -0.136 & ~0.159 \\
%\hline
 &  &  &  &  &  &  & \textbf{0.0640} & ~0.130 & -0.133 & -0.043 & ~0.146 & -0.173 \\
%\hline
 &  &  &  &  &  &  &  & \textbf{0.138} & -0.767 & ~0.396 & ~0.148 & -0.179 \\
%\hline
  &  &  &  &  &  &  & &  &\textbf{0.0163}  & -0.721 & -0.269 & ~0.318 \\
%\hline
 &  & &  &  &  &  &  & &  & \textbf{13.0} & ~0.288 & -0.336 \\
%\hline
 &  &  &  &  &  &  &  &  &  &  & \textbf{713.0} & -0.609 \\
%\hline
&  &  &  &  &  &  &  &  &  &  &  & \textbf{613.2} \\
\hline\hline
\end{tabular}
}
\end{center}
\end{table*}

In Tab. \ref{tab:syst_final_SPPprime}, systematic uncertainties
quoted in lines labelled III, VII and XI are the result of several
contributions, combined in quadrature. In Tab. 
\ref{tab:detailed_syst} these
components are detailed because each contribution can induce a positive
or a negative variation of the fitted quantities.

\begin{table*}[htbp!]
\caption[]{{ Systematic uncertainties  on parameters fitted using
a model for the signal which contains $S$-wave, $\akst$ and $\akstp$ components, expressed as $(x[0] - x[i])/\sigma_{stat}$: (IIIa) uncertainty associated with electron identification,
 (IIIb) uncertainty associated with kaon identification, 
(VIIa) : pole mass changed by -30 $\mevcc$ for the decay channel  $D^{0}\rightarrow K^{-}e^{+}\nu_{e}$, (VIIb) : pole mass changed by -100 $\mevcc$ for semileptonic decays of $D^{0}$ and $D^{+}$ into a pseudoscalar meson, (VIIc) : pole mass changed by 100 $\mevcc$ for $D_{s}$ meson semileptonic decays, (VIId-j) refer to  $D\rightarrow V~e^{+}\nu_{e} $ decays; (VIId) : $r_{2}$ changed from 0.80 to 0.85, (VIIe) : $r_{V}$ changed from 1.50 to 1.55, (VIIf) : $m_{A}$ changed from 2.5 to 2.2 $\gevcc$,  (VIIg) : $m_{V}$ changed from 2.1 to 1.9 $\gevcc$,  (VIIh) :$r_{BW}$ changed from 3.0 to 3.3 $\gev^{-1}$, 
 (VIIi) : $\Gamma_{\kst}$ varied by $-0.5~\mevcc$,
(XIa-d) : $\kstp$ mass and width, $\ksts$ mass and width in this order
and using the variations given in Tab. \ref{tab:kpistates} and, (XIe) : $m_V$ changed by
$100~\mevcc$.}
\label{tab:detailed_syst}}
{\small
\begin{tabular}{c c c c c c c c c c c c c c }
\hline\hline\noalign{\vskip1pt}
variation & $ \Delta M_{\kst}$  &  $ \Delta \Gamma_{\kst}$  &  $\Delta r_{BW}$  &  $\Delta m_{A}$  &  $ \Delta r_{V} $  &  $\Delta r_{2} $  &  $\Delta r_{S}$  & $\Delta r^{(1)}_{S}$  & $\Delta a_{S,BG}^{1/2}$  &  $\Delta r_{\kstp}$ &  $\Delta \delta_{\kstp}$ &  $\Delta N_{S}$  &  $ \Delta N_{B}$   \\
\noalign{\vskip1pt}
\hline
$III$ \\
\hline
a & ~0.02 & ~0.03 & -0.03 & ~0.70 & -0.73 & ~0.50 & ~0.12 & -0.07 & ~0.30 & ~0.07 & -0.20 & ~0.07 & -0.08 \\
b & ~0.00 & -0.10 & ~0.05 & -0.07 & -0.05 & -0.17 & -0.07 & ~0.01 & -0.28 & ~0.05 & ~0.10 & -0.07 & ~0.08 \\
\hline\noalign{\vskip1pt}
$VII$ \\
\hline
a & ~0.06 & ~0.06 & -0.05 & -0.10 & -0.05 & ~0.01 & ~0.17 & -0.15 & ~0.01 & ~0.07 & ~0.02 & ~0.05 & -0.06 \\
b & ~0.01 & ~0.01 & -0.01 & ~0.00 & -0.01 & ~0.01 & ~0.01 & -0.01 & ~0.00 & ~0.01 & -0.01 & ~0.00 & -0.01 \\
c & -0.01 & ~0.00 & ~0.01 & -0.01 & ~0.00 & ~0.00 & -0.01 & ~0.01 & ~0.00 & -0.01 & ~0.00 & ~0.00 & ~0.01 \\
d & -0.03 & ~0.00 & ~0.06 & ~0.00 & ~0.02 & ~0.05 & ~0.01 & -0.02 & ~0.02 & -0.02 & ~0.04 & ~0.06 & -0.07 \\
e & -0.03 & -0.01 & ~0.05 & ~0.03 & ~0.07 & ~0.03 & -0.02 & ~0.01 & ~0.01 & -0.03 & ~0.04 & ~0.01 & -0.01 \\
f & ~0.00 & ~0.03 & ~0.07 & -0.16 & ~0.07 & -0.03 & ~0.01 & -0.01 & ~0.05 & ~0.01 & ~0.04 & ~0.12 & -0.15 \\
g & -0.04 & -0.01 & ~0.06 & ~0.03 & ~0.08 & ~0.04 & -0.02 & ~0.01 & ~0.02 & -0.04 & ~0.05 & ~0.02 & -0.03 \\
h & -0.06 & -0.04 & ~0.02 & ~0.00 & ~0.00 & ~0.00 & ~0.00 & ~0.00 & ~0.02 & -0.04 & ~0.06 & ~0.00 & ~0.00 \\
i & -0.02 & -0.02 & ~0.01 & -0.00 & ~0.00 & ~0.01 & ~0.00 & -0.01 & ~0.00 & -0.00 & ~0.00 & ~0.01 & -0.01 \\
%j & -0.29 & -0.05 & ~0.00 & ~0.04 & -0.04 & -0.01 & ~0.02 & -0.01 & ~0.03 & -0.03 & -0.06 & -0.10 & ~0.12 \\
\hline\noalign{\vskip1pt}
$XI$ \\
\hline
a & -0.08 & -0.02 & ~0.08 & ~0.00 & ~0.01 & ~0.00 & -0.35 & ~0.34 & -1.00 & -0.34 & -0.20 & -0.04 & ~0.05 \\
b & ~0.20 & ~0.09 & -0.21 & ~0.03 & -0.03 & ~0.03 & ~0.44 & -0.41 & ~3.09 & ~0.62 & ~0.28 & ~0.07 & -0.08 \\
c & -0.06 & -0.02 & ~0.08 & ~0.01 & -0.01 & ~0.01 & ~0.08 & -0.10 & ~0.05 & -0.39 & -0.08 & ~0.02 & -0.02 \\
d & -0.16 & -0.07 & ~0.17 & ~0.01 & -0.01 & ~0.01 & ~0.05 & -0.06 & ~0.11 & ~0.37 & ~0.18 & ~0.02 & -0.03 \\
e & ~0.00 & ~0.00 & -0.01 & -0.06 & -1.15 & -0.07 & ~0.04 & -0.03 & -0.01 & ~0.00 & ~0.00 & ~0.01 & -0.01 \\
\hline\hline
 \end{tabular}
}
\end{table*}

The error matrix for systematic uncertainties on fitted parameters
is obtained using values of the variations given in Tab. 
\ref{tab:syst_final_SPPprime} and \ref{tab:detailed_syst}.
For each individual source of systematic uncertainty we create a matrix 
of elements equal to the product $\delta_i \,\delta_j$
of the variations observed on the values of the fitted parameters
$i$ and $j$. For systematic uncertainties IX and X, which have a statistical
origin, we multiply these quantities by the corresponding elements
of the statistical correlation matrix (Tab. \ref{tab:statcorrel}). 
These matrices are summed to obtain
the matrix given in Tab. \ref{tab:syst_matrix_SPPprime}. 

\begin{table*}[htbp!]
\begin{center}
\caption[]{{ Correlation matrix for systematic uncertainties of 
the $S+\akst +\kstp$ nominal fit.
On the diagonal, total systematic uncertainties 
($\sigma_i^{syst.}$) on fitted 
quantities ($i$), are given.}
\label{tab:syst_matrix_SPPprime}}
{\small
\begin{tabular}{c c c c c c c c c c c c c }
\hline\hline\noalign{\vskip1pt}
$ \Delta M_{\kst}$  &  $ \Delta \Gamma_{\kst}$  &  $\Delta r_{BW}$  &  $\Delta m_{A}$  &  $ \Delta r_{V} $  &  $\Delta r_{2} $  &  $\Delta r_{S}$  & $\Delta r^{(1)}_{S}$  & $\Delta a_{S,BG}^{1/2}$  &  $\Delta r_{\kstp}$ &  $\Delta \delta_{\kstp}$ &  $\Delta N_{S}$  &  $ \Delta N_{B}$   \\
\hline
\textbf{0.226} &  ~0.569 & -0.827 & -0.004 &  ~0.153 & -0.068&  -0.080 &  ~0.115 &  ~0.038&   ~0.731&  -0.384&    -0.422 &   ~0.439\\
 &\textbf{0.241}& -0.579 & ~0.081 & ~0.065 & ~0.049 & -0.113 & ~0.137 & ~0.001 &~0.436 & -0.400 & -0.147 & ~0.159\\
 &&\textbf{0.540} & -0.102 & -0.182 & -0.027 & ~0.125 & -0.153 & -0.045 & -0.728& ~0.470 & ~0.620 & -0.642\\
&&&\textbf{0.124} & -0.095 & ~0.697 &-0.279 & ~0.260 & ~0.071 & ~0.029 & ~0.044  & -0.216 & ~0.229\\
&&&&\textbf{0.0308} & -0.367 & -0.358 & ~0.360 & -0.080 & ~0.063 & ~0.179 & -0.264 & ~0.276\\
&&&&&\textbf{0.0197} & ~0.082 & -0.102 & ~0.110 & ~0.022 &  -0.186 & -0.067 & ~0.067\\
&&&&&&\textbf{0.0406} & -0.988 & ~0.407 & ~0.119 & -0.078 & ~0.461 & -0.484\\
&&&&&&&\textbf{0.0810} & -0.395 & -0.096 & ~0.035 & -0.448 & ~0.471\\
&&&&&&&&\textbf{0.474} & ~0.357 & ~0.335 & ~0.121 & -0.126\\
&&&&&&&&& \textbf{0.0222} & -0.137 & -0.346  & ~0.355\\
&&&&&&&&&& \textbf{13.2} & ~0.158 & -0.155\\
 &&&&&&&&&&&\textbf{1055.2} & -0.918\\
&&&&&&&&&&&& \textbf{1033.8}\\

\hline\hline
\end{tabular}
}
\end{center}
\end{table*}

%------------------------------------------------------------------------

\end{document}

%% file: authors_jul2010.tex
%% author list as of 08-Jul-2010 (435 authors)
%
\author{P.~del~Amo~Sanchez}
\author{J.~P.~Lees}
\author{V.~Poireau}
\author{E.~Prencipe}
\author{V.~Tisserand}
\affiliation{Laboratoire d'Annecy-le-Vieux de Physique des Particules (LAPP), Universit\'e de Savoie, CNRS/IN2P3,  F-74941 Annecy-Le-Vieux, France}
\author{J.~Garra~Tico}
\author{E.~Grauges}
\affiliation{Universitat de Barcelona, Facultat de Fisica, Departament ECM, E-08028 Barcelona, Spain }
\author{M.~Martinelli$^{ab}$}
\author{D.~A.~Milanes}
\author{A.~Palano$^{ab}$ }
\author{M.~Pappagallo$^{ab}$ }
\affiliation{INFN Sezione di Bari$^{a}$; Dipartimento di Fisica, Universit\`a di Bari$^{b}$, I-70126 Bari, Italy }
% ...
\collaboration{The \babar\ Collaboration}
\noaffiliation

%% file: acknowledgements.tex
We are grateful for the
extraordinary contributions of our \pep2\ colleagues in
achieving the excellent luminosity and machine conditions
that have made this work possible.
The success of this project also relies critically on the
expertise and dedication of the computing organizations that
support \babar.
The collaborating institutions wish to thank
SLAC for its support and the kind hospitality extended to them.
This work is supported by the
US Department of Energy
and National Science Foundation, the
Natural Sciences and Engineering Research Council (Canada),
the Commissariat \`a l'Energie Atomique and
Institut National de Physique Nucl\'eaire et de Physique des Particules
(France), the
Bundesministerium f\"ur Bildung und Forschung and
Deutsche Forschungsgemeinschaft
(Germany), the
Istituto Nazionale di Fisica Nucleare (Italy),
the Foundation for Fundamental Research on Matter (The Netherlands),
the Research Council of Norway, the
Ministry of Education and Science of the Russian Federation,
Ministerio de Ciencia e Innovaci\'on (Spain), and the
Science and Technology Facilities Council (United Kingdom).
Individuals have received support from
the Marie-Curie IEF program (European Union), the A. P. Sloan Foundation (USA)
and the Binational Science Foundation (USA-Israel).

%% file: paper.bbl
\begin{thebibliography}{99}
\bibitem{ref:angles}
The convention used by FOCUS to define the Breit-Wigner amplitude
differs by a (-) sign from the ``usual'' expression.
By ``usual'' it is meant a phase which grows from zero to $\pi$
when the mass increases and which is equal to $\pi/2$ at the pole.
In the present analysis we adopt this ``usual'' definition (Eq.~(\ref{eq:kstar2})) which is also
used in Ref. \cite{ref:pdg10}.
\bibitem{ref:hemass}
The hemisphere mass is
the mass of the system corresponding to the sum of the 4-vectors of
particles contained in that hemisphere.
\end{thebibliography}
